\documentclass[11pt,a4paper]{article}
\usepackage[a4paper,
            left=3cm,
            right=3cm,
            top=3cm,
            bottom=3cm]{geometry}

\usepackage{setspace}
\usepackage{mathtools}
\usepackage{chngcntr}
\usepackage{enumitem}
\usepackage{amssymb}
\usepackage{amsthm}

\usepackage[makeroom]{cancel}

\usepackage{multirow}

\usepackage{titling}
\usepackage{float}
\usepackage{pdfpages}
\usepackage[T1]{fontenc}
\usepackage{tikz}
\usetikzlibrary{arrows.meta, matrix, decorations.pathreplacing}
\usetikzlibrary{positioning}
\usetikzlibrary{calc}
\usepackage{array}
\newcolumntype{?}{!{\vrule width 1pt}}
\usepackage{caption}
\usepackage{subfig}
\usepackage[bottom]{footmisc}
\usepackage{comment}
\usepackage{accents}
\usepackage[
    backend=biber,
    style=authoryear,
    giveninits=true,
    eprint=true,
    dashed=false,
    url=false,
    doi=false
]{biblatex}

\DeclareFieldFormat[techreport, report]{title}{#1} 
\AtEveryBibitem{\clearfield{type}}
\DeclareFieldFormat{eprint:arxiv}{\mkbibemph{arXiv\addcolon\space#1}}
\renewbibmacro{in:}{} 

\AtEveryBibitem{\clearfield{note}}
\AtEveryBibitem{\clearfield{month}}
\AtEveryCitekey{\clearfield{month}}
\AtBeginBibliography{\setlength{\itemsep}{7pt}}

\usepackage[hidelinks]{hyperref}
\usepackage{bookmark}

\title{\LARGE Nonlinear Dynamic Factor Analysis\\ With a Transformer Network}
\author{\large Oliver Snellman\thanks{University of Helsinki, oliver.snellman@gmail.com. The work was partially conducted while visiting at the University of Pennsylvania. I thank Professor Francis X. Diebold for encouraging me to focus on dynamic factors, while I was developing a Transformer framework for estimating latent economic variables. I thank Professor Jesús Fernández-Villaverde for comments on the Monte Carlo design. Senior Lecturer Niko Hauzenberger and Assistant Professor Dominik Baumann provided helpful feedback on the manuscript. The idea of also evaluating Attention patterns implied by conventional models arose in a discussion with Professor Hannu Vartiainen. Professor Antti Ripatti suggested the empirical application. Funding from the Yrjö Jahnsson Foundation, the OP Research Foundation and the Emil Aaltonen Foundation is gratefully acknowledged. This manuscript still evolves; the first version is dated 26 October 2024 and was previously distributed under the title "Nowcasting with a Transformer Network".}}
\date{\today}

\begin{document}

\maketitle

\begin{center}
\textbf{Abstract}
\end{center}

\noindent
The paper develops a Transformer architecture for estimating dynamic factors from multivariate time series data under flexible identification assumptions. Performance on small datasets is improved substantially by using a conventional factor model as prior information via a regularization term in the training objective. The results are interpreted with Attention matrices that quantify the relative importance of variables and their lags for the factor estimate. Time variation in Attention patterns can help detect regime switches and evaluate narratives. Monte Carlo experiments suggest that the Transformer is more accurate than the linear factor model, when the data deviate from linear-Gaussian assumptions. An empirical application uses the Transformer to construct a coincident index of U.S. real economic activity.

\vspace{0.5em}

\vspace{0.3em}

\vspace{1cm}

\section{Introduction}

There are three methodological contributions. The first contribution is to design a Transformer architecture that can estimate latent dynamic factors from time series data. A Transformer neural network is a flexible machine learning algorithm for processing information, proposed by \textcite{Vaswani2017}. Time series Transformers have mostly been customized for predicting future values of observable variables. Even Transformers that relate to state-space modeling, such as \textcite{tang2021probabilistic}, or those that compress features from the input data, such as \textcite{Zhou2021informer}, tend to use these internal representations just to aid prediction. Going further, the Transformer in this study compresses the input data down to one real number per lag position, with an interpretation as a latent state variable or a dynamic factor. The factor estimates are then re-embedded and used to predict observable variables. The prediction errors are only used to condition the factor estimates on data during the training of the Transformer. To the best of my knowledge, Transformers have not been used for nonlinear dynamic factor analysis before.

The second contribution relates to improving the results on small datasets with prior information. A Transformer usually requires a large training or pre-training dataset to function properly, whereas typical macroeconomic datasets tend to be small, at least in the time dimension. As a remedy, I use conventional factor models as prior information to guide the training of the Transformer. Prior information is supplied through a regularization term, which is added to the prediction-error loss function. This regularization minimizes the distance between two factor estimates, one obtained from the Transformer and the other from a chosen conventional model. When a linear factor model is used as prior information, the Transformer is incentivized to learn the linear approximation. If the relative weight of prior information is modest, the Transformer can also find unspecified nonlinearities, which help minimize the prediction-error further. Training the Transformer with prior information helps narrow down its large parameter space in a useful manner, which would otherwise be hard with limited training examples. Prior information is found to improve the results on small datasets.

The third contribution is the development of time-series-specific tools for interpreting results with Attention patterns. The defining feature of a Transformer is the Attention mechanism, which connects each input token to all others to learn which relationships matter. With granular tokens, the Attention mechanism quantifies the contribution of every variable-lag pair to the factor estimate in each period. I analyze how the Attention patterns evolve over time and show how these changes can help detect regime switches, analyze shocks and assess narratives.

The Transformer is put to the test in a series of Monte Carlo experiments. Behind each simulated dataset, a state-transition equation determines the stochastic process for one latent factor, and a measurement equation describes the noisy mapping between the factor and several observed variables. The aim is to estimate the factor's values based on the observable variables, without any additional information about the underlying processes. Each subsequent dataset deviates from linearity and Gaussianity in different ways, cumulating complexity. Only 800 time periods are used for training per dataset, to reflect the typical scarcity of macro data. The Transformer is found to outperform the Kalman filter on most of the datasets, often by 20\% more accurate factor estimates. The results are summarized in Table~\ref{tab:results}.

In an empirical application, I use the Transformer to construct a coincident index of real business-cycle conditions in the U.S. economy. The dynamic factor is estimated from variables related to industrial production, sales, labor income, and hours worked, as proposed in a seminal study by \textcite{stock1989new}. I compare the resulting index to a linear factor model during three recent recessions, and to recession dates published by the NBER Business Cycle Dating Committee. The Transformer based index is mainly consistent with the linear benchmark and the NBER recession chronology.

Dynamic factors are important in macroeconomic analysis and policy work. Factors can be used for monitoring the economy, like the Aruoba--Diebold--Scotti business conditions index by \textcite{diebold2009real}, for compressing data, as with factor-augmented vector autoregression (FAVAR) by \textcite{bernanke2005measuring}, and for nowcasting real GDP, as in \textcite{giannone2008nowcasting}. It is typical to use a factor model with linear--Gaussian state space representation and the Kalman filter procedure by \textcite{kalman1960new} to estimate dynamic factors. This is also used as a baseline model in this study, referred to by the shorthand "Kalman filter factor estimate". Macroeconomic processes are rarely linear or Gaussian. The Kalman filter can be augmented in several ways to relax the linearity assumption \parencite{sarkka2023bayesian} and the particle filter is capable of dealing with nonlinear and non-normal processes, see \textcite{fernandez2005estimating} for the macroeconomic context. Nonetheless, it holds for all conventional methods that the model has to be specified before it can be estimated. This gives rise to the \textit{model selection problem}: How to specify the functional form and stochasticity, so that they reasonably describe the data generating process?

The Transformer offers an attractive alternative to conventional factor models by improving on the model selection problem. Given a misspecified baseline model that is used as prior information, the Transformer can find \textit{unspecified} complex nonlinear conditional patterns between variables and across time to improve the accuracy beyond the baseline. The functional forms of the state-transition and measurement equations do not require any rigid assumptions beyond the number of factors to be estimated. Because all models are misspecified in the real world, the chosen conventional model's accuracy could potentially be improved by using it in the Transformer framework as prior information.

There are several benefits to using Transformers for factor analysis, compared also to other machine learning alternatives. The Attention mechanism makes the output of a Transformer more transparent and easier to interpret than the output from a recurrent network, such as the Long Short Term Memory (LSTM) by \textcite{Hochreiter1997LSTM}. An Autoencoder based on a vanilla feed forward network lacks the power of Transformer modules and interpretability of the Attention mechanism. I propose new time series specific Attention based tools in Subsection~\ref{subsec:attention} to help analyze the results. A Transformer can also find dependencies over longer horizons than the LSTM network, because the Attention mechanism does not suffer from vanishing gradients. Additionally, it is possible to specify a sharp lag length for the Transformer, which is useful when the results are compared to other models with the same information set, such as the vector autoregression.

The rest of the Introduction highlights the different choices in the proposed Transformer to other existing Transformers, and explains how the differences allow and improve upon estimating factors from small macroeconomic datasets.

The definition of a token, the partition of data into basic computational units, is chosen to be one variable's one lag. A similar choice was made in \textcite{zerveas2020multivariate}. This choice is maximally granular and improves the interpretability of results at the expense of computational cost, differing from what is popular in the literature on time series Transformers. Patching by \textcite{nie2022time} treats a set of subsequent values of one variable as a token, which allows increasing the length of the input sequences processed at once. Patching typically leads to univariate analysis, where the cross-correlations between variables are not accounted for explicitly. On the other hand, typical feature-mixing models treat all variable values from one time period as one token, as in \textcite{Zhou2021informer}. Both token definitions make it possible to pack more information into the context window of a Transformer, at the cost of interpretability. This is desirable when there is an abundance of data, and when only the prediction accuracy matters. The situation is very different with scarce macroeconomic data, where the computational constraints are hardly ever a problem. This allows the use of more granular tokens to capture multivariate dynamics and to improve the researcher's ability to interpret the results. With the token definition used in this study, the Transformer resembles a nonlinear vector autoregression model. It is possible to analyze how much the Transformer paid attention to each variable's every lag, when it created the factor estimate.

The Transformer has an Encoder--Encoder architecture; see Figure~\ref{fig1:transformer}. In the beginning, an initial representation is created for the factor estimate series. The first, State Encoder, then refines this factor series using information in the data. The refined factor estimate is then used in the second, Measurement Encoder, to predict the data onward. It is a crucial feature of the algorithm that the prediction error heavily relies on the factor estimate. This informational bottleneck is further enforced by omitting the residual skip around the Measurement Attention. Minimizing the prediction error loss in training teaches the Transformer how to construct the factor estimate from data. More precisely, the State Encoder functions as a nonlinear filter, which infers a state-transition dynamics from the data. The Measurement Encoder uses the estimated factor series and the observable variables up to period t, to predict the values of the observable variables at period $t+1$. The outcome of interest comes from the State Encoder of a trained Transformer, the dynamic factor estimate. The Monte Carlo only includes one-factor datasets, but it is technically straightforward to extend the Transformer to estimate multiple factors at the same time, albeit at the cost of additional identification challenges.

The key benefit of creating an initial factor representation prior to the State Encoder is the enhanced interpretability of results. The State Attention gets the Queries from the factor series and the Keys and Values from the data. Therefore the Attention matrices directly describe the construction of the factor estimate, based on the observable variables and their lags. Additionally, the output layers after both of the Encoder stacks only compress each token representation from a vector back into a real number, without combining information across tokens. This ensures that most of the heavy lifting is done by the Encoders, especially the Attention mechanism. Attention is the only mechanism, where different variables and their lags are mapped directly into each other, to find out which connections matter.

The reason for specifying two Encoder stacks instead of a more typical Encoder-Decoder architecture stems from the specialties of macroeconomic data, and from the emphasis on estimating factors instead of predicting observables. Transformers operate on fixed-size context windows, meaning that a Transformer receiving $P$ lags will output $P$ next token predictions simultaneously for the $k$ observable variables. This means that when the context window includes positions $t, \dots, t-P$, then the outputs from the Measurement Encoder correspond to predictions for $t+1,\dots,t-P+1$. Only the last $t+1$ period predictions are used in the loss function during training, as they are based on the whole $P$-lag information set. The rest of the $P-1$ predictions are discarded. If the input sequences are long, conducting the forward pass this way becomes computationally costly, and instead it would be reasonable to use all $P$ predictions in the loss function, called teacher forcing. Teacher forcing requires using a measurement Decoder instead of an Encoder to generate the predictions, which uses a mask to guarantee that each prediction is only based on past information. Additional to the computational benefit of a Measurement Decoder, the Transformer could also learn to work with variable length information sets, but this will not be required at the inference stage in macroeconomic factor analysis. Furthermore, the predictions based on smaller information sets are likely inferior in quality to those predictions that are based on the full information set, introducing noise to the training process. Because of the relatively small input sequences in typical macroeconomic applications, it is computationally feasible to only use the last prediction for position $t+1$ in the loss function, to improve the quality of the parameter updates in training. The last prediction has only attended to past values, and therefore it can be obtained from an unmasked Encoder. Using a Measurement Encoder allows for richer dynamic interaction between the factor estimate and observable variables across the whole $P$ lag context window.

The next Subsection~\ref{sec:literature} gives a general overview of dynamic factor models and Transformers. Section~\ref{sec:model} details the proposed Transformer algorithm and the training process. Subsection~\ref{subsec:loss} describes how prior information is supplied to the loss function to guide the training. Subsection~\ref{subsec:attention} presents the tools for analyzing the results, based on the Attention mechanism. Section~\ref{sec:montecarlo} conducts a Monte Carlo experiment to evaluate the performance of the Transformer, and analyzes the impact of prior information on the results. Section~\ref{sec:empirical} applies the Transformer to U.S. data. Section~\ref{sec:discussion} discusses the upsides and improvement areas of conducting factor analysis with the Transformer and the Section~\ref{sec:conclusion} concludes.

\subsection{Related literature} \label{sec:literature}

The idea of distilling information from multiple relevant variables into a latent factor that represents an underlying phenomenon was formalized by \textcite{spearman1904iq}, teasing out a measure for general intelligence in humans. The concept of a factor has since become a central tool in many fields, including economics. The importance of using the co-movement across many macroeconomic variables in assessing the business cycle was emphasized already by \textcite{burns1946measuring}. Early dynamic factor models were developed for time series analysis by \textcite{geweke1977dynamic}, \textcite{sargent1977business} and \textcite{stock1989new}. Some recent efforts aim to increase the flexibility of the conventional filtering approach \parencite{Villaverde2024filtering}.

The earlier machine learning paradigm for sequential data was centered around the Recurrent neural networks, Long Short Term Memory (LSTM) and Gated Recurrent Unit (GRU), going back to \textcite{Hochreiter1997LSTM}. LSTM networks performed well with larger datasets and GRU networks were functional even with smaller datasets. The paradigm changed quickly after the publication of the Transformer white paper by \textcite{Vaswani2017}. Currently most contemporary large language models are based on Transformers, including the Generative Pre-trained Transformer (ChatGPT) by \textcite{gpt2023}, which has excelled in natural language translation, sentiment analysis and text production. It speaks to the flexibility of Transformers that they have been found useful in modeling spatial inputs such as pictures and video, sequential inputs such as language, sound and time series data, and even tabular data, consisting of discrete, ordinal and categorical variables \parencite{Huang2020tabtransformer}. Due to the lack of recurrence, the estimation of Transformers can be parallelized, enabling extensive pre-training using GPUs. This has been one of the key elements behind the ability of Transformers to generalize their success to new datasets with little to no additional fine-tuning. The recurrent structure made it computationally infeasible to pre-train the LSTM and GRU models in a similar manner, contributing to their inability to generalize.

The particular architecture for a Transformer depends on the task at hand. The key operations of a Transformer can be collected into Decoder and Encoder blocks, where the main difference is analogous to filtering and smoothing, respectively; whether it is permitted or not for an input to access its successors within a context window in the Attention mapping. Many well known generative Transformers such as LLaMA by Meta AI, \parencite{touvron2023llama}, use a Decoder-only structure. As the idea of those models is to extrapolate the input prompt, only one backwards looking Decoder is needed. In tasks involving time series prediction where the same sequence is extrapolated further, a Decoder-only architecture can also suffice. The Bidirectional Encoder Representations from Transformers (BERT by Google, \textcite{Devlin2018bert}) is an Encoder-only. This structure focuses on distilling information from the whole input sequence at once for the output, like in sentiment analysis or imputing missing values. The original Encoder-Decoder structure by \textcite{Vaswani2017} was meant for a language translation task, where the whole input sequence in one language can be first analyzed in the Encoder and then the output sequence is generated based on the Encoder's instructions word by word in the Decoder. This architecture was the inspiration for the Transformer presented here.

\textcite{gruver2023large} show that large language models like LLaMA-2 and GPT-3, which have been pre-trained using natural language data, can nonetheless be used as such also for predicting time series data. Google research improved the results by training a Transformer foundation model specifically for time series prediction, with 200 million parameters and 100 gigabytes of pre-training data \parencite{das2023decoder}.

In economics Transformers have been so far mainly applied to univariate problems, like predicting the S\&P 500 in \textcite{Wang2022stock}, where the Transformer is found to surpass the LSTM network. Economic multivariate applications include a Transformer for portfolio allocation by \textcite{lezmi2023time}.

There are some other time series Transformers, which also use a dual structure of Encoder and Decoder stacks, such as the Informer by \textcite{Zhou2021informer}. The Encoder is used to compress features from data that help in long range predictions. The number of features is typically much larger than the number of potential factors driving the data. These vector-valued features are not assigned with an interpretation as latent variables or extracted out to be used in themselves as objects of interest. \textcite{tang2021probabilistic} combine a state-space structure with a Transformer network, but yet again, with the aim of improving forecasts instead of estimating scalar latent factors. Additionally, \textcite{sarapisto2024subsystem} use a masked-autoencoder Transformer to cluster variables in a high-dimensional dataset into functional subsystems. They compress the data into latent embeddings which summarize the clustering structure, but not into dynamic factors which summarize the evolution of the system over time. I am not aware of prior work that uses dual-stack Transformers for explicitly estimating dynamic factors as latent real-valued time series variables.

The proposed prior information regularizer relates to the subspace shrinkage literature, which penalizes high-dimensional models typically towards a linear subspace within their parameter space, see for example \textcite{Shin2020}, \textcite{Huber2023}, and \parencite{Sewell2025}. The prior information regularizer can use any (nonlinear) pre-estimated factor model as a subspace prior. To the best of my knowledge, conventional models have not been utilized in this way before to train Transformers.

Recently there have been innovative efforts to make better use of the machine learning approach in macroeconometric modeling. For example, \textcite{chernis2025bayesian} uses Gaussian process priors to estimate time-varying factor loadings with a dynamic factor model. \textcite{Klieber2024} uses an autoencoder to augment a vector autoregression model with a nonlinear factor. \textcite{coulombe2022neural} estimates latent inflation expectations and output gap using a neural network.
\textcite{hauzenberger2025machine} use a Bayesian neural network to estimate nonlinear impulse responses to structural shocks in macro data. My aim is to establish foundations for using Transformers in macroeconomic modelling, with an emphasis on latent-state estimation and interpretable dynamics.

\section{Methodology} \label{sec:model}

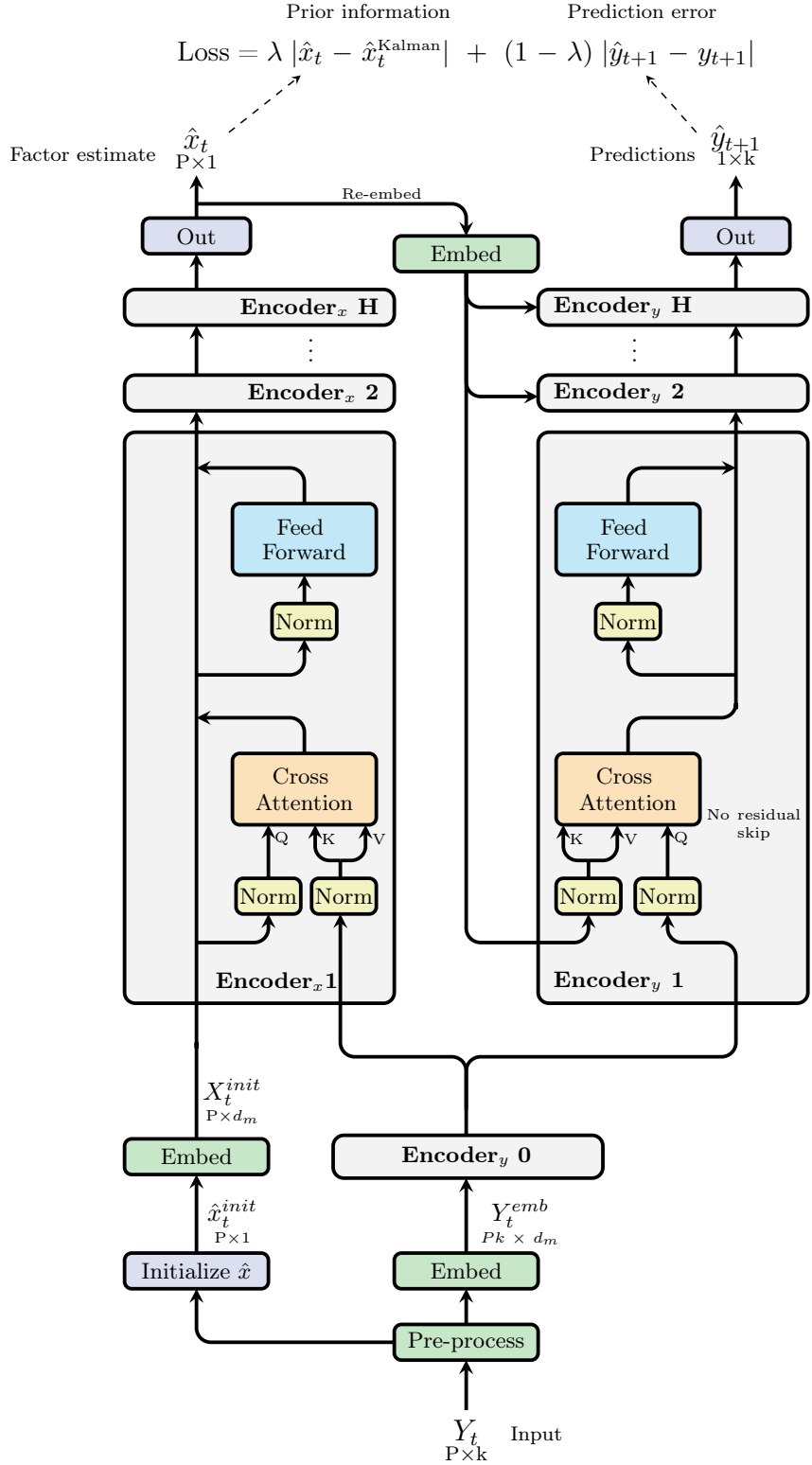
\begin{figure}[!htbp]
    \centering

\begin{tikzpicture}

\definecolor{emb_color}{RGB}{200,230,200}
\definecolor{multi_head_attention_color}{RGB}{252,226,187}
\definecolor{add_norm_color}{RGB}{242,243,193}
\definecolor{ff_color}{RGB}{194,232,247}
\definecolor{softmax_color}{RGB}{203,231,207}
\definecolor{linear_color}{RGB}{220,223,240}
\definecolor{encoder_color}{RGB}{200,230,240}
\definecolor{grey_color}{RGB}{243,243,244}

\footnotesize


\node[anchor=west] at (5.5, -1.05) {\scriptsize Input};
\node[text width=2.5cm, anchor=south, align=center] at (5,-1.3) {\large $Y_t$};
\node[text width=2.5cm, anchor=south, align=center] at (5,-1.55) {\scriptsize P$\times$k};

\draw[line width=0.05cm, fill=emb_color, rounded corners=0.1cm] (4, 0.5) -- (6, 0.5) -- (6, 0) -- (4, 0) -- cycle;
\node[text width=2.5cm, align=center] at (5, 0.25) {Pre-process};

\draw[line width=0.05cm, fill=emb_color, rounded corners=0.1cm] (4, 1.5) -- (6, 1.5) -- (6, 1) -- (4, 1) -- cycle;
\node[text width=2.5cm, align=center] at (5,1.25) {Embed};

\node[text width=2.5cm, anchor=south, align=center] at (5.75, 1.75) {\small $Y_t^{emb}$};
\node[text width=2.5cm, anchor=south, align=center] at (5.75, 1.5) {\tiny $Pk\times d_m$};

\draw[line width=0.05cm, fill=grey_color, rounded corners=0.15cm] (3.15, 3.15) -- (6.9, 3.15) -- (6.9, 2.55) -- (3.15, 2.55) -- cycle;
\node[text width=2.5cm, align=center] at (5, 2.85) {$\textbf{Encoder}_{y}$ \textbf{0}};

\draw[line width=0.05cm, fill=emb_color, rounded corners=0.1cm] (0.25, 3.1) -- (2.25, 3.1) -- (2.25, 2.6) -- (0.25, 2.6) -- cycle;
\node[text width=2.5cm, align=center] at (1.25, 2.85) {Embed};

\node[text width=2.5cm, anchor=south, align=center] at (1.75, 1.75) {\small $\hat x_t^{init}$};
\node[text width=2.5cm, anchor=south, align=center] at (1.75, 1.5) {\tiny P$\times$1};

\draw[line width=0.05cm, fill=linear_color, rounded corners=0.1cm] (0.25, 1.5) -- (2.25, 1.5) -- (2.25, 1) -- (0.25, 1) -- cycle;
\node[text width=2.5cm, align=center] at (1.25,1.25) {Initialize $\hat x$};

\node[text width=2.5cm, anchor=south, align=center] at (1.75, 3.45) {\small $X_t^{init}$};
\node[text width=2.5cm, anchor=south, align=center] at (1.75, 3.2) {\tiny P$\times$$d_m$};

\draw[line width=0.05cm, -stealth] (5, -0.7) -- (5, 0);
\draw[line width=0.05cm, -stealth] (5, 0.5) -- (5, 1);
\draw[line width=0.05cm, -stealth] (5, 1.5) -- (5, 2.55);

\draw[fill=grey_color, line width=0.05cm, rounded corners=0.15cm] (0.25, 13) -- (4, 13) -- (4, 5) -- (0.25, 5) -- cycle;
\draw[fill=grey_color, line width=0.05cm, rounded corners=0.15cm] (0.25, 13.8) -- (4, 13.8) -- (4, 13.3) -- (0.25, 13.3) -- cycle;
\draw[fill=grey_color, line width=0.05cm, rounded corners=0.15cm] (0.25, 15) -- (4, 15) -- (4, 14.5) -- (0.25, 14.5) -- cycle;

\node[anchor=east] at (3.335, 5.3) {\textbf{Encoder$_x$1}};
\node[anchor=east] at (3.9, 13.55) {\textbf{Encoder$_x$ 2}};
\node[anchor=east] at (3, 14.25) {$\vdots$};
\node[anchor=east] at (3.9, 14.75) {\textbf{Encoder$_x$ H}};

\draw[-stealth, line width=0.05cm, rounded corners=0.2cm] (4, 0.25) -- (1.25, 0.25) -- (1.25, 1);
\draw[-stealth, line width=0.05cm, rounded corners=0.2cm] (1.25, 1.5) -- (1.25, 2.6);
\draw[-stealth, line width=0.05cm, rounded corners=0.2cm] (5, 3.15) -- (5, 4.25) -- (3.25, 4.25) -- (3.25, 6.25);

\draw[-stealth, line width=0.05cm, rounded corners=0.2cm] (1.25,15) -- (1.25, 15.5);
\draw[-stealth, line width=0.05cm, rounded corners=0.2cm] (1.25, 3.1) -- (1.25, 4.25) -- (1.25, 4.25) -- (1.25, 13.3);

\draw[fill=grey_color, line width=0.05cm, rounded corners=0.15cm] (6, 13) -- (9.75, 13) -- (9.75, 5) -- (6, 5) -- cycle;
\draw[fill=grey_color, line width=0.05cm, rounded corners=0.15cm] (6, 13.8) -- (9.75, 13.8) -- (9.75, 13.3) -- (6, 13.3) -- cycle;
\draw[fill=grey_color, line width=0.05cm, rounded corners=0.15cm] (6, 15) -- (9.75, 15) -- (9.75, 14.5) -- (6, 14.5) -- cycle;

\node[anchor=west] at (6.1, 5.3) {\textbf{Encoder$_y$ 1}};
\node[anchor=west] at (6.1, 13.55) {\textbf{Encoder$_y$ 2}};
\node[anchor=east] at (7.5, 14.25) {$\vdots$};
\node[anchor=west] at (6.1, 14.75) {\textbf{Encoder$_y$ H}};

\draw[-stealth, line width=0.05cm, rounded corners=0.2cm] (5, 3.5) -- (5, 4.25) -- (8.75, 4.25) -- (8.75, 5.85) -- (7.8, 5.85) -- (7.8, 6.25);
\draw[-stealth, line width=0.05cm, rounded corners=0.2cm] (8.75,13.8) -- (8.75, 14.5);
\draw[-stealth, line width=0.05cm, rounded corners=0.2cm] (8.75,15) -- (8.75, 15.5);

\draw[-stealth, line width=0.05cm, rounded corners=0.2cm] (1.25,13.8) -- (1.25, 14.5);


\draw[-stealth, line width=0.05cm, rounded corners=0.2cm] (1.25, 5.85) -- (2.25, 5.85) -- (2.25, 6.25);

\draw[line width=0.05cm, fill=add_norm_color, rounded corners=0.1cm] (1.8, 6.75) -- (2.7, 6.75) -- (2.7, 6.25) -- (1.8, 6.25) -- cycle;
\node[text width=2.5cm, align=center] at (2.25,6.5) {Norm};

\draw[line width=0.05cm, fill=add_norm_color, rounded corners=0.1cm] (2.85, 6.75) -- (3.75, 6.75) -- (3.75, 6.25) -- (2.85, 6.25) -- cycle;
\node[text width=2.5cm, align=center] at (3.3,6.5) {Norm};

\draw[-stealth, line width=0.05cm, rounded corners=0.2cm] (2.25, 6.75) -- (2.25, 7.5);
\draw[-, line width=0.05cm, rounded corners=0.2cm] (3.25, 6.75) -- (3.25, 7);
\draw[-stealth, line width=0.05cm, rounded corners=0.2cm] (3.25, 7) -- (2.9, 7) -- (2.9, 7.5);
\draw[-stealth, line width=0.05cm, rounded corners=0.2cm] (3.25, 7) -- (3.6, 7) -- (3.6, 7.5);

\node[anchor=east] at (2.65,7.3) {\tiny Q};
\node[anchor=east] at (3.3,7.3) {\tiny K};
\node[anchor=east] at (4,7.3) {\tiny V};

\draw[line width=0.05cm, fill=multi_head_attention_color, rounded corners=0.1cm] (1.75, 8.5) -- (3.75, 8.5) -- (3.75, 7.5) -- (1.75, 7.5) -- cycle;
\node[text width=2.5cm, align=center] at (2.75,8) { Cross \vspace{-0.02cm} \linebreak Attention};

\draw[-stealth, line width=0.05cm, rounded corners=0.2cm] (2.75, 8.5) -- (2.75, 9) -- (1.25, 9);

\draw[-stealth, line width=0.05cm, rounded corners=0.2cm] (1.25, 9.6) -- (2.75, 9.6) -- (2.75, 10.1);

\draw[line width=0.05cm, fill=add_norm_color, rounded corners=0.1cm] (2.3, 10.6) -- (3.2, 10.6) -- (3.2, 10.1) -- (2.3, 10.1) -- cycle;
\node[text width=2.5cm, align=center] at (2.75,10.35) {Norm};

\draw[-stealth, line width=0.05cm, rounded corners=0.2cm] (2.75, 10.6) -- (2.75, 11);

\draw[line width=0.05cm, fill=ff_color, rounded corners=0.1cm] (1.75, 12) -- (3.75, 12) -- (3.75, 11) -- (1.75, 11) -- cycle;
\node[text width=2.5cm, align=center] at (2.75,11.5) {Feed \vspace{-0.05cm} \linebreak Forward};

\draw[-stealth, line width=0.05cm, rounded corners=0.2cm] (2.75, 12) -- (2.75, 12.5) -- (1.25, 12.5);

\draw[line width=0.05cm, fill=linear_color, rounded corners=0.1cm] (0.5, 16) -- (2, 16) -- (2, 15.5) -- (0.5, 15.5) -- cycle;
\node[text width=2.5cm, align=center] at (1.25,15.75) {Out};

\draw[line width=0.05cm, fill=emb_color, rounded corners=0.1cm] (4, 15.75) -- (6, 15.75) -- (6, 15.25) -- (4, 15.25) -- cycle;
\node[text width=2.5cm, align=center] at (5, 15.5) {Embed};

\draw[-stealth, line width=0.05cm, rounded corners=0.2cm] (1.25, 16.2) -- (5, 16.2) -- (5, 15.75);
\draw[-stealth, line width=0.05cm, rounded corners=0.2cm] (5, 15.25) -- (5, 5.85) -- (6.7, 5.85) -- (6.7, 6.25);
\node[anchor=east] at (4.5, 16.35) {\tiny Re-embed};

\draw[-stealth, line width=0.05cm, rounded corners=0.2cm] (5, 15.25) -- (5, 13.5) -- (6, 13.5);

\draw[-stealth, line width=0.05cm, rounded corners=0.2cm] (5, 15.25) -- (5, 14.75) -- (6, 14.75);

\draw[-stealth, line width=0.05cm] (1.25, 16) -- (1.25, 16.6);

\node[text width=2.5cm, anchor=south, align=center] at (1.25,16.8) {\large $\hat x_{t}$};
\node[text width=2.5cm, anchor=south, align=center] at (1.25,16.55) {\scriptsize P$\times$1};
\node[anchor=east] at (0.8, 16.9) {\scriptsize Factor estimate};




\draw[line width=0.05cm, fill=add_norm_color, rounded corners=0.1cm] (7.35, 6.75) -- (8.25, 6.75) -- (8.25, 6.25) -- (7.35, 6.25) -- cycle;
\node[text width=2.5cm, align=center] at (7.8, 6.5) {Norm};

\draw[-, line width=0.05cm, rounded corners=0.2cm] (6.7, 6.75) -- (6.7, 7);
\draw[-stealth, line width=0.05cm, rounded corners=0.2cm] (6.7, 7) -- (6.35, 7) -- (6.35, 7.5);
\draw[-stealth, line width=0.05cm, rounded corners=0.2cm] (6.7, 7) -- (7.1, 7) -- (7.1, 7.5);

\draw[line width=0.05cm, fill=add_norm_color, rounded corners=0.1cm] (6.25, 6.75) -- (7.15, 6.75) -- (7.15, 6.25) -- (6.25, 6.25) -- cycle;
\node[text width=2.5cm, align=center] at (6.7,6.5) {Norm};

\draw[-stealth, line width=0.05cm, rounded corners=0.2cm] (7.8, 6.75) -- (7.8, 7.5);

\node[anchor=east] at (6.75,7.3) {\tiny K};
\node[anchor=east] at (7.5,7.3) {\tiny V};
\node[anchor=east] at (8.2,7.3) {\tiny Q};

\draw[line width=0.05cm, fill=multi_head_attention_color, rounded corners=0.1cm] (6.25, 8.5) -- (8.25, 8.5) -- (8.25, 7.5) -- (6.25, 7.5) -- cycle;
\node[text width=2.5cm, align=center] at (7.25,8) {Cross \vspace{-0.02cm} \linebreak Attention};

\draw[-stealth, line width=0.05cm, rounded corners=0.2cm] (7.25, 8.5) -- (7.25, 9) -- (8.75, 9) -- (8.75, 9) -- (8.75, 13.3);

\draw[-stealth, line width=0.05cm, rounded corners=0.2cm] (8.75, 9.6) -- (7.25, 9.6) -- (7.25, 10.1);

\draw[line width=0.05cm, fill=add_norm_color, rounded corners=0.1cm] (6.8, 10.6) -- (7.7, 10.6) -- (7.7, 10.1) -- (6.8, 10.1) -- cycle;
\node[text width=2.5cm, align=center] at (7.25, 10.35) {Norm};

\draw[-stealth, line width=0.05cm, rounded corners=0.2cm] (7.25, 10.6) -- (7.25, 11);

\draw[line width=0.05cm, fill=ff_color, rounded corners=0.1cm] (6.25, 12) -- (8.25, 12) -- (8.25, 11) -- (6.25, 11) -- cycle;
\node[text width=2.5cm, align=center] at (7.25,11.5) {Feed \vspace{-0.05cm} \linebreak Forward};

\draw[-stealth, line width=0.05cm, rounded corners=0.2cm] (7.25, 12) -- (7.25, 12.5) -- (8.75, 12.5);

\draw[line width=0.05cm, fill=linear_color, rounded corners=0.1cm] (8, 16) -- (9.5, 16) -- (9.5, 15.5) -- (8, 15.5) -- cycle;
\node[text width=2.5cm, align=center] at (8.75,15.75) {Out};

\draw[-stealth, line width=0.05cm, rounded corners=0.2cm] (8.75, 16) -- (8.75, 16.6);

\node[text width=2.5cm, anchor=south, align=center] at (8.75,16.8) {\large $\hat y_{t+1}$};
\node[text width=2.5cm, anchor=south, align=center] at (8.75,16.6) {\scriptsize 1$\times$k};
\node[anchor=east] at (8.3, 16.9) {\scriptsize Predictions};

\draw[-stealth, line width=0.02cm, dashed, rounded corners=0.2cm] (1.7, 17.2) -- (2.7, 18);
\draw[-stealth, line width=0.02cm, dashed, rounded corners=0.2cm] (8.2, 17.2) -- (7.5, 18);
\node[text width=9cm, anchor=south, align=center] at (5,18) {\normalsize Loss = $\lambda \ |\hat x_{t} - \hat x_{t}^{\text{\tiny Kalman}}| \ + \ (1-\lambda) \ | \hat y_{t+1} - y_{t+1} |$};
\node[anchor=east] at (4.9, 18.9) {\scriptsize Prior information};
\node[anchor=east] at (8.6, 18.9) {\scriptsize Prediction error};

\node[text width=2.5cm, align=center] at (9,7.5) {\tiny{ No residual \vspace{-0.13cm} \linebreak skip}};

\end{tikzpicture}
    \captionsetup{width=\linewidth}
    \caption{The State Encoder stack on the left constructs a factor estimate from the data. The Measurement Encoder stack on the right is forced to rely heavily on the factor estimate when predicting the observable variables. The factor estimate is conditioned on data during training. Performance on small datasets is improved by using the linear factor model with the Kalman filter as prior information regularizer in the loss function with a weight $\lambda$. Figure by the author, adapted from an MIT-licensed \href{https://github.com/negrinho/sane_tikz/blob/fd6f291d9815613594d724678cb91ac9d412fbb7/examples/transformer.tex}{TikZ template}.}
\label{fig1:transformer}
\end{figure}

This section presents the Encoder-Encoder Transformer in detail\footnote{I chose a procedural way of explaining the algorithm to reduce the prerequisite knowledge requirements about machine learning and Transformers.}, and points out the differences from other Transformers. An outlook of the architecture is presented in Figure~\ref{fig1:transformer}. Subsections~\ref{subsec:pre}--\ref{subsec:measure} describe the forward pass -- how data is transformed into factor estimates and predictions. Subsection~\ref{subsec:loss} introduces a way to use prior information in the loss function, and Subsection~\ref{subsec:train} discusses the process of training the Transformer. Finally, Subsections~\ref{subsec:modelavg}--\ref{subsec:residual} detail the tools for analyzing the results.

\subsection{Pre-processing} \label{subsec:pre}

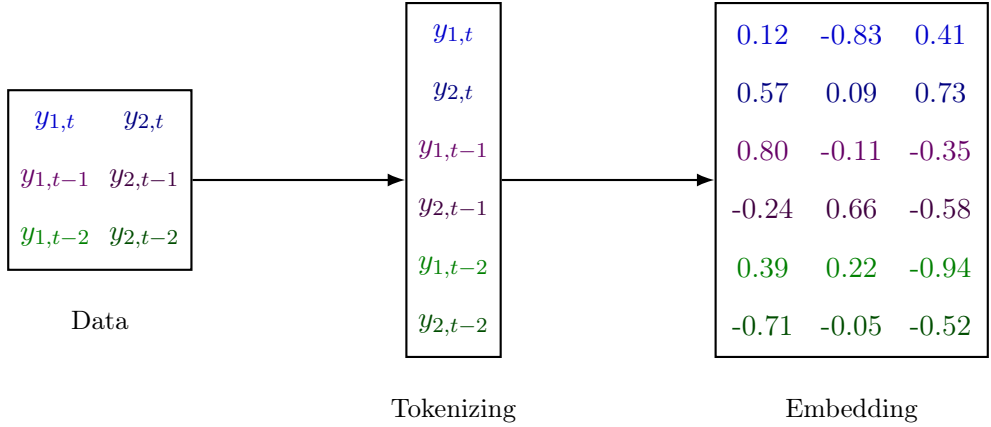
\begin{figure}[!htbp]
    \centering
\begin{tikzpicture}[
  >=Latex,
  lab/.style = {font=\small},
  mcell/.style = {minimum width=1.1cm, minimum height=7mm, anchor=center, inner sep=1pt},
  box/.style   = {draw, thick, inner sep=2pt}
]

\colorlet{tok1}{blue!80!black}
\colorlet{tok2}{blue!50!black}
\colorlet{tok3}{violet!80!black}
\colorlet{tok4}{violet!50!black}
\colorlet{tok5}{green!50!black}
\colorlet{tok6}{green!30!black}

\matrix (data) [matrix of nodes, nodes={mcell}, column sep=2pt, row sep=2pt, box]
{
  \textcolor{tok1}{$y_{1,t}$}   & \textcolor{tok2}{$y_{2,t}$}   \\
  \textcolor{tok3}{$y_{1,t-1}$} & \textcolor{tok4}{$y_{2,t-1}$} \\
  \textcolor{tok5}{$y_{1,t-2}$} & \textcolor{tok6}{$y_{2,t-2}$} \\
};
\node[lab, below=4mm of data.south] {Data};

\matrix (tok) [matrix of nodes, nodes={mcell}, row sep=2pt, right=2.8cm of data, box]
{
  \textcolor{tok1}{$y_{1,t}$}   \\
  \textcolor{tok2}{$y_{2,t}$}   \\
  \textcolor{tok3}{$y_{1,t-1}$} \\
  \textcolor{tok4}{$y_{2,t-1}$} \\
  \textcolor{tok5}{$y_{1,t-2}$} \\
  \textcolor{tok6}{$y_{2,t-2}$} \\
};
\node[lab, below=4mm of tok.south] {Tokenizing};

\draw[->, thick] (data.east) -- node[lab, above] {} (tok.west);

\matrix (emb) [matrix of nodes, nodes={mcell}, column sep=2pt, row sep=2pt, right=2.8cm of tok, box,
  row 1/.style={nodes={text=tok1}},
  row 2/.style={nodes={text=tok2}},
  row 3/.style={nodes={text=tok3}},
  row 4/.style={nodes={text=tok4}},
  row 5/.style={nodes={text=tok5}},
  row 6/.style={nodes={text=tok6}}
]
{
   0.12 & -0.83 &  0.41  \\
  0.57 &  0.09 &  0.73   \\
   0.80 & -0.11 & -0.35  \\
  -0.24 &  0.66 & -0.58  \\
   0.39 &  0.22 & -0.94  \\
  -0.71 & -0.05 &  -0.52 \\
};
\node[lab, below=4mm of emb.south] {Embedding};

\draw[->, thick] (tok.east) -- node[lab, above] {} (emb.west);

\end{tikzpicture}
    \caption{Example of pre-processing: Matrix of input data is vectorized into granular tokens, which are then embedded with representation vectors.}
    \label{fig1:tokenizing}
\end{figure}

Let a time series dataset consist of $i=1,2,\dots,k$ variables and $t=1,2,\dots,N$ time periods. The data is \textit{standardized} variable-wise to remove the impact of scale differences, $y_{ti} = \frac{\text{y}_{ti}-\bar {\textbf{y}_i}}{\sigma_{i}}$, with variable-wise mean $\bar {\textbf{y}_i} = \sum_{t=1}^N\frac{\text{y}_{ti}}{N}$ and standard deviation obtained from $\sigma_{i}^2 = \sum_{t=1}^N\frac{(\text{y}_{ti}-\bar {\textbf{y}_i})^2}{N}$. The data is tokenized, embedded, and augmented with positional and variable information. Figure~\ref{fig1:tokenizing} offers an intuition on the procedure.

\textit{Tokenizing} partitions a context window $Y_t$ of data into computational units, tokens. Defining what a token is has a large impact on the computational efficiency and interpretability of the results. Here a token is defined as the value of one variable on one time period, $y_{ti} \in \mathbb R$. Hence, the input block $Y_t$ is transposed and vectorized $\text{vec}(Y_t^\prime)$ (flattened) into a vector of $Pk$ tokens.

\subsection{Embedding} \label{subsec:embedding}

Embedding is a process of creating initial representations for tokens, which include relevant information about them. These representations are spacious information containers, into which the Transformer can encode more nuanced information, based on the dependencies and conditional patterns uncovered from the data.

\textit{Time series Embedding} is a linear projection $\mathbb R \rightarrow \mathbb R^{d_m}$ that maps each real number into a unique $d_m \in \mathbb N^{+}$ dimensional vector. Each token is embedded separately using the same one--row weight matrix $W_{\text{Embed}} \in \mathbb R^{1 \times d_m}$
\begin{align}
TE_{yt} = \text{vec}(Y_t^\prime) W_{\text{Embed}} \label{eq:embed}
\end{align}

to create a matrix of value--embedded representations for the block of input data, $TE_t \in \mathbb R^{Pk \times d_m}$. The parameters in $W_{\text{Embed}}$ are learned during the training. The elements $z\in \{1,\dots,d_m\}$ of an embedding vector are called \textit{features}. This step is loosely similar to how natural language data is embedded \footnote{Given a lexicon of Unicode tokens of size N, the input sequence of $P$ tokens is one-hot-encoded into a sparse $P \times N$ matrix Y. This matrix is projected into an embedding space by a dense weight matrix $W^E \in \mathbb R^{N\times d_m}$, where $d_m$ gives the dimensionality of the representations in the embedding space. In large language models, $d_m$ is typically on the order of thousands. It is typical to use a trained embedding projection $W^E$ such as Word2vec by \textcite{Mikolov2013}, which organizes the vector representations of tokens in a way that preserves syntactic and semantic properties.}. It is noteworthy that all value--embedded tokens are initially in the same span in the $d_m$-dimensional representation space. The representations are pushed off from this span by adding positional- and variable encoding information to them.

\textit{Positional Encoding} vectors $pe_p \in \mathbb R^{d_m}$ for p=$1,\dots,P$ add sequential information to the embeddings. There are no recurrent elements in the Transformer network and it is permutation invariant with respect to the token sequence. For this reason the sequential structure of the data is supplied directly to the embedding vectors to differentiate the meaning of values from different time periods. All $k$ tokens at lag $p$ share the same positional encoding vector $pe_p$, which differs from positional encoding vectors added to tokens at other lags. The values of the elements in these encoding vectors are learned similarly to the embedding weights\footnote{It is also common to insert fixed positional information to the tokens is by using sinusoidal functions, $PE_{t,2i} = \sin(\frac{t}{10000^\frac{2i}{d_m}})$ and $PE_{t,2i+1} = \cos(\frac{t}{10000^\frac{2i}{d_m}})$ for $t=1, \dots, P$ and $i=1,\dots, d_m/2$, with $PE_{t,2i} \in [-1,1]$.}. The positional encoding vectors are collected into a matrix by repeating each positional vector p=$1,\dots,P$ for $k$ times $[pe_1, pe_1, \dots, pe_P]^\prime =: PE_y \in \mathbb R^{Pk \times d_m}$, where the prime stands for a transpose. The positional encoding matrix is added to the embedding matrix.

\textit{Variable Encoding} vectors $ve_i \in \mathbb R^{d_m}$ for $i=1,\dots, k, k+1$, amend the token representations with information about which variable the token is standing for. There is a unique vector for each of the $k$ observable variables and one for the estimated dynamic factor. Similarly to the positional encoding matrix, the variable encoding vectors are used to construct a matrix $[ve_1, ve_2, \dots, ve_k] =: VE_y \in \mathbb R^{Pk \times d_m}$, this time repeating the cycle of i=$1,\dots, k$ vectors $P$ times. Variable encoding with a separate vector for the factor also conveys the difference between latent and observable variables, when only one factor is estimated. If multiple factors were to be estimated instead of one, then additional binary \textit{type encoding} could be added to supply additional latency information.

All of the different characteristics are added together to create the initial token representations for the Transformer to analyze
\begin{align}
Y_t^{\text{emb}} = TE_{yt} + \gamma(PE_y + VE_y)
\label{eq:initialRepresentation}
\end{align}

It is useful to scale down the positional, variable and type-encoding matrices with $\gamma<1$ to prevent them from overriding the value embedding information, especially with small variable values. The representations in matrix $Y_t^{\text{emb}}$ will be called as the "embedded token representations" for short.

\subsection{Factor initialization} \label{subsec:factorinit}

An initial crude factor estimate is created from the pointwise means of the block of input data, $\hat x_{\text{init}} = mean(Y_t)$, resulting in a vector with $P$ real numbers. The factor representation can also be initialized with the values of one of the observable variables, or with a factor estimate from another model. It is potentially useful to initialize the factor with a different model than what is used as prior information, to encourage exploration.

The initialized factor estimate series is embedded with the same time series embedding projection as the observable variables, $W_{\text{Embed}}$, and augmented with the same position and variable encoding vectors $pe_p, ve_x$ as in equation~\eqref{eq:initialRepresentation}.
\begin{align}
X_{\text{init}} = \hat x_{\text{init}} W_{\text{Embed}} + \gamma(PE_x + VE_x) \label{eq:initx}
\end{align}

with $X_{\text{init}} \in \mathbb R^{P \times d_m}$. Initializing the factor series before the State Encoder, and treating it as a latent variable, is one of the contributions in this study.

The benefit of initializing the factor estimate right in the beginning is in the enhanced interpretability. The State Encoder's Cross-Attention mechanism now directly describes the construction of the factor estimate. An alternative option would be to compress the factor estimate on an output layer after the State Encoder. In that case the State Encoder's Self-Attention would describe only the construction of intermediate representations for the input tokens, which only later are used to compress the factor. Additionally, as the output compression would combine information from several tokens, it could potentially dilute the information in the self-Attention matrices. Instead, the initial factor estimate is refined in the State Encoder and the output layer merely projects each period's representation of a factor estimate into a real number. The Cross-Attention matrices describe the relative importance of each variable and their lag for the factor estimate.

\subsection{Initial Encoder} \label{subsec:initEnc}

The raw input representations $Y_t^{\text{emb}}$ are first refined in an initial Encoder, to augment the token representations with cross-dependencies in the data. The Encoder has standard Self-Attention and Feed Forward network modules, with residual connections and normalization steps.
\begin{align}
    Y_t^{\text{init}} &= \text{Encoder}_{y0}(Y_t^{\text{emb}})
\end{align}

Without an initial Encoder, the Cross-Attention of the State Encoder could not use direct information about the cross-dependencies across input variables and time, to refine the factor representations. Because the initial Encoder is very similar to the State Encoder, the operations will not be presented here in detail, but instead the differences will be pointed out in the next subsection.

\subsection{State Encoder}

There is a stack of $H\in \mathbb N^+$ identical State Encoders, which refine the factor estimate further. The first State Encoder takes in the data $Y_t^{\text{init}}$ and the initial factor estimates $X_{\text{init}}$. The rest of the State Encoders take the refined factor estimate from their predecessor as their input, but use the same data $Y_t^{\text{init}}$. The final State Encoder sends its output to the output layer, equation~\eqref{eq:factorEstimate}, which projects each period's factor estimate $\hat x_{t}$ into a real number. Specifying a deeper architecture with multiple subsequent Encoders can help improve the results, especially if there is a lot of data, but it can also make the training process harder. The indices for the different State Encoders in the stack, $j = \{1, 2, \dots, H\}$, are omitted for notational brevity. Also, only one State and Measurement Encoder is used in the Monte Carlo experiment.

The inputs to the State Attention are first normalized token-wise.
\begin{align}
        X_0 &= \text{Norm}_{1x}(X_{\text{init}}) \label{eq:norm11} \\
        Y_0 &= \text{Norm}_{1y}(Y_t^{\text{init}}) \label{eq:norm1}
\end{align}

Layer Normalization, $\text{Norm}_1(z)=$ $\gamma \frac{z-\bar z}{\sigma_z} + \beta$, standardizes token representations. Each \textit{feature} z of a token's representation vector is expressed as units of standard deviation from the mean of that vector, and then scaled and shifted, with trained parameters $\gamma, \beta \in \mathbb R$. Each token is normalized separately and the operation is applied to all of its $d_m$ features. The normalization step is a solution to the internal covariance shift problem in layered Transformers, where the distribution of the inputs to the subsequent layers changes during training \parencite{ioffe2015batch}. The scale $\gamma$ and the bias vector $\beta$ are trainable parameters, which re-scale and shift the input representations. \textcite{ba2016layernorm} show that the normalization improves training speed and generalization accuracy. According to \textcite{Lu2021}, $\gamma$ and $\beta$ of a pre-trained Transformer are the most important parameters to tweak during the fine tuning stage. In the original article \textcite{Vaswani2017} normalization was conducted in the residual stream, after the two operations, Attention and Feed forward network. \textcite{Xiong2020} argues that the normalization should instead be applied before the operations, for a copy of data read from the residual stream. Especially with subsequent Encoders it is useful to normalize the previous Encoder's output before applying Attention. In the Initial Encoder there is only one input $Y_t^{\text{emb}}$ which is layer normalized. A copy of the un-normalized inputs $X_{\text{init}}$ also bypass the State Attention in the residual stream, to be combined with the output from the Attention later on.

\textit{The Attention mechanism} is at the heart of the Transformer network. At this step the inputs communicate with each other. Each State Encoder has a Multi-Headed Scaled Dot-Product Cross-Attention module, with h parallel Attention heads. An Attention head first performs three projections, creating Query, Key, and Value representations for the input.\footnote{An information retrieval system ranks the Values (movies) of different options in the database based on the similarity between their Key (description) to the Query (search words). In time series context, for intuition, think of a Query as representing the situation of a certain input, Keys representing what information other inputs have to offer, and Values representing for different interpretations about the future.}.
\begin{align}
    Q_i&=X_0 W_i^Q \\
    K_i&=Y_0 W_i^K \\
    V_i&=Y_0 W_i^V
\end{align}

where each $i=1, \dots, h$ Attention head has separate learned weight matrices $W_i^Q,W_i^K, W_i^V \in \mathbb R^{d_m\times d_k}$. The Transformer learns through training to create the Query, Key and Value representations for the input data in such a way that allows the Attention mechanism to uncover useful information for constructing the series of factor estimates. The Query $Q_i$, Key $K_i$, and Value $V_i$ matrices compress information by representing the data in a lower dimensional vector space, where typically $d_k\approx d_m/h$. The qualifier "Cross" in Cross-Attention means that the Query representation is derived from a different source than the Key and Value representations. Importantly, the initial factor estimate "queries" the observable data, and its refined values are based on the "answers" (keys and values) provided by the data. Projections in the Attention mechanism don't include biases, as the subsequent normalization in equation \eqref{eq:norm2} will de-mean the representations anyway. The difference to the Initial Encoder's Self-Attention is that there all three representations Q, K and V are derived from the same input $Y_t^{\text{emb}}$, hence the qualifier "Self".

The following is the most important equation of a Transformer. Using the three representations, the head i calculates a Scaled Dot-Product Attention
\begin{align}
    \text{head}_i &= \text{Softmax}\left( \frac{Q_i K_i^\prime}{\sqrt{d_k}} \right) V_i \label{eq:StateAttention}
\end{align}

The matrix multiplication $Q_i K_i^\prime$ measures the cosine similarity between the Query and Key representations of input tokens, where the prime denotes a transpose. The j$^{\text{th}}$ row of the outcome signifies the relevance of other k*P tokens for the j$^{\text{th}}$ token. These scores are normalized with the dimensionality of the Key representation, $\frac{1}{\sqrt{d_k}}$, to help avoid the consequent Softmax function from converging towards a max function, with large choices of $d_k$. The vector-valued function, Softmax($z_j$) $=e^{z_j} / \sum_{h=1}^{k*P} e^{z_h}$, is then applied to each row, enforcing the rows of the Attention matrix to sum up to unity.

The matrix of \textit{Attention scores}, $\text{Softmax}\left( \frac{Q_i K_i^\prime}{\sqrt{d_k}} \right)$, is the central tool for interpreting the results of a Transformer. The elements on row j of the matrix are weights for how to construct a new representation for token j, from alternative options presented in the Value matrix $V_i$. The output from the Attention head, matrix $\text{head}_i \in \mathbb R^{Pk \times d_k}$, consists of updated representations for the factor estimates, based on the contextual information of how they relate to the input tokens.

In other words, the Attention scores convey how much attention was paid to each variable's every lag when creating the factor estimate. Notice that the matrix for State Attention scores is not a square, $\text{Softmax}\left( \frac{Q_i K_i^\prime}{\sqrt{d_k}} \right) \in \mathbb R^{P\times Pk}$, due to $Y_0$ and $X_0$ having different number of rows. Subsection~\ref{subsec:attention} shows how the Attention scores can be used to create Attention matrices and how to use them to interpret the results.

Multi-Head Attention concatenates the outputs of each h parallel Attention heads and combines their information with a linear projection
\begin{align}
    X_1&=\text{MultiHead}(Q,K,V) \nonumber \\
    &=\text{Concatenate}(\text{head}_1, \dots, \text{head}_h) W^O
\end{align}

where the projection $W^O \in \mathbb R^{h*d_k\times d_m}$ retains the model dimensionality $d_m$ for the new representation matrix $X_1\in \mathbb R^{P \times d_m}$ of the factor estimate.

It is important to have separate Attention heads for two reasons. They learn to focus on uncovering different relations between the inputs, and they allow for further parallelization to improve training speed. With time series data the heads can for example specialize in observing different frequencies, conditional patterns, rare events or structural breaks. The original Transformer article by \textcite{Vaswani2017} used eight parallel Attention heads, but \textcite{Voita2019} suggests that more than three might be redundant. Also \textcite{Cordonnier2021} finds that different heads independently tend to focus on the same representational subspaces, making them redundant.

Output of the Multi-Head Attention is added back to the residual stream. Dropout de-activates a fraction of the features randomly.
\begin{align}
    X_2 = X_{\text{init}} + \text{Dropout}(X_1) \label{eq:skip1}
\end{align}

It is useful to think of the input representations of the data as flowing through the Transformer in a (residual) stream. Information is read from and written back into the residual stream during operations such as the Attention. Instead of overriding the previous representations with the new ones, the representations are refined, encoding them with new information. \textcite{Dong2021a} finds that without the residual stream, the Transformer's output degenerates to a rank 1 matrix, creating a uniformity between token representations. Residual stream is also called a skip connection, as a copy of the token representations skip an operation. Dropout is applied directly to the input representations in Transformers. Dropout is only applied during training to help avoid overfitting, but disabled during inference.

The representation $X_2$ is normalized for the second time, similarly as in equation~\eqref{eq:norm1}, but with different parameters $\gamma$ and $\beta$.
\begin{align}
        X_3 &= \text{Norm}_2(X_2) \label{eq:norm2}
\end{align}

A Feed Forward Network (FFN) processes each token representation in $X_3$ individually, through a fully connected expand and contract network with a GELU activation on the first layer
\begin{align}
        X_4 &=\text{FFN}(X_3) \nonumber\\
        &=GELU(X_3W_1^{ff} + b_1)W_2^{ff} + b_2 \label{eq:ffn}
\end{align}

with $W_1^{ff}\in \mathbb R^{d_m\times d_{ff}}$ and $W_2^{ff}\in \mathbb R^{d_{ff}\times d_m}$, keeping the output dimensions the same as the input, $X_3,X_4\in \mathbb R^{P\times d_m}$, but expanding in the middle, typically $d_{ff}\approx 4*d_m$. The biases have dimensions $b_1\in \mathbb R^{1\times d_{ff}}$ and $b_2\in \mathbb R^{1\times d_{m}}$. The Gaussian Error Linear Units (GELU) activation function is applied elementwise to introduce nonlinearities, by $GELU(x)=x \Phi(x)$, where $\Phi$ is the Gaussian CDF. The GELU activation was proposed by \textcite{hendrycks2016gelu}, who find it to improve performance compared to the popular alternative Rectified Linear Units (ReLU) activation, $ReLU(x)=\max\{0, x\}$. The FFN is called the computation step, in contrast to the communication step of the Attention mechanism, because in FFN each token is processed independently.

After the FFN the output is subjected to another dropout and added back to the residual stream (second skip connection)
\begin{align}
    X_5 = X_2 + \text{Dropout}(X_4) \label{eq:skip2}
\end{align}

The output $X_5$ is sent to the next State Encoder in the stack as an input. From the last State Encoder the output is sent to the output layer.

The \textit{factor estimate} is created from the output $X_5$ of the last State Encoder, with a linear projection
\begin{align}
        \hat x_{\text{t}} &= X_5 W_{\text{factor}} \label{eq:factorEstimate}
\end{align}

with matrix $W_{\text{factor}} \in \mathbb R^{d_m \times 1}$ producing the local P-lag time series for the dynamic factor $\hat x_{t} \in \mathbb R^{P \times 1}$. The projection $W_{\text{factor}}$ compresses a $d_m$-element vector representation of one factor at one time period into a real number. The projection is done separately for each lag position in the context window, $\{t-P, \dots, t\}$. Factor estimate series $\hat x_{t}$, and particularly its most recent value, is the main output of interest from the Transformer proposed in this study. The benefit of only conducting a simple representation-wise output projection is that most of the analysis is left to be done by the operations in the Encoder. The aim is to maximize interpretability of results with Attention matrices, and therefore it is useful to prevent the Transformer from relying too heavily on other means of combining information across inputs, such as a pooling output layer.

The loss function uses the distance between $\hat x_{t}$ and the factor estimate from the Kalman filter (or any other factor model), $\hat x_{t}^{\text{Kalman}}$, as prior information regularizer, equation~\eqref{eq:loss}, to guide the training.

The factor estimate $\hat x_{\text{t}}$ is re-embedded and augmented with the same positional and variable information as the input data in equation~\eqref{eq:initialRepresentation}, before being sent to the Measurement Encoders as
\begin{align}
\hat X_{\text{t}} = \hat x_{\text{t}} W_{\text{Embed}} + \gamma(PE_x + VE_x)
\end{align}
resulting in the matrix of re-embedded representations $\hat X_{\text{t}} \in \mathbb R^{P \times d_m}$ for the final factor estimates.

The key idea is to compress the factor representation all the way down to one real number per lag position, interpret it as a latent variable, and then re-embed it again to be used as Keys and Values in the Measurement Encoder's Attention. This conditions the factor estimate $\hat x_{t}$ with data during training, while posing minimal strict identifying assumptions for the functional form of the implicit state space system.

\subsection{Measurement Encoder} \label{subsec:measure}

The first Measurement Encoder receives the same input data $Y_t^{\text{init}}$ as the State Encoder, and the re-embedded factor representations $\hat X_{\text{t}}$ coming out of the State Encoder stack. The State Encoders and Measurement Encoders have separate weight matrices and operations, but stack-specific indices are omitted for simplicity. A Measurement Encoder conducts separate Layer Normalizations for the data and for the factor.
\begin{align}
        Y_0 &= \text{Norm}_{1,y}(Y_t^{\text{init}})\\
        \hat X_{\text{final}} &= \text{Norm}_{1,x}(\hat X_{t})
\end{align}

There are two differences in the Measurement Encoder's Cross-Attention mechanism, compared to that of the State Encoder. The Query matrix is obtained from the input data $Y_0$, but the Key and Value matrices are constructed from the factor representation $\hat X_0$.
\begin{align}
    Q_i &= Y_0 W_i^Q \\
    K_i &= \hat X_{\text{final}} W_i^K \\
    V_i &= \hat X_{\text{final}} W_i^V
\end{align}

where $W_i^Q, W_i^V, W_i^K \in \mathbb R^{d_m\times d_k}$ are otherwise similar to the State Encoder. There is only one Cross-Attention module in the Measurement Encoder, without an initial Self-Attention module as in \textcite{Vaswani2017}, as the input representations have already been refined in the Initial Encoder's self-Attention.

The rest of the steps in the Attention module are mostly identical to the State Encoder
\begin{align}
    \text{head}_i &= \text{Softmax}\left( \frac{Q_i K_i^\prime}{\sqrt{d_k}} \right) V_i \label{eq:MeasureAttention} \\
    Y_1&=\text{Concatenate}(\text{head}_1, \dots, \text{head}_h) W^O \\
    Y_2 &= \text{Dropout}(Y_1) \label{eq:no_res}
\end{align}

Masking is not needed in the Measurement Encoder's Attention, because only the fully backwards-looking last period's prediction will be used in the loss function.\footnote{If a measurement Decoder was specified instead with teacher forcing, the Attention scores would have to be censored with an upper-block-diagonal mask of large negative values, with zeros elsewhere. The mask has to be block-recursive due to the granular definition of a token.} Importantly, the residual stream passes only through the Attention module with no skip connection, in equation~\eqref{eq:no_res}. This is to enforce the key information bottleneck of the Transformer: The predictions for the observable variables must rely on the factor estimate via the cross-Attention with no bypass. Like the State Attention, the Measurement Attention matrix is also not a square, $\text{Softmax}\left( \frac{Q_i K_i^\prime}{\sqrt{d_k}} \right) \in \mathbb R^{Pk\times P}$.

This is followed by the second Layer Normalization, Feed Forward Network, and this time a standard skip connection
\begin{align}
Y_3 &= \text{Norm}_2(Y_2)\\
Y_4 &= GELU(Y_3 W_1^{ff} + b_1)W_2^{ff} + b_2 \\
Y_5 &= Y_2 + \text{Dropout}(Y_4) \label{eq:skip3}
\end{align}

The Measurement Encoder's output $Y_5$ becomes an input for the next Measurement Encoder in the stack. Every subsequent Measurement Encoder derives its Queries from the previous Encoder's output, but uses the same factor representation $\hat X_{\text{t}}$ to get Keys and Values. Last Measurement Encoder's output $Y_5$ is sent to the output layer of the Measurement Encoder stack to create predictions for the next period values of the observable variables.
\begin{align}
        \hat Y_{\text{t+1}} &= Y_5 W_{\text{predict}} \label{eq:prediction}
\end{align}

with $W_{\text{predict}} \in \mathbb R^{d_m \times 1}$, producing the prediction output matrix $\hat Y_{\text{t+1}} \in \mathbb R^{Pk \times 1}$. Each real number is a next period's predicted value for the corresponding input token. Similarly to the State Encoder's output projection, also here the output projection only compresses individual token representations into real numbers. This forces the Transformer to rely more on the Measurement Attention to identify dependencies between the inputs and the factor.

Only the last $k$ values of $\hat Y_{\text{t+1}}$ are saved in a k-value vector $\hat y_{\text{t+1}} \in \mathbb R^k$. These values are the t+1 period predictions for the $k$ observable variables. These predictions are based on the full $P$ lag information set and only include past information. The prediction errors between $\hat y_{\text{t+1}}$ and the real values $y_{t+1}$ are used in the loss function, equation~\eqref{eq:loss}, as the main objective for the training.

\subsection{Loss function and prior information} \label{subsec:loss}

The loss function used for training the Transformer is comprised of two main parts, Prior Information and Prediction Error
\begin{align}
    Loss =&  \lambda \ \text{PI}_t + \ (1-\lambda) \ \text{PE}_t \label{eq:loss}
\end{align}

where
\begin{align}
    \text{PI}_t &= \frac{1}{P}\sum_{p=0}^{P-1} |\hat x_{T,t-p} - \hat x_{K,t-p}| \\
    \text{PE}_t &= \frac{1}{k}\sum_{i=1}^{k} | \hat y_{t+1, i} - y_{t+1, i} |
\end{align}

and $\lambda \in [0,1]$ controls for the relative weight of prior information to the prediction errors in training. The loss function measures Mean Absolute Errors (MAE). MAE loss is less sensitive to non-Gaussian errors, which are very typical in macro data, than the Mean Squared Error (MSE) loss function.\footnote{MAE loss is optimal for random variables that follow the Laplace distribution, $f(x|\mu, b)=\frac{1}{2b}e^{-\frac{|x-\mu|}{b}}$, where $\mu\in \mathbb R$ is the location parameter and $b>0$ is the scale parameter. The loss function can be upgraded to also estimate b as a measure of uncertainty and to discount large deviations if they occur under heightened uncertainty.}

\textit{Prior Information}, PI$_t$, is the proposed new regularization term, which minimizes the distance between the factor estimates from the Transformer and the Kalman filter. The Transformer's factor estimate $\hat x_{T}\in\mathbb R^P$ is the output from the State Encoder, equation~\eqref{eq:factorEstimate}, and $\hat x_{\text{K}}\in \mathbb R^P$ is the factor estimate generated with the Kalman filter, as outlined in Subsection~\ref{subsec:kalman}. Prior information pulls the training process towards a reasonable direction in the Transformer's parameter space. This helps the Transformer avoid getting stuck in a bad local optimum. The simulated real factor is never observed at any stage of the training. Any factor model, other than the Kalman filter, can also be easily plugged in as prior information.

\textit{Prediction Errors}, PE$_t$, measure the distance between the observable variables and the Transformer's next token predictions $\hat y_{t+1}\in \mathbb R^k$, coming from the Measurement Encoder in equation~\eqref{eq:prediction}.

\begin{figure}[!htbp]
    \centering
    \includegraphics[width=1\linewidth]{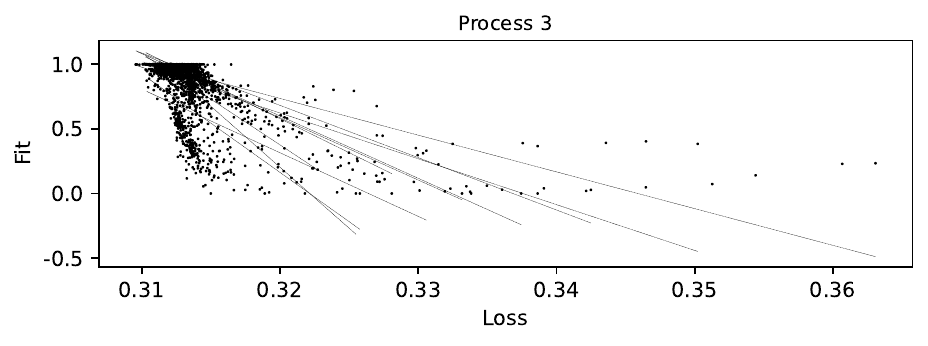}
    \caption{The effectiveness of the training in improving the factor accuracy can be analyzed by visualizing the loss-accuracy pairs as points from different epochs, and fitting linear curves over them. Each line in the figure represents a separate training run with different initial parameters on the same data.}
    \label{fig1:fitLoss}
\end{figure}

The objective in minimizing the loss function is to improve the accuracy of the Transformer's factor estimates. Whether reducing the loss actually corresponds to factor accuracy can be studied using simulated data, where the true factor is known. During each epoch, both the loss and an accuracy metric is documented. A particular metric \textit{Fit} is introduced in equation~\eqref{eq:fit}, but other metrics can also be used. Figure~\ref{fig1:fitLoss} regresses the resulting factor accuracies against the accompanying validation losses. Negative regression coefficients indicate that the Transformer improves the accuracy in training, as the loss diminishes.

\subsection{Training process} \label{subsec:train}

The data is divided into non-overlapping training, validation and test sets across the time dimension. This division has to be done with care when dealing with real-world macroeconomic time series data, to get representative samples to each set.
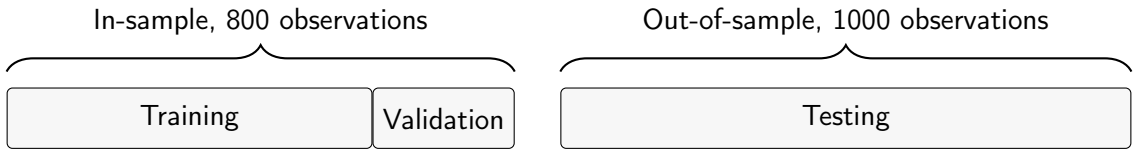
\begin{figure}[!htbp]
    \centering
\begin{tikzpicture}[font=\sffamily]
\node[draw, rounded corners=2pt, fill=black!3,
      minimum height=8mm, minimum width=4.8cm,
      align=center, anchor=west] (train) at (0,0) {Training};

\node[draw, rounded corners=2pt, fill=black!3,
      minimum height=8mm, minimum width=1.2cm,
      align=center, anchor=west] (val) at (train.east) {Validation};

\node[draw, rounded corners=2pt, fill=black!3,
      minimum height=8mm, minimum width=7.5cm,
      align=center, right=6mm of val, anchor=west] (test) at (val.east) {Testing};

\draw[decorate, decoration={brace, amplitude=10pt,raise=6pt}, thick]
      (train.north west) -- (val.north east)
      node[midway, yshift=25pt] {In-sample, 800 observations};

\draw[decorate, decoration={brace, amplitude=10pt,raise=6pt}, thick]
      (test.north west) -- (test.north east)
      node[midway, yshift=25pt] {Out-of-sample, 1000 observations};
\end{tikzpicture}
    \caption{The simulated datasets in the Monte Carlo study of this study are divided into training, validation and testing sets. The out-of-sample test set contains more data than training and validation sets combined, to guarantee that the results generalize.}
    \label{fig1:division}
\end{figure}

Training data is used for updating parameters, validation data for monitoring how well the results generalize out-of-sample during the training and test set for analyzing the out-of-sample results after the training has concluded. When the Transformer starts to overfit the training data, the prediction error in the validation loss begins to increase.

The Transformer processes the data in blocks $Y_t\in\mathbb R^{P\times k}$, where the number of lags $P\leq N$ defines the context window, or information set, of time periods $\{t-P, \ t-P+1, \ \dots, \ t\}$. The time index $t$ denotes the most recent time period in the context window $Y_t$.

\begin{figure}[!htbp]
    \centering
    \begin{tikzpicture}[x=0.6cm,y=0.6cm, scale=0.85]

\draw[thick] (0,0) rectangle (15,1.5);

\def\winW{5}      
\def\winH{1.5}    
\def\dx{2.0}      
\def\dy{-0.25}     

\def\x0{0.1}
\def\y0{1.4}
\foreach \k in {0,1,2,3,4,5}{
  \draw[dashed,rounded corners=2pt,thick]
    ({\x0+\k*\dx},{\y0+\k*\dy}) rectangle
    ({\x0+\k*\dx+\winW},{\y0+\k*\dy-\winH});
}

\draw[thick] (15.5,0) rectangle (24.5,1.5);

\def\x0{15.6}
\def\y0{1.4}
\foreach \k in {0,1,2}{
  \draw[dashed,rounded corners=2pt,thick]
    ({\x0+\k*\dx},{\y0+\k*\dy}) rectangle
    ({\x0+\k*\dx+\winW},{\y0+\k*\dy-\winH});
}

\draw[decorate, decoration={brace, mirror, amplitude=10pt,raise=6pt}, thick]
      (0.1,-0.7) -- (2.1,-0.7)
      node[midway, yshift=-25pt] {Stride of 1 time period};

\draw[decorate, decoration={brace, mirror, amplitude=10pt,raise=6pt}, thick]
      (10.1,-1.3) -- (15.1,-1.3)
      node[midway, yshift=-25pt] {P-lag context window};

\node[] (train) at (7.5,2) {Training};
\node[] (train) at (20,2) {Validation};

\end{tikzpicture}
    \caption{The training, validation and test sets are partitioned into P--lag context windows using a rolling window with a stride of 1.}
    \label{fig1:rollingwindow}
\end{figure}
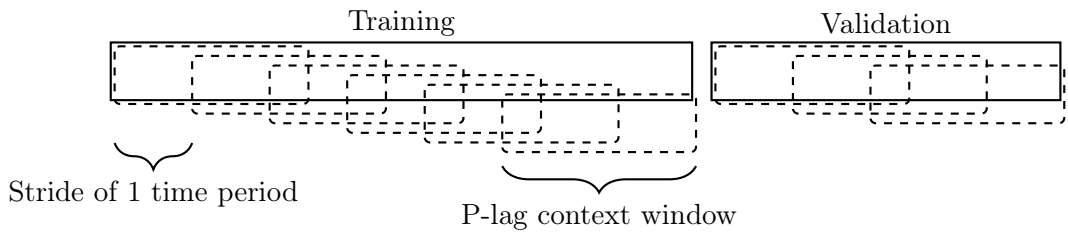

For each of the sets, a rolling window with a stride of one is applied to create input matrices with all possible P-lag information sets. However, there is no overlap between training, validation, and test sets. Figure~\ref{fig1:rollingwindow} illustrates the outcome. These matrices are also called training examples and their size corresponds to the context window of the Transformer. Each matrix consists of P+1 consecutive time periods, where the first $P$ periods are used to predict the most recent time period. Because only the last period's predictions for the observable variables are used in the loss function, it is important to use these partially overlapping context windows in training.

The training examples are further grouped into batches, each consisting of several P-lag training examples, $\{Y_{1}^{\text{emb}}, Y_{2}^{\text{emb}}, \dots, Y_{\text{batchsize}}^{\text{emb}}\}$. All training examples in one batch are processed in parallel through the Transformer. The context windows in the validation set are treated as one big batch -- same for the test set.

The Transformer calculates the loss for each training example in a batch according to equation~\eqref{eq:loss}. The average loss across a batch is backpropagated to the Transformer's parameters. The gradients are calculated using the automatic differentiation tools of Pytorch. After all of the training data has been processed, Transformer's performance is evaluated on the validation data without updating parameters, after which an epoch concludes. Those parameter values are retrospectively chosen for the Transformer, after the training has concluded, which resulted in the lowest validation loss during the training. There is a risk that the Transformer found a way to overfit the training data in a way that also happens to overfit the validation data. For this reason there is also a separate test data which is analyzed only after the training process has concluded.

\textit{Local filtering in parallel.} The series of factor estimates for the $N^{\text{test}}$ period test set are created in parallel. Each P-lag context window is processed in parallel, and only the last period's factor estimate from each of them is appended into $\hat x^{(i)} \in \mathbb R^{N^{\text{test}}}$. This sequence is the outcome from one training run i=$1,\dots,r$ of the Transformer, used in the Monte Carlo in Section~\ref{sec:montecarlo}.
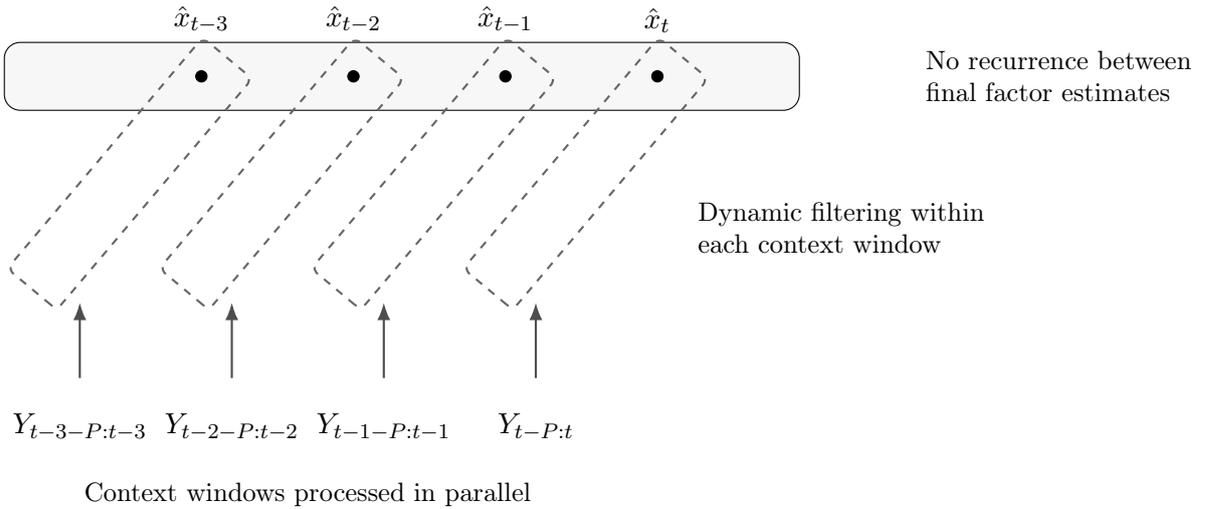
\begin{figure}[!htbp]
\centering
\begin{tikzpicture}[%
>={Latex[length=2.8mm]},
token/.style={circle,fill=black!70,inner sep=1.1pt},
est/.style={circle,fill=white,draw=black,inner sep=1pt},
win/.style={rounded corners=3pt,draw=black!60,dashed,thick},
paral/.style={-Latex,thick,black!70},
note/.style={font=\small,align=left}
]

\path let \p1=(2,4), \p2=(15,4) in
  node[draw,rounded corners=6pt,fill=black!3,minimum height=0.9cm,
       anchor=west,inner xsep=6pt] (band) at (1,4) {\phantom{XXXXXXXXXXXXXXXXXXXXXXXXXXXXXXXXXXX}};

\foreach \x/\lab in {3/$\hat x_{t-3}$,5/$\hat x_{t-2}$,7/$\hat x_{t-1}$,9/$\hat x_{t}$}
{
  \node[circle, fill=black, draw=black, inner sep=1.5pt] (e\x) at (\x+0.6,4) {};
\node[above=15pt, font=\normalsize] at (e\x.south) {\lab};
}

\foreach \c in {3,5,7,9}{
  \draw[win,rotate around={50:(\c-2.3,1.35)}] (\c-2,0.3) rectangle (\c+2,1.15);
}

\foreach \x in {2,4,6,8}{
  \draw[paral] (\x,0) -- (\x,1);
}

\foreach \x/\lab in {2/$Y_{t-3-P:t-3}$,4/$Y_{t-2-P:t-2}$,6/$Y_{t-1-P:t-1}$,8/$Y_{t-P:t}$}
{
  \node[] (y\x) at (\x,-0.1) {};
  \node[below=3pt, font=\normalsize] at (y\x.south) {\lab};
}

\node[note,anchor=west] at (13,4) {No recurrence between\\ final factor estimates};
\node[note,anchor=west] at (10,2) {Dynamic filtering within\\ each context window};
\node[note] at (5,-1.55) {Context windows processed in parallel};

\end{tikzpicture}

\caption{Training and inference is conducted in parallel, with local dynamic filtering within each context window.}
\label{fig:dyn_filtering}
\end{figure}

\subsection{Model averaging and identification} \label{subsec:modelavg}

The Transformer is trained multiple times for the same data to counter the path dependent impact of random parameter initialization. The actual series of estimated factor values $\hat x$ is constructed from the pointwise averages over the runs. Model averaging evens out some of the biases of individual models. All individual factor estimates are also included in the background of the visualizations in thin lines. The consistency of the results is proxied by the standard deviation of the accuracy of the model average $\hat x$ across different seeds of the same process. The within-seed variation between training runs is also a measure of uncertainty of the factor value.

\textit{Factor identification}, the scale and sign of the factor, cannot be identified from the data alone. Because the factor describes relative changes, the scale is not important. To enable comparison, the Transformer's factor estimates can be scaled to best match the estimated series from the model under comparison. The sign of the factor can be identified with narratives based on the data. For example, a coincident index for the business cycle should decline during historic recessions.

\textit{Real time factor projections} extend the factor estimate over the next n time periods with only current information at hand. The forward pass is repeated for n times recursively, each time adding the previous predictions $\hat y_{\text{t+1}}$ to the end of the input data, and saving the last factor estimate $\hat x_{t}$ from each round. This procedure is conducted separately for each time period in the test set, creating the real time predictions. The resulting "tentacle plot" was inspired by a similar type of plot in \textcite{Diebold2020tentacle}, but instead of looking backwards, here the factor estimates are extended into the future. The benefit of looking forward is that it is possible to see what kind of dynamics the Transformer has learned for the state-transition. The real time projections use one trained Transformer, so the results may deviate from the average of training runs.
\begin{figure}[!htbp]
\centering
\begin{tabular}{cc}
    \hspace{-0.65cm}
     \includegraphics[scale=1.1]{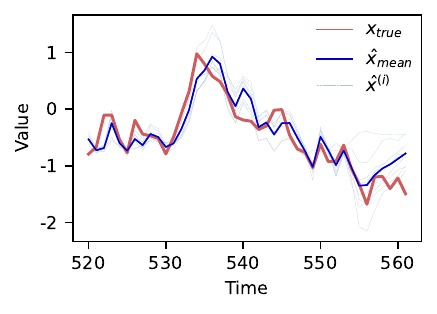}\hspace{-0.65cm} &  \includegraphics[scale=1.1]{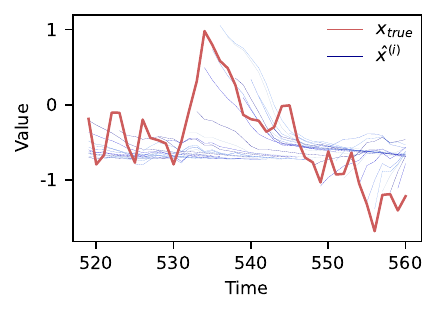}
     \end{tabular}
\caption{The left panel displays the estimated factor series against a simulated ground truth. The model average factor estimate $\hat x_{\text{mean}}$ is the pointwise mean over $i=1,\dots,r$ training runs with different initial parameters $\hat x^{(i)}$ from equation~\eqref{eq:factorEstimate}. The panel on the right visualizes real time factor projections, where the factor estimates are extended into the future, by using one trained model recursively with a fixed information set.}
\label{fig:estimate_projections}
\end{figure}

\subsection{Interpreting results with Attention} \label{subsec:attention}

The matrix of State Attention scores, $\text{Softmax}\left( \frac{Q_i K_i^\prime}{\sqrt{d_k}} \right) \in \mathbb R^{P \times Pk}$, equation~\eqref{eq:StateAttention}, is used to interpret how important different variables and their lags were in the process of refining the Transformer's factor estimate. As the Transformer is not restricted only to search for constant or linear relations, the impact of different inputs can change through time. Using the Attention scores, it is possible to analyze which variables and their lags were paid most attention to at any given moment. The last row of the Attention score matrix describes the most recent factor estimate (period $t$ in the context window), with an Attention distribution over the $k$ variables and their $P$ lags summing up to unity.

\begin{figure}[H]
    \centering
    \includegraphics[width=0.6\linewidth]{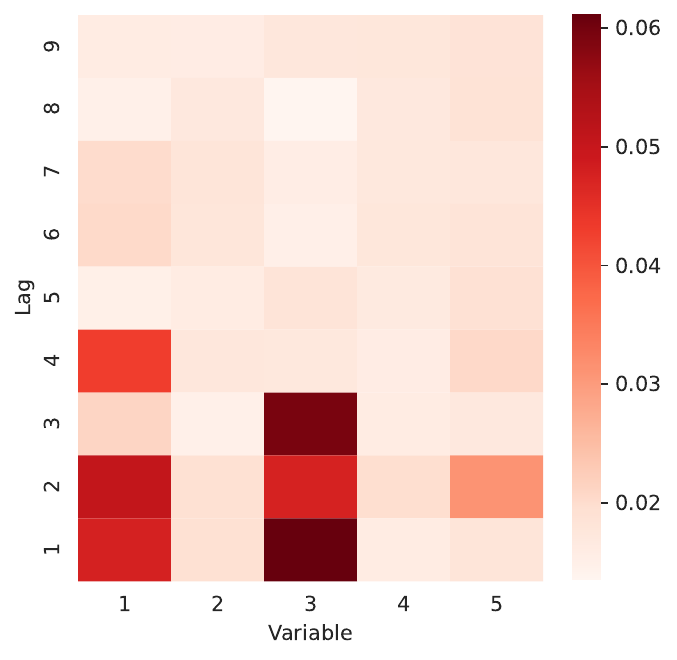}
    \caption{State Attention matrix describes the relative importance of each variable and their lag for the factor estimate, one real number, at period t. The matrix highlights variables 1 and 3, and few of their recent lags.}
    \label{fig1:stateMatrix}
\end{figure}

\textit{State Attention matrix} organizes these last-row Attention scores into a $P\times k$ matrix, where each column corresponds to P-lags of one input variable. The State Attention matrix is visualized with a heatmap, Figure~\ref{fig1:stateMatrix}, where darker colors reflect higher importance of the corresponding variable's lag. Only the Attention scores from the last State Encoder in the stack are considered. The scores are averaged over the Attention heads for simplicity, although more information could be obtained by observing each of the heads separately. The whole $P \times k$ matrix describes the construction of one real number -- the factor estimate at time t.

All of the visualizations in this subsection are examples of a process with k=5 observable variables and P=9 lags.

\begin{figure}[H]
    \centering
    \includegraphics[width=\linewidth]{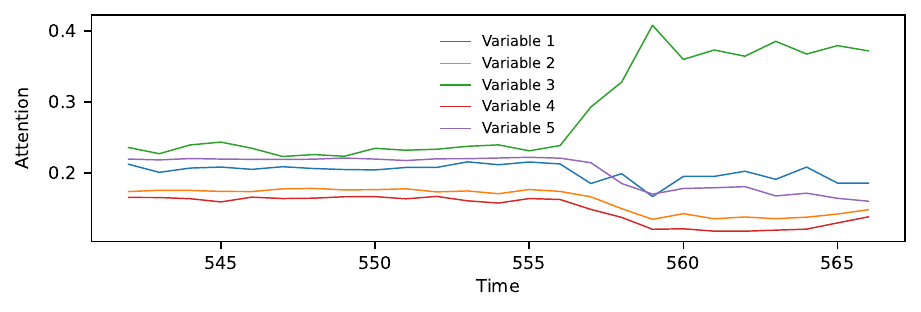}
    \caption{Visualizing how the Attention patterns change over time reveals useful information. Here the steady Attention pattern changes abruptly, signaling accurately about a regime switch in the process. The change also shows which variables convey the information about  the regime switch. Here the importance of variable 3 increases in the new regime.}
    \label{fig1:stateVar}
\end{figure}

Using the State Attention matrix, I construct new time series specific visualizations, which show how the Attention patterns change over time. The columns of the State Attention matrix are summed to get a vector of variable contributions. This vector describes the relative importance of the variables at time t. These variable contributions can be obtained similarly for each time period in the test set. Visualizing these contributions as time series for each variable helps to analyze how the importance of different variables change over time, Figure~\ref{fig1:stateVar}. Changes in the relative importance of suitable macroeconomic variables can potentially help to evaluate whether a sudden change is a demand or a supply shock.

\begin{figure}[H]
    \centering
    \includegraphics[width=\linewidth]{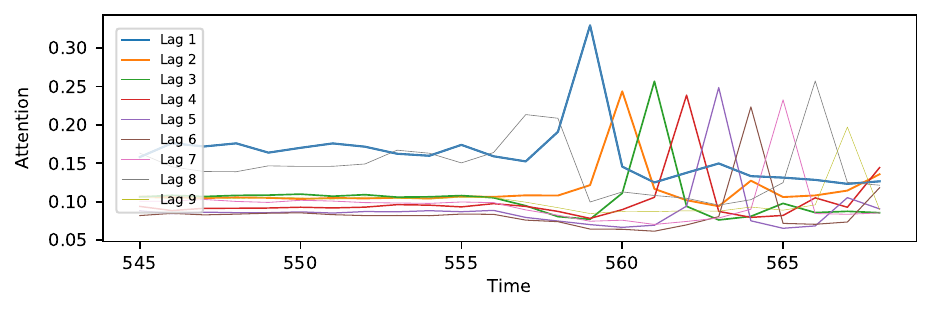}
    \caption{Analyzing the Attention patterns for lags helps to identify influential moments in time, like regime switches and large shocks. Here the period 558 propagates through the lag structure for the next 9 periods.}
    \label{fig1:stateLag}
\end{figure}

Similarly, the State Attention matrices for each time period can be summed row-wise into lag contributions, Figure~\ref{fig1:stateLag}. Changes in the allocation of Attention over lags can inform about important events in time. When a regime switch or a large shock occurs in the factor process, the Attention to most recent lag position spikes. The same influential moment can still remains useful in the near future for constructing the factor estimate and predicting the observables. In that case the Attention to each of the lags elevate one after another as time moves on. This also helps to detect heavy-tailed stochasticity in the state-transition shocks, which causes these spike patterns to emerge in the Attention plot at uneven intervals. Gaussian shocks on the other hand rarely cause such disruptions. The Attention patterns over lags can potentially also signal about increased uncertainty for example during a transition between steady states, where the Attention to more recent information increases permanently, or the volatility of the pattern increases.

\textit{Measurement Attention matrix} is composed of the last $k$ rows of the Measurement Attention score matrix in equation~\eqref{eq:MeasureAttention}. These rows include the $k$ next period predictions for the period $t+1$, after the context window. The matrix is transposed to preserve a similar  $P\times k$ look as with the State Attention matrix. However, the interpretation is different, in that each column i of the Measurement Attention matrix corresponds to a prediction for a different variable $\hat y_{i,t+1}$, for $i=1,\dots,k$. The Measurement Attention matrix is used to evaluate how relevant each lag of the factor estimate was for the predictions of the observable variables. The matrix conveys information about the temporal factor loadings.
\begin{figure}[!htbp]
    \centering
    \includegraphics[width=0.6\linewidth]{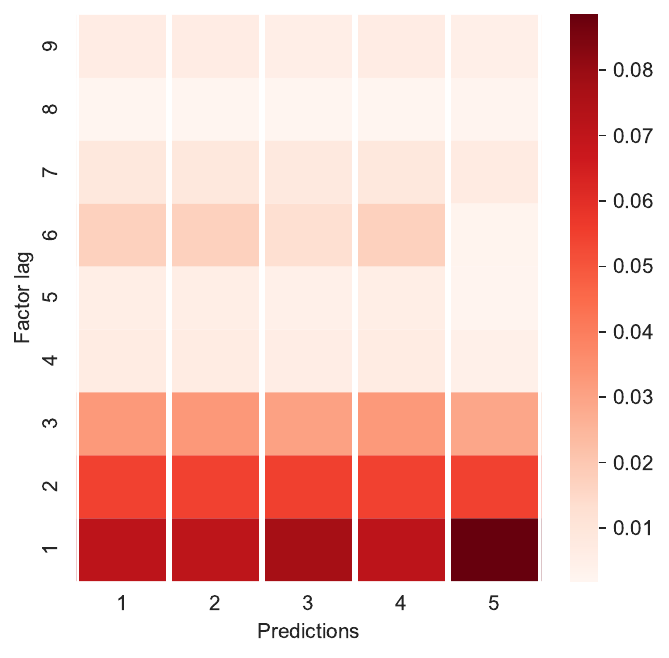}
    \caption{The columns of the Measurement Attention matrix describe the relevance of each lag of the factor estimate, on predicting the observable variables. Column i describes what the prediction $\hat y_{i,t+1}$ was based upon.}
    \label{fig1:measureMatrix}
\end{figure}

\begin{figure}[H]
    \centering
    \includegraphics[width=\linewidth]{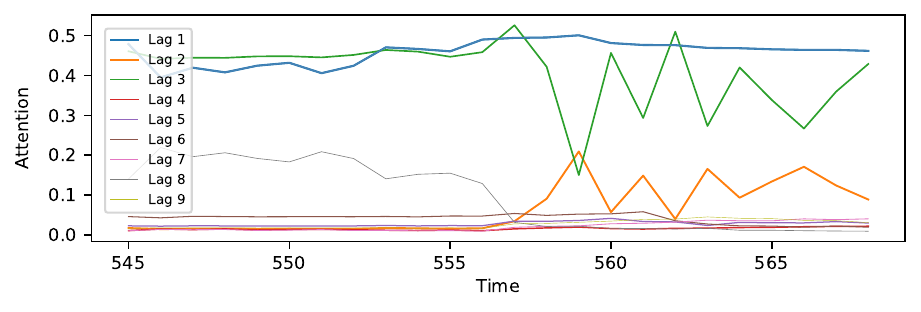}
    \caption{Time-evolution of importance of different factor lags for predicting observables. The Measurement Encoder seems to have found a way to rely on the first, third and eighth lag of the factor during one regime, and combining information differently in another regime.}
    \label{fig1:measureLag}
\end{figure}

By summing the rows of the Measurement Attention matrices over time, it is possible to describe how the relevance of factor lags changes over time, Figure~\ref{fig1:measureLag}. The analysis of these patterns can reveal similar information as the State Attention patterns over lags, explained above.

\begin{figure}[!htbp]
\centering
\begin{tabular}{cc}
    \hspace{-0.65cm}
     \includegraphics[scale=1]{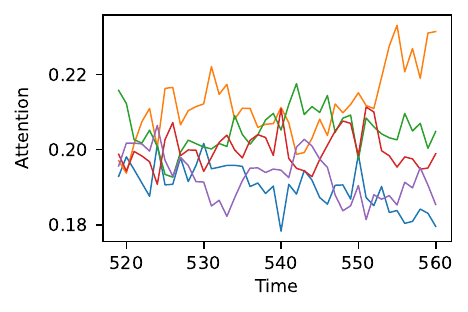}\hspace{-0.65cm} &  \includegraphics[scale=1]{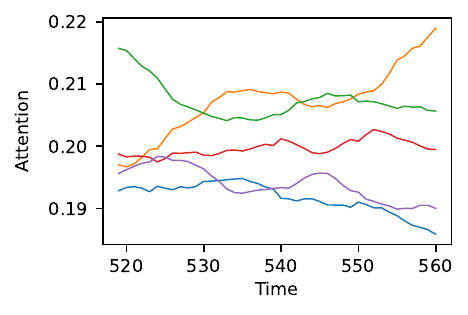}
     \end{tabular}
\caption{The smoothed attention patterns on the right show more clearly how the relative importance of the input variables change over a longer time span, but misses some of the potentially important local variation.}
\label{fig1:attn_smoothing}
\end{figure}

Attention patterns can be smoothed to highlight the signal from noise. The ragged State Attention patterns $a_{it}$ for variables $i=1,\dots,5$ on the left panel of Figure~\ref{fig1:attn_smoothing} have been processed through an exponential smoother, $\tilde a_{it} = \alpha a_{it} + (1-\alpha)b_{t-1}$, where $b_0=a_{i0}$ and $\alpha=0.1$. The starting point bias of the smoother can be overcome by smoothing both ways and taking the average. It can be easier to observe long term changes in the Attention patterns with smoothed curves in the panel on the right. However, smoothing also loses some information about local attention patterns and volatility changes in those patterns, so it will not be used in this study.

There is a caveat for interpreting results with Attention matrices. The factor estimate is constructed in the Transformer during several subsequent stages, only one of which is the Attention mechanism. This means that the impact of the Attention on the factor estimate cannot be completely isolated. However, the Attention is the only operation, where different inputs can directly impact each other, so it is plausible that what is found important in the Attention mechanism describes also the end result of the respective Encoder stack. The numeric values of the Attention matrix are also not to be interpreted as coefficients, as they do not reflect the sign or magnitude of the impact channel between the factor and the observable input, only its relative importance in that moment.

\subsection{Analyzing impact of Attention with residual stream} \label{subsec:residual}

\begin{figure}[H]
  \centering
\includegraphics[width=\linewidth]{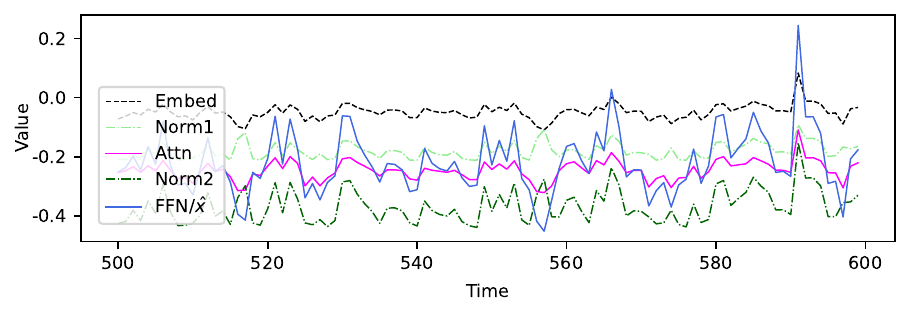}
  \caption{Visualizing intermediate representations before and after operations helps analyze their impact on the factor estimate.}
  \label{fig1:residualstream1}
\end{figure}

\begin{figure}[H]
  \centering
\includegraphics[width=\linewidth]{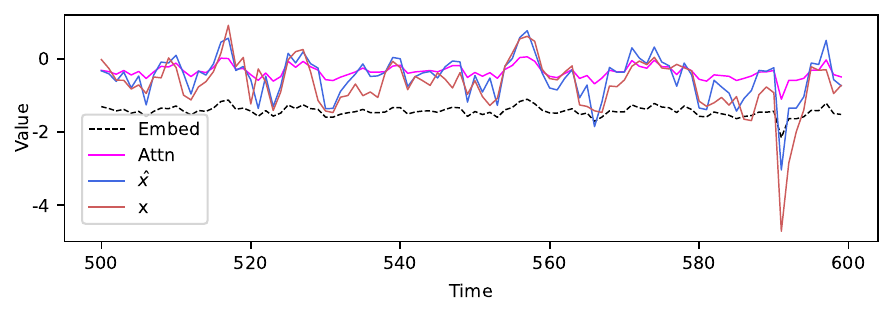}
  \caption{A closer look at the changes Attention and the Feed Forward network cause for the factor estimate. The lines are scaled to the true factor x in red.}
  \label{fig1:residualstream2}
\end{figure}

The residual stream refers to the sequence of representations for the factor estimate in one forward pass of the Transformer's State Encoder. Operations, such as the Attention and the Feed Forward Network, read information from the residual stream and write back into it, updating the representations. By visualizing the representations from different stages of the residual stream, it is possible to analyze how the Attention and other operations impact the factor estimate. To do this, I map all of the intermediate representations described in Table~\ref{tab:residual}, through the output projection $W_{\text{factor}}$ from equation~\eqref{eq:factorEstimate}. For the initial representation $X_{init}$, this means bypassing the whole Encoder machinery, whereas the output from FFN corresponds to the final estimate. Figure~\ref{fig1:residualstream1} visualizes all of the intermediate representations and Figure~\ref{fig1:residualstream2} focuses on the two key operations, Attention and Feed Forward, scaled based on the true factor for comparison.
\begin{table}[!htbp]
    \centering
    \caption{Internal factor representations from different stages of residual stream of the Transformer.}
    \begin{tabular}{l|cc}
      Label & Symbol & Equation \\
      \hline
      Embed & $X_{\text{init}}$ & \eqref{eq:initx} \\
      Norm1 & $X_0$ & \eqref{eq:norm11} \\
      Attn & $X_2$ & \eqref{eq:skip1} \\
      Norm2 & $X_3$ & \eqref{eq:norm2} \\
      FFN/$\hat x$ & $X_5$ & \eqref{eq:skip2} \\
    \end{tabular}
    \label{tab:residual}
\end{table}

\section{Monte Carlo Analysis} \label{sec:montecarlo}

In this section, the performance of the Transformer is evaluated on several simulated datasets against a Kalman filter baseline model, and an oracle particle filter. The datasets are designed to impose a series of different accumulating challenges, while remaining somewhat representative of typical macroeconomic data. The functional form and parameterization of the underlying state-space processes deviate from linear--Gaussian, but not excessively so. The results from the Monte Carlo are summarized in Table~\ref{tab:results}, and Figure~\ref{fig:all_results}, demonstrating that the Transformer can outperform the Kalman filter baseline when the process deviates from linear--Gaussian. This section presents the benchmark models, evaluation criteria, Transformer hyperparameters, data generating processes, and finally the results.

The Kalman filter is purposefully misspecified for the datasets. If the conventional baseline model matched the data generating process perfectly, it would be an "oracle" and the Transformer couldn't consistently outperform it. Furthermore, if we knew how to specify perfect models for real-world tasks, we would not need Transformers to begin with. Instead, the Monte Carlo is a simplified representation of the real-world situation, where all models are misspecified to a different degree. The Monte Carlo study explores whether a somewhat misspecified baseline model can guide an overparameterized Transformer, so that the Transformer not only performs reasonably but also surpasses the baseline in accuracy. An "oracle filter" with a functional form matching the underlying process and true parameters is used to evaluate how much of the explanatory gap between the Kalman filter and the oracle can the Transformer cover.

The Monte Carlo results should not be interpreted as conclusive evidence of the Transformer's general performance on arbitrary nonlinear datasets. Rather, the Monte Carlo showcases some particular interesting cases where the Transformer can improve upon the accuracy of a factor series estimated with a Kalman filter.
\begin{table}[!htbp]
    \centering
    \caption{The first four processes introduce nonlinearity and non-normality into the state-transition and measurement equations of a simple state-space system. Processes 5 and 6 are more complicated, with multiple lags, moving average dynamics, regime switching, and conditional heteroskedasticity.}
    \begin{tabular}{l|c|c|c|c}
         & State-transition & Measurements & Shocks & Errors \\
         \hline
        Process 1  & linear & linear & Gaussian & Gaussian \\
        Process 2 & linear & nonlinear & Gaussian & t(df=10) \\
        Process 3 & nonlinear & linear & t(df=10) & Gaussian \\
        Process 4 & nonlinear & nonlinear & t(df=10) & t(df=10) \\
        Process 5 & linear ARMA(3,1) & nonlinear AR(3)  & t(df=10, skewed) & t(df=10) \\
        Process 6 & Markov switching & Markov switching & t(df=10, skewed) & t(df=10)
    \end{tabular}
    \label{tab:firstfour}
\end{table}

\subsection{Benchmark models} \label{subsec:kalman}

The benchmarks include a Kalman filter with and without the estimation step, and an oracle particle filter with the correct functional form and stochasticity for each process, the simple piecewise mean of observable variables, and finally a Transformer which "peeks under the hood" to stop the training optimally. Only the estimated Kalman filter and oracle filter are visualized alongside with the Transformer and the true factor, to keep the figures readable.

\begin{table}[!htbp]
    \centering
    \caption{Transformer's results (first row) are compared to several benchmarks, the main comparison being with an estimated Kalman filter.}
    \begin{tabular}{c|c}
        Transformer & Training stopped at lowest validation loss \\
        Transformer max & Training stopped at highest test accuracy \\
        Mean of data & Pointwise average of observables \\
        Kalman filter & Estimated from the training data \\
        Kalman max & Uses true linear parameters \\
        Oracle filter & Particle filter with true DGP and parameters \\
    \end{tabular}
    \label{tab:placeholder}
\end{table}

\textit{Kalman filtering} procedure, proposed by \textcite{kalman1960new}, is an optimal way to filter the latent dynamic factor(s), when the underlying process is fully linear and Gaussian. All but one of the processes violate either the linearity or the Gaussian assumption. The following linear state space model is estimated for all of the datasets with the Kalman filter, with a \textit{state-transition} equation for the dynamic factor $x_t\in\mathbb R$
\begin{align}
    x_t &= \mu_x + \alpha x_{t-1} + \varepsilon_t && \varepsilon_t \sim \mathcal N(0, \sigma_x^2)
\end{align}

and a \textit{measurement equation} for the i=$1,\dots,k$ observable variables
\begin{align}
    y_{i,t} &= \mu_i + \beta_i x_t + u_{i,t} \\
    u_{t} &= (u_{1t}, \dots, u_{kt})^\prime \sim \mathcal N(0, R)
\end{align}

The maximum likelihood estimation is conducted on the combined training and validation data, totaling 800 time periods, to estimate $\mu_x, \mu_y \in \mathbb R$, $\alpha \in [0,1]$, $\sigma_x \in \mathbb R^+$, $\beta \in \mathbb R^{k}$, and a symmetric and positive semi-definite measurement error covariance $R \in \mathbb R^{k \times k}$. The Kalman filter is then used with the estimated parameters to obtain the factor estimate $\hat x_K$ on the 1000 period test data. For simplicity, the results from this estimation procedure will be referred to as the factor estimate by the "Kalman filter" or just "Kalman". Additionally to the estimated Kalman results, also the results from the Kalman filter with true linear parameters are reported with a shorthand "Kalman max".

The estimation of the Kalman filter benchmark is conducted in R using the 'kalmanfilter' package. For a detailed description of the Kalman filtering procedure, see for example \textcite{sarkka2023bayesian}.\\

\textit{Oracle particle filter} represents the best possible situation, where the optimal factor estimate is constructed using the true data generating process with the correct parameters. For the linear--Gaussian process 1 the Kalman filter is the oracle. For the rest of the processes the oracle is the \textit{auxiliary particle filter}, in pseudocode:

\begin{enumerate}
    \item Look-ahead (auxiliary) weights $\tilde w_{t-1}^{(i)} \propto w_{t-1} p(y_t | \tilde x_t^{(i)})$ using a deterministic predictor $\tilde x_t^{(i)}=\mathbb E[x_t | x_{t-1}^{(i)}]$ obtained with the true state-transition without stochasticity from old particles $x_{t-1}^{(i)}$. \parencite{PittShephard1999}.
    \item Systematic resampling of particle indices using the look-ahead weights $\tilde w_{t-1}^{(i)}$ to get a refined pool of particles $x_{t-1}^{(j)}$. \parencite{Kitagawa1996}.
    \item Propagation with true state-transition $x_t^{(j)} \sim p(x_t | x_{t-1}^{(j)})$ and the refined pool of particles.
    \item Weight correction by likelihood ratio $\frac{p(y_t | x_t^{(j)})}{p(y_t | \tilde x_t^{(i)})}$
    \item Adaptive resampling, when effective sample size (ESS) falls too low, ESS/(\# particles) < threshold. ESS is the inverse of sum of squared weights. \parencite{KongLiuWong1994}.
    \item Roughening after resampling adds versatility to the particles by adding small random numbers to them
    \item The factor estimate $\hat x_t^{O}=\sum_j w_t^{(j)}x_t^{(j)}$ is the pointwise weighted mean of particles
\end{enumerate}

For more information, see for example \textcite{DoucetJohansen2011}. See \textcite{fernandez2005estimating} for an example on using the particle filter in economic modeling.

\subsection{Evaluation criteria} \label{subsec:eval}

The performance of the Transformer is evaluated using several metrics.

In order to make the results comparable, all factor estimates are first oriented and scaled optimally based on the true latent factor series. This is done by minimizing the mean absolute errors with the true factor
\begin{align}
\min_{\{\gamma_0, \gamma_1\}} \sum_{t=1}^{N^{train}} |\gamma_0 + \gamma_1 \tilde x_{t,model} - x_t|
\end{align}

where $\tilde x_{t,model}$ is the raw output from a factor model, model $\in$ \{Transformer, Kalman, Oracle\}. Mean Absolute Errors match the loss function, and the assumption of non-Gaussian data. The optimal scaling parameters $\gamma_0, \gamma_1$ are calculated on the combined training and validation data of 800 periods. These parameters are then used to scale the factor estimate of the model in the 1000 period test set.
\begin{align}
\hat x_{t,model} = \gamma_0 + \gamma_1 \tilde x_{t,model}
\end{align}

\textit{Mean squared errors} $\text{MSE}_{\text{model}} = \sum_{t=1}^{N^{test}} \frac{(\hat x_{model,t} - x_{t})^2}{N^{\text{test}}}$ are used to calculate further metrics, such as the correlation coefficient $R^2$.
\begin{align}
    R_{\text{model}}^2 &= 1 - \frac{\text{MSE}_{\text{model}}}{Var(x)}
\end{align}

$R^2$ describes how much of the variation of the factor a model can explain. The rest of the metrics are relative to a baseline model.

\textit{Fit} is a comparative metric of accuracy, defined as the relative change in explanatory power, when using the Transformer instead of the baseline Kalman filter.
\begin{align}
    Fit = \frac{\text{MSE}_{\text{Kalman}} - \text{MSE}_{\text{Transformer}}}{\text{MSE}_{\text{Kalman}}}*100 \in (-\infty,100] \label{eq:fit}
\end{align}

where $\hat x_T$ is the average of Transformer's factor estimates from equation~\eqref{eq:factorEstimate} over r training runs with different initial parameters. Kalman filter's factor estimate is denoted by $\hat x_K$. Best possible result is $100\%$, meaning that the Transformer estimated the factor series perfectly. This is infeasible due to stochasticity. Fit of 0\% means a tie and a negative value indicates that the Kalman filter was more accurate. The Fit is calculated only for the test sets of each process. The standard deviation in Fit across seeds is considered as a measure of consistency of the results.

Two different Fit scores are reported for the Transformer, one attainable with real data and one currently unattainable. The realistic result comes from early stopping the training based on validation loss, and the best achievable result comes from early stopping based on test Fit. The previous result is reported under "Fit" and the latter under "Fit max".

\textit{Gain} describes the share of the explanatory gap between the Kalman baseline and the oracle filter, which the Transformer manages to cover up.
\begin{align}
    Gain = \frac{\text{MSE}_{\text{Kalman}} - \text{MSE}_{\text{Transformer}}}{\text{MSE}_{\text{Kalman}} - \text{MSE}_{\text{Oracle}}}*100 \in (-\infty,100]
\end{align}

with a similar interpretation of the number as with Fit.

\textit{Correlation} measures the scale- and shift-invariant linear alignment between the factor and the factor estimate, $corr(x, \hat x)=\frac{cov(x, \hat x)}{sd(x) sd(\hat x)}$. Also the \textit{validation loss} is reported.

Further analysis could be done to evaluate whether the prediction accuracies differ meaningfully, for example by using the Diebold-Mariano test \parencite{diebold2002comparing}.

\subsection{Hyperparameters} \label{subsec:hyper}

Hyperparameters govern the Transformer's complexity and training process, and they can have a large impact on the outcome. Hyperparameters can be tuned for each dataset using a search algorithm to improve performance, with some added complications. Only one set of hyperparameters is used for all processes for simplicity. The hyperparameters were chosen while keeping an eye on the accuracy of the factor estimate. Further work is needed to create a tuning protocol based only on validation loss for real data, where the accuracy is unobservable.
\begin{table}[!htbp]
    \centering
    \caption{The same hyperparameters are used for all datasets in the Monte Carlo. Potentially substantial gains can be realized by tuning the hyperparameters separately for every dataset.}
    \begin{tabular}{lccl}
        & Hyperparameter & Value & Explanation \\
        \hline
        & $N^{\text{train}}$ & 800 & Training + validation observations \\
        & $N^{\text{test}}$ & 1000 & Test observations \\
        & $k$ & 5 & Observable variables \\
        & $P$ & 9 & Context window, lags \\
        & $d_m$ & 32 & Embedding dim \\
        Architecture & $n_{\text{head}}$ & 4 & Number of Attention heads \\
         & $d_k$ & 8 & Dimension of Q, K and V \\
        & $d_{ff}$ & 64 & FFN dim \\
        &  & Gelu & FFN activation \\
        & H & 1 & Encoders in a stack \\
        &  & 26k & Number of parameters \\
        \hline
        & b & 32 & Batch size \\
       & $lr$ & 1e-4 & Initial learning rate \\
       & $T_0$ & 100 & Scheduler cycle, epochs \\
      Training &  & 1000 & Max epochs \\
       & $dr$ & 0.15 & Dropout \\
       & $L2$ & 0.015 & Weight decay (Ridge)\\
       & $\lambda$ & 60\% & Relative weight on prior regularizer in loss\\
       & $Loss$ & MAE & Mean absolute error loss function \\
        \hline
    \end{tabular}
    \label{tab:hyper}
\end{table}

I chose a modest and simple, but reasonably powerful size for the Transformer. The recent double descent literature highlights the benefits of large overparameterized models \parencite{nakkiran2021deep}. Even though the Transformer with 26 000 parameters is overparameterized for the training data, it is orders of magnitude smaller than what is typical. The length of the context window is 9 lags, corresponding to three quarters of a year with monthly data, being long enough to observe short term dependencies but short enough to enable using granular tokens, which improve interpretability. The batch size is set to a small value of 32, to not miss rare events which reveal important information about the processes. With a large batch size the average gradient can lose nuances.

The weights are initialized using the \textcite{Glorot2010} scheme. Learning rates are adjusted during training for each parameter separately with standard Adaptive Moment (AdamW) optimizer with a decoupled weight decay. Learning rate scheduler conducts a linear warmup followed by a cosine decay in cycles of $T_0=100$ epochs, changing the maximum allowed learning rate to avoid rushing past a good local optimum or getting stuck in a bad one. The linear warmup dampens the initial large gradients, to give the optimizer time to calibrate the momentum estimates for reasonable parameter updates. This helps to prevent off-shooting a good local optimum or getting stuck in a bad local optimum. Restarting the cycle gives the Transformer a possibility to continue improving, with early stopping patience set to the length of one cycle.

Overfitting is reduced with dropout and L2 regularization. Dropout is a form of model averaging, which disables a fraction of the parameters during each training iteration to prevent the Transformer from relying on memorizing idiosyncrasies with individual parameters. The L2 regularization penalizes excess model complexity, similarly to ridge regression when used with AdamW. Both dropout and regularization are used sparingly, in order not to waste too much information in the small datasets.

\subsection{Simulated processes}

There are six processes, which are thematically organized into three groups. Each process is used to generate 10 datasets, where only the random shocks and errors change between the 10 seeds.\footnote{Out of the 10 seeds, 0-2 were discarded per process due to the Kalman filter baseline being clearly mis-estimated, with a flatlining optimally scaled factor series. The reported results are means over the 8-10 reasonable seeds per process.} The Transformer is trained 10 times for each seed, each time with different randomly initialized parameters. The Transformer's accuracy for one seed is calculated from the mean of the 10 runs.

All of the six processes have one state variable (dynamic factor) and k=5 observable variables. The training data consists of N=800 observations, of which 20\% are used for validation.\footnote{All datasets are simulated with R, with a burn-in of the first 1000 observations. Each dataset is stationary, non-cyclical and standardized. All observable variables have about the same signal-to-noise ratio. The level of nonlinearity and heavy tails were chosen to make the estimated Kalman filter baseline miss but not completely off.} The small amount of training data was chosen to reflect the typical situation with scarce macroeconomic data, corresponding to monthly observations over 65 years. The trained Transformer is evaluated on a test data of 1000 periods to guarantee that the results generalize. All of the reported statistics are calculated only from the test set, which was not involved in the parameter updates during training.

\textit{Sign preserving power function} (spow) is used to introduce nonlinearities to all of the processes
\begin{align}
f(z, \gamma, c) &= \ c_{}* \ \text{sign}(z) \left(\text{abs}\left(\frac{z}{c}\right)\right)^{\gamma_{}} \label{eq:spow}
\end{align}

Spow is just a power function $z^\gamma$ for positive inputs, but with capability of dealing also with negative inputs z. The scale c controls the exposure of the input to the nonlinear regions where the slope changes fast, making it possible to calibrate the level of challenge for each process.\footnote{For computational stability, the spow is smoothed with $\epsilon=0.0001$, to remove a kink in the derivative at 0, with $spow(z, \gamma, c) = \ c_{}* \ \text{sign}(z) \left[ (\text{abs}(\frac{z}{c_{}}) + \epsilon)^{\gamma_{}} - \epsilon^{\gamma_{}} \right ]$.}

\begin{figure}[H]
    \centering
    \includegraphics[width=\linewidth]{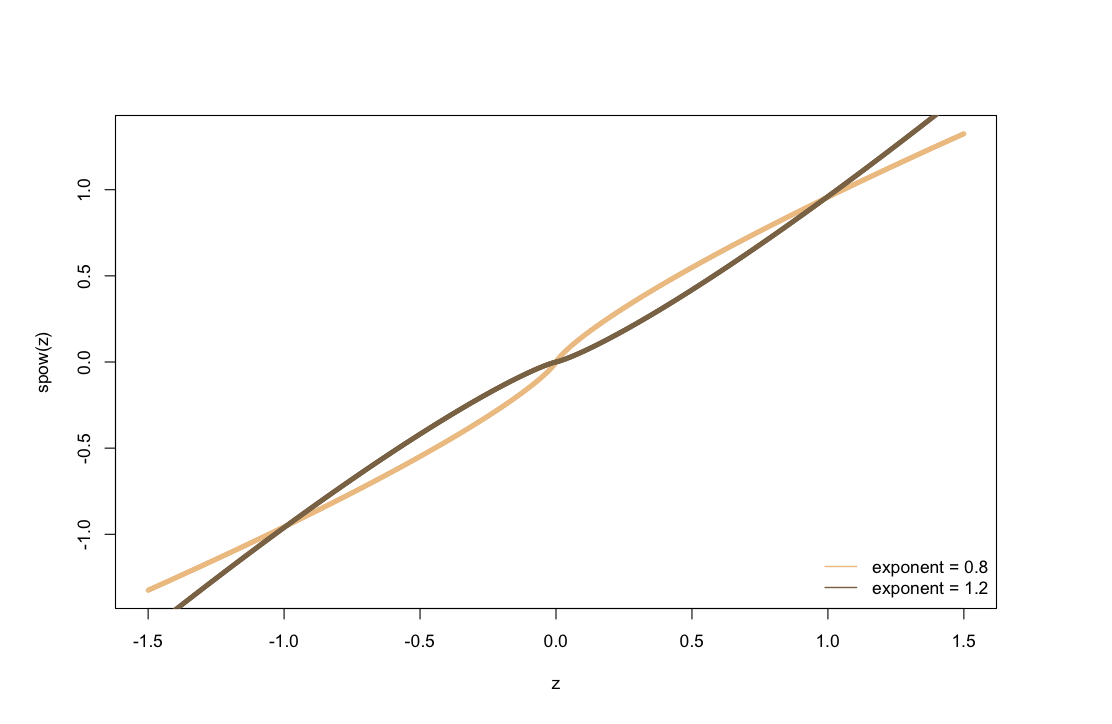}
    \caption{Sign preserving power function (spow) is used as the nonlinear function.}
    \label{fig:spow}
\end{figure}

\textit{Parameters.} State-transition is defined to be persistent with $\alpha=0.96$ to reflect typical macro data. State curvature is set with $\phi=0.8$ spow-exponent. Constant $\mu_x$ is set to zero so that the state travels across the nonlinear areas of the spow-function, without needing to use mean reverting structure for the state-space. The state is initialized as $x_0=0$. The linear coefficients related to measurements were drawn randomly from $\beta_i \sim \mathbb U(0.3, 1.7)$ and the exponents from $\gamma_i \sim \mathbb U(0.6, 1.4), i=1,\dots,k$, with $\mu_{yi} \sim \mathbb U(-1,1)$. Drawing only positive linear coefficients makes the system co-monotone, which enables using the mean of data to initialize the Transformer's factor series in the beginning of the State Encoder. The choice of having only positive $\beta$-coefficients is not very restrictive. In many cases a system with mixed negative and positive loadings of monotone observable variables can be transformed into a co-monotone system by re-orienting signs. However, it is noteworthy that the mean of the data has a relatively high correlation coefficient $R^2$ with the true factor in many of the processes, which might add positive bias to the results.

The \textit{measurement error covariance} matrix 
\begin{align}
    \Sigma_{y} = DRD \in \mathbb R^{k \times k} \label{eq:covariance}
\end{align}

is constructed from a vector of standard deviations $D=\text{diag}(\sigma_{y1}, \dots, \sigma_{yk})$ and a random correlation matrix R=($p_{i,j}$) with a unit diagonal $p_{i,i}=1$, and 0.3 average cross-correlation and 0.15 jitter off-diagonal $p_{i,j}, i\neq j$. The error covariance matrix is positive semi-definite with serially uncorrelated errors which are independent of the state shock. The volatility for the measurement error $\sigma_{y,i}$ of variable $i=1,\dots,k$ is chosen to enforce a uniform signal-to-noise ratio over the variables. The measurement error covariance $\Sigma_y$ is constructed separately for each of the six processes with process specific D and a random R.

\subsection{Results} \label{subsec:results}

\begin{figure}[H]
\begin{tabular}{ccc}
    \hspace{-1.2cm}
     \includegraphics{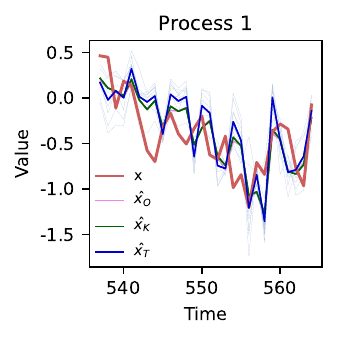}\hspace{-0.6cm} &  \includegraphics{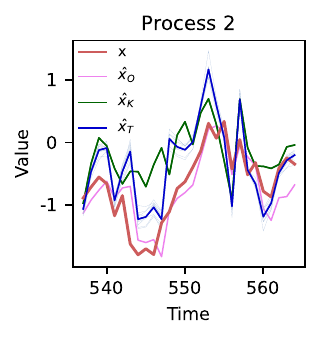}\hspace{-0.6cm} & 
     \includegraphics{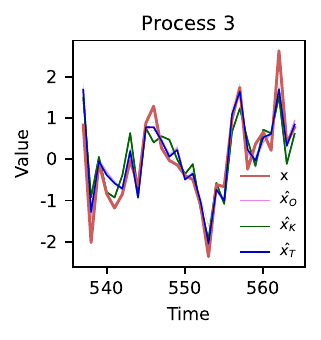}\hspace{-0.6cm}\\
     Fit -25.8\% (26.2) & Fit 25.0\% (4.3) & Fit 45.7\% (5.5) \\
    \hspace{-1.1cm}
    \includegraphics{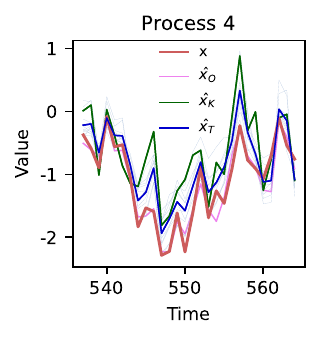}\hspace{-0.8cm} &  \includegraphics{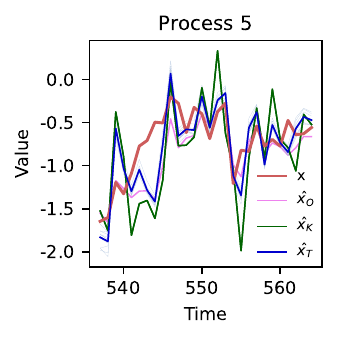}\hspace{-0.8cm} &  \includegraphics{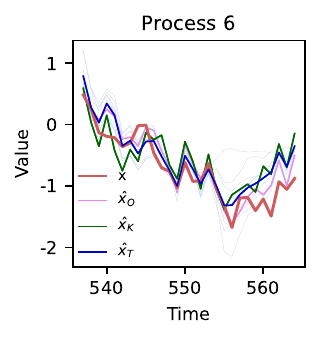}\hspace{-0.8cm} \\
     Fit 49.2\% (23.3) & Fit 28.2\% (8.1) & Fit 25.7\% (21.9)\\
\end{tabular}
\caption{Snapshots from the out-of-sample test set for one seed per process. Accuracy measure Fit describes the Transformer's relative performance against the Kalman filter. The values are averages over training runs, with standard deviations in brackets. Lines $x, \hat x_O, \hat x_K$, and $\hat x_T$ correspond to the true factor, oracle filter, estimated Kalman filter and the Transformer, respectively. Transformer's all training runs are depicted with thin lines and the piecewise mean with a thick blue line.}
\label{fig:all_results}
\end{figure}

The results are summarized in Table~\ref{tab:results}. The reported values for a process are averages over seeds, with standard deviations in brackets. Figure~\ref{fig:all_results} shows a 30-period test set snapshot of factor estimates from each process with one seed. The value for one seed is an average over the training runs with different initial parameters. Additionally, processes 2, 4, and 6, are looked upon more closely in Subsubsections~\ref{results:set1}, \ref{results:set2} and \ref{results:set3}, where I chose one representative seed and visualize the data, factor estimates, real time factor projections, training process, residual stream, and Attention patterns. All of the 1000-period test set factor estimates can be found in the Appendix with zoomable vector graphics, alongside with information about the training process. The Appendix also holds 20-period look-ahead prediction plots for the observable variables of processes 2, 4 and 6. These visualizations can be used to assess what kind of dynamics the Transformer learns for the measurement equation. Table~\ref{tab1:results_all_seeds} in the Appendix documents the results from all seeds.

\begin{table}[H]
    \centering
    \caption{Results from six simulated processes. The Transformer outperforms the Kalman filter baseline on average by 20\% out-of-sample, on datasets which deviate from the linear--Gaussian. The Transformer fares well especially with processes 2--4, with more variable results with processes 5 and 6.}
    \small
    \begin{tabular}{l?ccccc}
        Process & Fit\% & Fit max\% & Gain & R$^2$ & Loss  \\
       \hline
       1: linear--Gaussian  &  -38.3 (19.8) & -12.2 & -  & 0.779 (0.03) & 0.2437   \\
       2: Nonlinear measurements  & 32.5 (12.6) & 57.7 & 41.5 (14.7) & 0.636 (0.07) & 0.2656  \\
       \hline
       3: Nonlinear state & 42.8 (6.6) & 46.6 & 88.1 (22.8) & 0.801 (0.17) & 0.3153  \\
       4: Fully nonlinear  &  45.3 (23.6) & 65.7 & 50.8 (28.9) & 0.841 (0.03) & 0.2593   \\
       \hline
       5: ARMA(3,1) & -8.8 (56.8) & 30.3  & -8.1 (94.8) & 0.772 (0.15) & 0.1707 \\
       6: Markov switching & 4.6 (29.1) & 30.3  & 7.3 (46.4) & 0.876 (0.04) & 0.1035
    \end{tabular}
    \label{tab:results}
\end{table}

\subsubsection{Nonlinear measurements} \label{results:set1}

\textit{Process 1}
\begin{align}
x_t &= \mu_x + \alpha x_{t-1} +  \varepsilon_t, &&\varepsilon_t \sim \mathcal N(0, \sigma_{x}^2) \label{eq:pros1a} \\
y_{i,t} &= \mu_{i,t} + \beta x_t + u_{i,t}, && u_{t} \sim \mathcal N(0, \Sigma_y) \label{eq:pros1b}
\end{align}

is linear--Gaussian in both state-transition and measurements.

\noindent \textit{Process 2}
\begin{align}
x_t &= \mu_x + \alpha x_{t-1} +  \varepsilon_t, &&\varepsilon_t \sim \mathcal N(0, \sigma_{x}^2) \label{eq:2a} \\
y_{i,t} &= \mu_{i,t} + \beta_i f(x_t, \gamma_i, c_y) + u_{i,t}, && u_{t} \sim \mathcal N(0, \Sigma_y) \label{eq:2b}
\end{align}

has linear--Gaussian state-transition, but nonlinear and non-Gaussian measurements. Error vector $u_{t} = (u_{1t}, \dots, u_{kt})^\prime$, $f(\cdot)$ is the spow function described in equation~\eqref{eq:spow} and the p.s.d. measurement error covariance matrix $\Sigma_y$ is defined in equation~\eqref{eq:covariance}.\footnote{Randomly drawn parameters used in simulating Process 1 are $\mu_y$: c(0.11, 0.61, 0.70, -0.74, 0.65) and $\beta$: c(1.01, 1.25, 0.60, 0.98, 0.91), while Process 2 has $\mu_y$ = c(0.79, -0.47, -0.256, 0.146, 0.82), $\beta$ c(0.58, 1.56, 1.62, 1.23, 1.18), $\gamma$ = c(0.55, 1.37, 0.57, 1.48, 0.61), $c_y=0.77$.}

Table~\ref{tab1:results1and2} summarizes the results from comparison to the estimated Kalman filter on one seed. As expected, the Transformer clearly loses to the estimated Kalman filter with the linear--Gaussian Process 1. It should not be possible for the Transformer to outperform the Kalman filter when the filtering is conducted with true parameters without estimation. Indeed, the Transformer did not exceed Kalman max once even when the early stopping was executed based on test Fit. However, the estimation process with Kalman filter can still sometimes create inadequate factor estimates due to small sample noise, persistent factor and a sensitive scaling procedure. The Transformer won the estimated Kalman once out of 10 seeds, with 5\% better accuracy. The Transformer's performance can be easily improved on process 1 by increasing the weight on prior information, as can be seen in Subsection~\ref{subsec:priorimpact}.

In Process 2 the Transformer managed to outperform the estimated Kalman filter by 25\% (Fit), explaining about 32\% (Gain) of what is left to explain between the Kalman filter and the oracle particle filter.

Table~\ref{tab1:results1and2comp} compares the explanatory power between benchmark models, in terms of R$^2$. In Process 1 the oracle particle filter, estimated Kalman filter and the Kalman max all reached the same result of $R^2= 0.82$ with the chosen representative seed. Using a different early stopping criterion did not make much difference for the Transformer either. In Process 2 the gap is substantial between the oracle and the two Kalman filters. Also the gap between the two Transformers with different early stopping criteria grew similarly.
\begin{table}[!htbp]
    \centering
    \caption{Comparing the explanatory power R$^2$ of baseline models.}
    \small
    \begin{tabular}{c?cc?cccc}
        R$^2$ & Transformer & Transformer max & Kalman & Kalman max & Oracle & Mean(y) \\
       \hline
       Process 1  &  0.77 & 0.79 & 0.82 & 0.82 & 0.82 & 0.66 \\
       Process 2  &  0.6 & 0.75 & 0.31 & 0.79 & 0.87 & 0.63
    \end{tabular}
    \label{tab1:results1and2comp}
\end{table}
\begin{table}[!htbp]
    \centering
    \caption{Comparing Transformer's performance to the estimated Kalman filter. The results characterize one seed and therefore can deviate from the averages in Table~\ref{tab:results}. Gain is not reported for process 1, because of a very small denominator.}
    \small
    \begin{tabular}{c?ccccc}
        Transformer & Fit & Fit max & Gain & Corr & Loss \\
       \hline
       Process 1  &  -25.8 (26.2) & -12.2 (7.1) & - & 0.88 & 0.2473 \\
       Process 2  &  25.0 (4.3) & 57.7 (8.0) & 32.2 & 0.8 & 0.2570
    \end{tabular}
    \label{tab1:results1and2}
\end{table}

\begin{figure}[!htbp]
  \centering
  \subfloat[Data. Latent factor x depicted in red, five observable variables in grey and their mean in dashes.]{\includegraphics{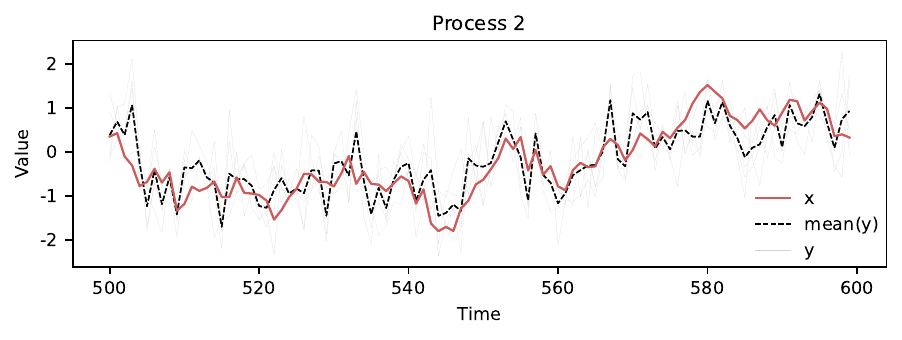}}  \\
  \subfloat[Factor estimates. Each period's estimate is based on a 9 lag look-back window.]{\includegraphics{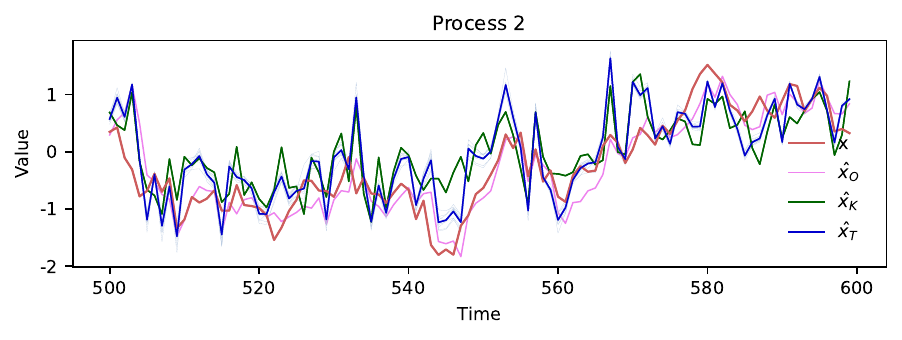}} \\
  \subfloat[Real time projections extend the factor estimate 20 periods ahead with each information set.]{\includegraphics{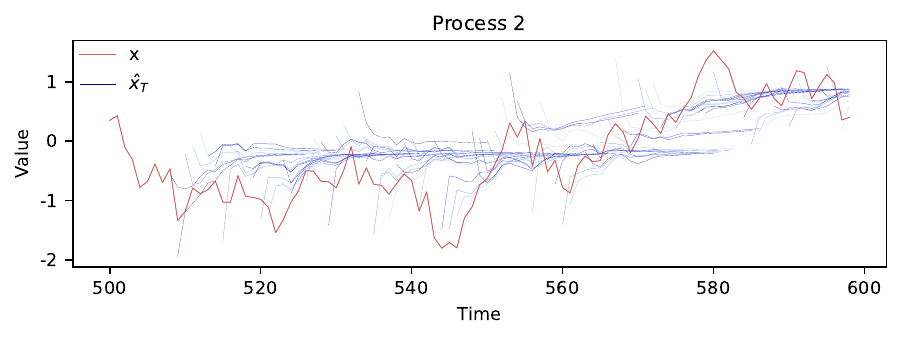}}
  \caption{Visualizing 100 periods from the out-of-sample test set.}
  \label{fig1:process2}
\end{figure}

The results for Process 2 are analyzed further by focusing on a representative segment from the middle of the test set. Figure~\ref{fig1:process2} shows that the Transformer's factor estimate moves quite similarly to the Kalman filter, but deviates constantly towards the true factor. The variation across training runs is decently small, depicted in thin blue lines.

The 20-periods-ahead real time factor projections in panel (a) of Figure~\ref{fig1:process2} use the Transformer with best Fit across training runs, and therefore the results deviate somewhat from the average factor estimate depicted in panel (b). In applications where Fit is unobservable, a random training run or the run with lowest validation loss could be chosen instead. The factor projections mostly predict convergence towards the mean, which is appropriate for a stationary and simple underlying state-transition. The projections start to predict an upward trend by the end of the visualized segment, when the estimated factor series has multiple subsequent positive values. Real time predictions for the observable variables can be found in the Appendix, Figure~\ref{fig:pred2}. The Transformer is modified for estimating factors, and the predictions come as a side effect, but they can still be useful for deducing what kind of dynamics the Transformer has found for the measurements.

\begin{figure}[!htbp]
  \centering
  \subfloat[Residual stream]{\hspace{-0.5cm}\includegraphics[scale=1]{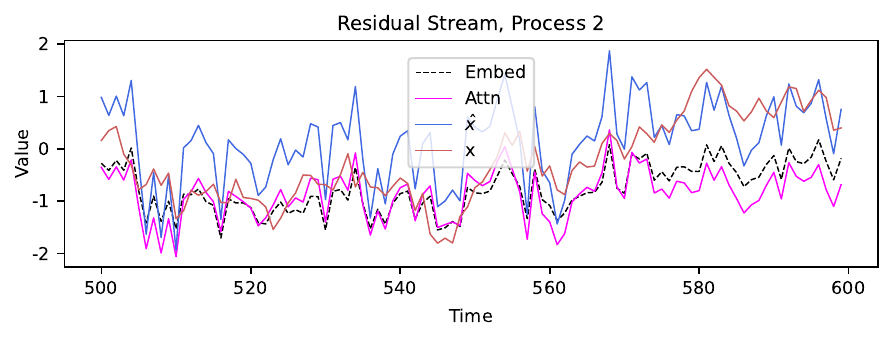}}  \vspace{-0.4cm}\\
  \subfloat[Learning through epochs, Fit normalized]{\hspace{-0.5cm}\includegraphics[scale=1]{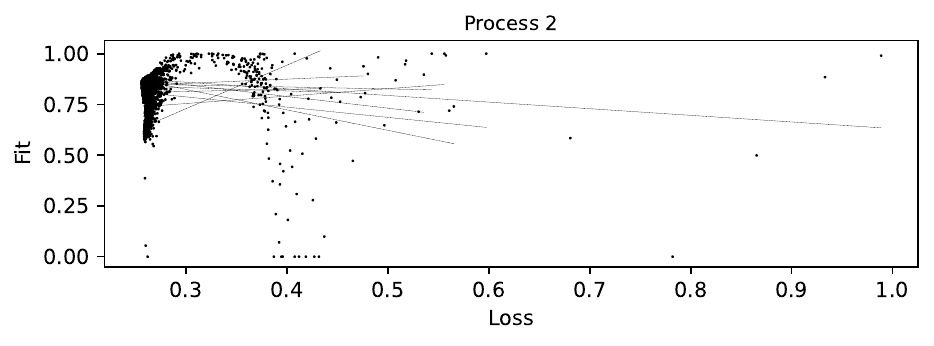}}
  \caption{A look into the internal functioning and the training process of the Transformer.}
  \label{fig1:process2res}
\end{figure}

Figure~\ref{fig1:process2res} visualizes the residual stream in panel (a). In this case the Attention (Attn) and FFN ($\hat x$) steps don't dramatically change the shape of the initial factor series, but there are some changes beyond scaling and shifting. Panel (b) shows that the Transformer learns more accurate representations for the factor estimate through training. On most training runs the lowering validation loss corresponds to a higher Fit, as can be seen from the declining regression lines between loss and fit. The few runs where the regression line is inclining got likely stuck in some bad local optima. The Y-axis is standardized for the plot. Further aspects of the training process are visualized in Figure~\ref{fig1:loss_1to4} in the Appendix.

\begin{figure}[!htbp]
  \centering
  \hspace{-0.5cm}\includegraphics[scale=1]{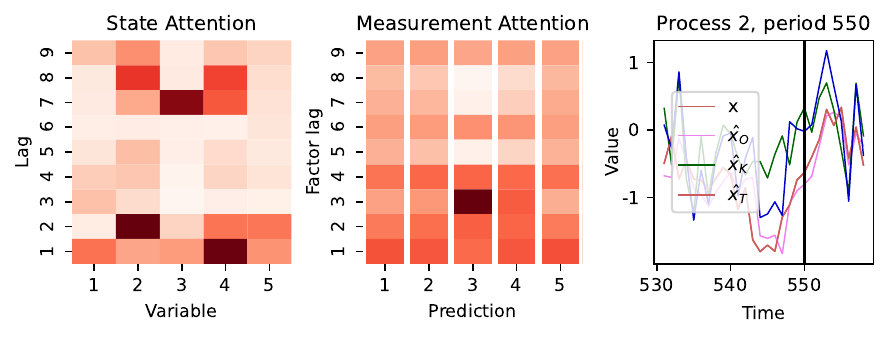}
  \caption{Snapshot of the Attention matrices describing one period.}
  \label{fig1:process2_attention}
\end{figure}

Figure~\ref{fig1:process2_attention} prints out the State- and Measurement Attention matrices, describing the factor estimate on period 550 and the predictions for period 551. The State Attention matrix seems to regard variables two and four most important at that period. Higher attention to some of the lags further in the past indicate that an important moment occurred seven or eight periods ago. Seemingly, this moment was when the factor began a strong incline from a low starting point, as an aftermath of several positive shocks. It is possible that variables two and four delivered stronger local information about the shocks than the rest of the variables in that moment, based on how their individual measurement curvature reacts to the location of the state x via the measurement link, sign preserving power function. The Measurement Attention uses the whole factor series in predicting the observables, with more weight on the recent values.

\begin{figure}[H]
  \centering
  \subfloat[State Attention for variables over time]{\hspace{-0.5cm}\includegraphics[scale=1]{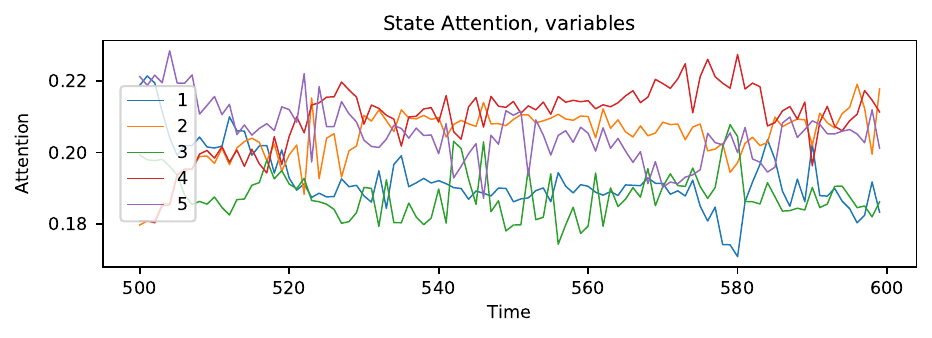}}  \vspace{-0.4cm}\\
  \subfloat[State Attention for lags over time]{\hspace{-0.5cm}\includegraphics[scale=1]{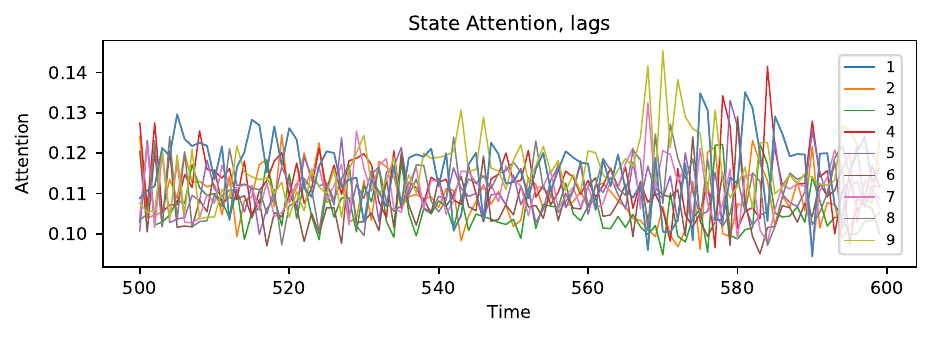}} \vspace{-0.4cm}\\
  \subfloat[Measurement Attention for lags over time]{\hspace{-0.5cm}\includegraphics[scale=1]{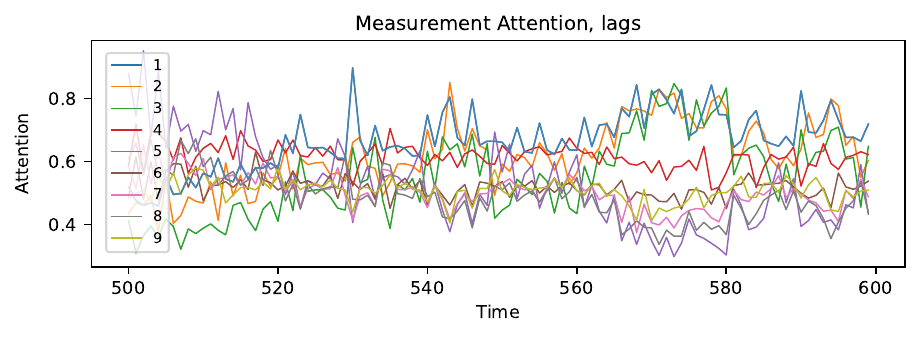}}
  \caption{Evolution of Attention patterns over time.}
  \label{fig1:process2Attn}
\end{figure}

Figure~\ref{fig1:process2Attn} combines snapshots from every period to visualize how the Attention patterns change over time. The construction of these figures from Attention matrices is explained in Subsection~\ref{subsec:attention}. Panel (a) shows the relative importance of different variables for the factor estimate. The importance of variables 2 and 4 develop similarly from low to high over the time span. This is likely because they both have a similar curvature governed by an exponent larger than one, $\gamma_2=1.37$ and $\gamma_4=1.48$, with respect to the factor. With the sign preserving power function this means that they convey more information about the factor when the factor's value deviates further from zero. Panel (b) shows the importance of different lags over time. The Attention pattern seems to vary a lot locally, but the pattern is actually quite stable in that only relatively small changes occur between the lag weights. The reason for this is probably related to the Gaussian stochasticity in the state-transition, as heavier tails create a very different looking lag-Attention-pattern for Process 4, Figure~\ref{fig1:proc4_4}. The volatility of the pattern does increases after period 560, corresponding to the upward trend of the factor, potentially signaling about increased uncertainty. Smoothing the curves could help detect a possible hierarchy of importance between the lags and long term changes in it, as shown in Figure~\ref{fig1:attn_smoothing}. Panel (c) describes the importance of factor lags on the predictions. The pattern shows that the most recent factor estimate is typically most relevant, with other lags supplementing information to varying degrees.

\subsubsection{Nonlinear state} \label{results:set2}

\textit{Process 3}
\begin{align}
x_t &= \mu_x + \alpha f(x_{t-1}, \phi, c_x) +  \varepsilon_t, \hspace{0.5cm}  &&\varepsilon_t \sim t(0, df_p, \sigma_{x}^2) \label{eq:pros2a} \\
y_{i,t} &= \mu_{i,t} + \beta_i x_t + u_{i,t}, && u_{t} \sim \mathcal N(0, \Sigma_y) \label{eq:pros2b}
\end{align}

has nonlinear and non-Gaussian state-transition but linear--Gaussian measurements.
\\

\noindent \textit{Process 4}
\begin{align}
x_t &= \mu_x + \alpha f(x_{t-1}, \phi, c_x) +  \varepsilon_t, \hspace{0.5cm}  &&\varepsilon_t \sim t(0, df_p, \sigma_{x}^2) \label{eq:prosAx} \\
y_{i,t} &= \mu_{i,t} + \beta_i f(x_t, \gamma_i, c_y) + u_{i,t}, && u_{t} \sim \mathcal N(0, \Sigma_y) \label{eq:prosAy}
\end{align}

is fully nonlinear and non-Gaussian in both state-transition and measurements. Function $f(\cdot)$ is the sign preserving power function from equation~\eqref{eq:spow} and measurement error covariance matrix $\Sigma_y$ is given by equation~\eqref{eq:covariance}.\footnote{Parameters used in simulating Process 3 datasets: $\mu_y$ = (0.79, -0.47, -0.26, 0.15, 0.82), $\beta$ = (0.58, 1.56, 1.62, 1.23, 1.18), $\gamma$ = (0.61, 1.24, 0.63, 1.34, 0.68) and $\phi$=0.36 and $\sigma_x=1.2$ and $c_x=0.13$. Process 4: $\mu_y$=(-0.46, -0.43, 0.24, 0.85, 0.10), $\beta$=(1.41, 1.50, 1.60, 0.94, 0.51) $\gamma$=(1.08, 0.67, 1.03, 1.02, 1.06), $\phi=0.8$, $\sigma_x=0.65$, $c_x=1$ and $c_y=15$.}

The Transformer's accuracy in process 3 is high and its variability is low. Gain of 94\% tells that the Transformer managed to explain most of what could be explained beyond the estimated Kalman filter. Fit in process 4 is very high, but also the variability over training runs is considerable.

\begin{table}[!htbp]
    \caption{Comparison to baseline models}
    \small
    \begin{tabular}{c?cc?cccc}
        R$^2$ & Transformer & Transformer max & Kalman & Kalman max & Oracle & Mean(y) \\
       \hline
       Process 3  &  0.87 & 0.87 & 0.76 & 0.87 & 0.87 & 0.87 \\
       Process 4  &  0.84 & 0.89 & 0.68 & 0.95 & 0.96 & 0.89
    \end{tabular}
    \label{tab:results3and4comp}
\end{table}

\begin{table}[!htbp]
    \centering
    \caption{Comparison to Kalman filter.}
    \small
    \begin{tabular}{c?ccccc}
        Performance & Fit & Fit max & Gain & Corr & Loss \\
       \hline
       Process 3  &  45.7 (5.5) & 46.6 (1.1) & 94.0 & 0.93 & 0.3102 \\
       Process 4  &  49.2 (23.3) & 65.7 (11.7) & 55.7 & 0.92 & 0.2340 
    \end{tabular}
    \label{tab:results3and4}
\end{table}

\begin{figure}[!htbp]
  \centering
  \subfloat[Data]{\includegraphics{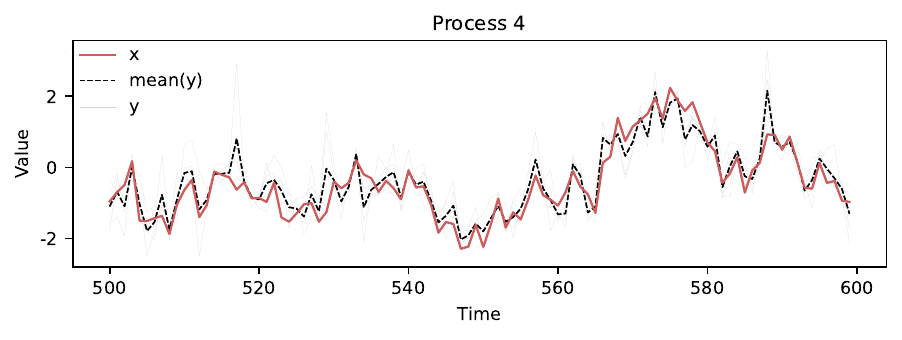}}  \\
  \subfloat[Factor estimates]{\includegraphics{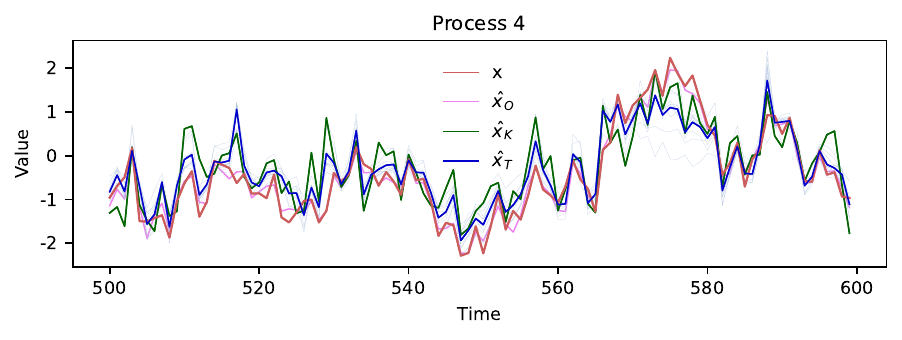}} \\
  \subfloat[Real time projections]{\includegraphics{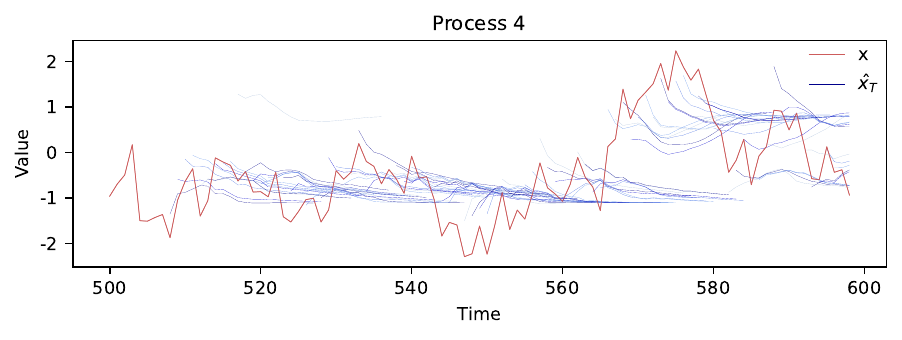}}
  \caption{Visualizing the Transformer's outcomes.}
  \label{fig1:proc4_1}
\end{figure}

Figure~\ref{fig1:proc4_1} shows that the Transformer moves somewhat similarly to the Kalman filter, with constant small departures towards the true factor. The Transformer also departures in the wrong direction during periods 570--580, when the factor peaks. Real time predictions for the observable variables can be found in the Appendix, Figure~\ref{fig1:proc4_1}.

\begin{figure}[!htbp]
  \centering
  \subfloat[Residual stream]{\hspace{-0.5cm}\includegraphics[scale=1]{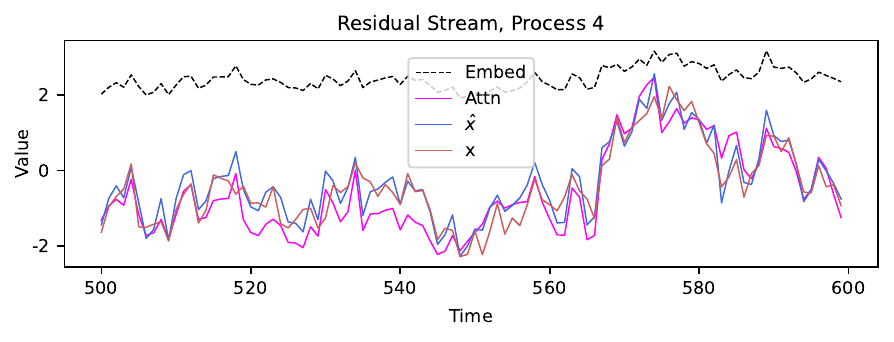}}  \vspace{-0.4cm}\\
  \subfloat[Learning through epochs]{\hspace{-0.5cm}\includegraphics[scale=1]{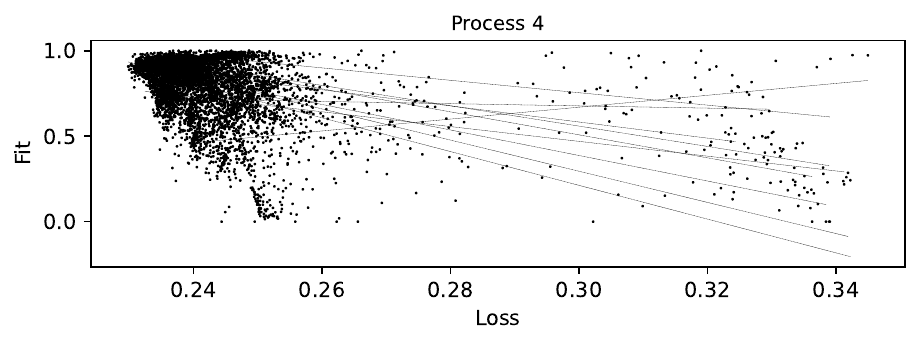}}
  \caption{Visualizing the Transformer's internal processing and training quality.}
  \label{fig1:proc4_2}
\end{figure}

\begin{figure}[!htbp]
  \centering
  \hspace{-0.5cm}\includegraphics[scale=1]{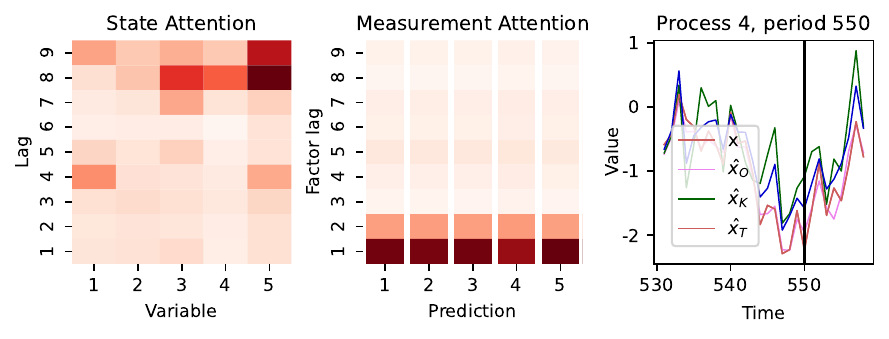}
  \caption{Snapshot of Attention matrices.}
  \label{fig1:proc4_3}
\end{figure}

\begin{figure}[H]
  \centering
  \subfloat[State Attention for variables over time]{\hspace{-0.5cm}\includegraphics[scale=1]{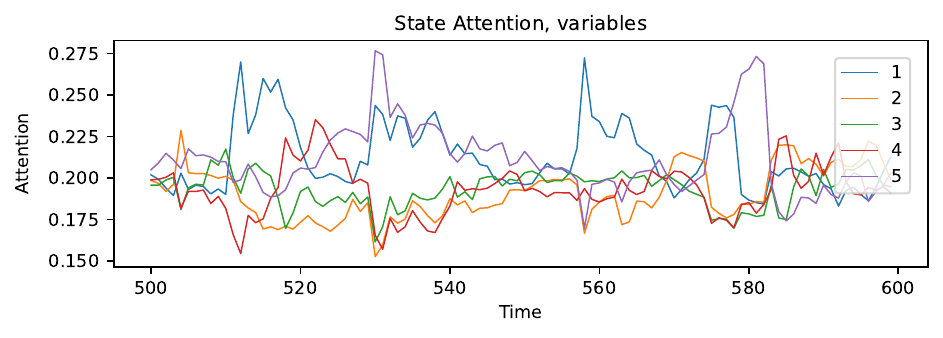}}  \vspace{-0.4cm}\\
  \subfloat[State Attention for lags over time]{\hspace{-0.5cm}\includegraphics[scale=1]{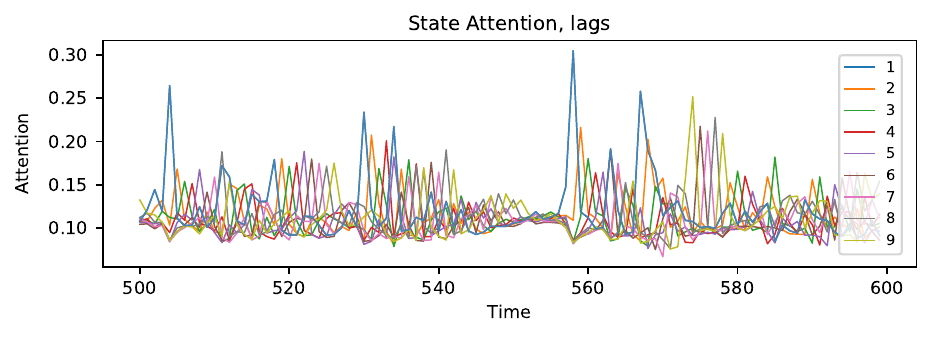}} \vspace{-0.4cm}\\
  \subfloat[Measurement Attention for lags over time]{\hspace{-0.5cm}\includegraphics[scale=1]{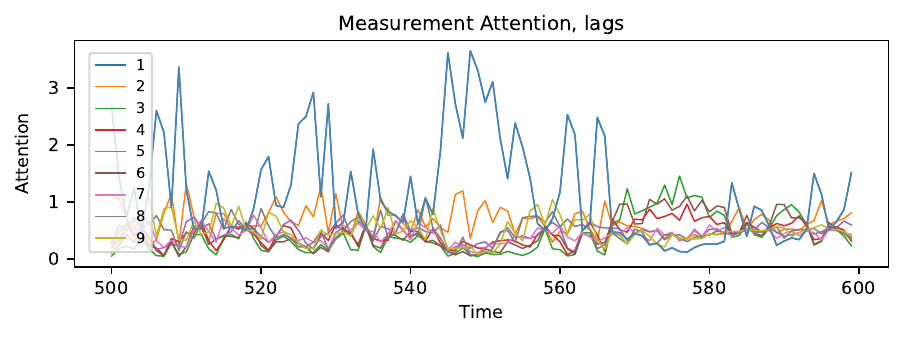}}
  \caption{Panel (a) shows that the variables have different information profiles regarding the factor. Depending on how far the factor is from zero, different variables increase in importance. The variables could be further clustered into information groups based on the co-movements of their Attention patterns. Panel (b) reflects the impact of non--Gaussian shocks in the state-transition, where disproportionately large jumps propagate through the lag structure, causing spike-patterns to the Attention plot.}
  \label{fig1:proc4_4}
\end{figure}

In Figure~\ref{fig1:proc4_3} the State Attention matrix emphasizes the past lags more than the recent ones. It might be the case that there had been a sequence of less consequential shocks up to that point in the context window, which contribute very little to the system dynamics. The lags 8 and 9 seem to correspond to the time when the factor stopped declining. Most of the relevant information is inferred from the two most recent factor lags in the Measurement Attention.

\subsubsection{Markov switching conditional volatility} \label{results:set3}

\noindent\textit{Process 5}
\begin{align}
x_t &= \mu_x + \sum_{i=1}^3 \alpha_i x_{t-i} + \varepsilon_t + \alpha_2 \varepsilon_{t-1}, \hspace{1.6cm}  \varepsilon_t \sim t(df=10, skew=-2, var=\sigma_x^2) \\
y_{ti} &= \mu_{yi} + \beta_i f(x_t, \gamma,c_y) + \sum_{l=1}^3 \gamma_{i l} y_{i,t-l} + u_{it}, \hspace{0.5cm} u_t \sim t(df=10, var=\Sigma_y)
\end{align}

has three lags in both the state and measurement equations, and one lag for the state shock, where $\alpha=[0.74, 0.15, 0.074]$ and $\gamma=[0.2, 0.05, 0.02]$. Otherwise the parameters are the same as before, equations \eqref{eq:prosAx} and \eqref{eq:prosAy}.\\

\noindent\textit{Process 6}
\begin{align}
x_t &=
\begin{cases}
    \mu_x + \alpha_1^{s} f(x_t, \phi^{s},c_x) + \sum_{l=2}^3 \alpha_{l} x_{t-l} + \varepsilon_t, & \text{when } s=0 \\
    \mu_x + \alpha_1^{s} f(x_t, \phi^{s},c_x) + \sum_{l=2}^3 \alpha_l x_{t-l} +  \varepsilon_t, & \text{when } s=1
\end{cases} \\
\varepsilon_t &=
\begin{cases}
    \sim \mathcal N(0,\sigma_x^2), & \text{when } s=0 \\
    \sim t(df=10, skew=-2, var=\sigma_x^2), & \text{when } s=1
\end{cases} \\
\nonumber\\
y_{ti} &= 
\begin{cases}
    \mu_y + \beta_i^s f(x_t, \gamma_i^s, c_y) + \sum_{i=1}^3 \gamma_{i} \beta y_{t-i} + u_t, &\text{when } s=0 \\
    \mu_y +  \beta_i^s f(x_t, \gamma_i^s, c_y) + \sum_{i=1}^3 \gamma_{i} \beta y_{t-i} + u_t, &\text{when } s=1
\end{cases} \label{eq:prosCy}\\
u_{t} &=
\begin{cases}
    \sim \mathcal N(0, \Sigma_y), & \text{when } s=0 \\
    \sim t(df=10, var=\Sigma_y), & \text{when } s=1
\end{cases}
\end{align}

 switches between two regimes $s\in\{0,1\}$ corresponding to downturn and growth periods, with different dynamics and volatility in both, inspired by \textcite{hamilton1989new} and \textcite{diebold1996measuring}. Parameters are increased to $\alpha_1^s=1.01 \alpha_1$ for s=0 and decreased to $\alpha_1^s=0.98 \alpha_1$ for s=1, with the same scalars 1.01 and 0.98 used similarly for the rest of the regime-dependent parameters $\phi^s, \beta_i^s$ and $\gamma_i^s$.\footnote{Parameters for Process 5: $\phi$ 0.8, $\sigma_x$ 0.387, $c_y=1.28$, $\mu_y$ c(0.79, -0.47, -0.26, 0.15, 0.82), $\beta$ c(0.58, 1.56, 1.62, 1.23, 1.18), $\gamma$ c(0.68, 1.12, 0.70, 1.21, 0.75). Process 6: $\sigma_x$ 0.51, $c_x=2.24$, $c_y=1.83$ $\phi$ 0.8, $\mu_y$ c(0.79, -0.47, -0.26, 0.15, 0.82), $\beta$ c(0.58, 1.56, 1.62, 1.23, 1.18), $\gamma$ c(0.57, 1.34, 0.59, 1.45, 0.62)}
 The Markov switching to growth occurs during each downturn period with 3\% probability, $\mathbb E [s=1|s=0]=0.03$, and going the other way around with 1\% probability, $\mathbb E [s=0|s=1]=0.01$. Function $f(\cdot)$ is the sign preserving power function, equation~\eqref{eq:spow}, and measurement error covariance matrix $\Sigma_y$ comes from equation~\eqref{eq:covariance}.

Processes 5 and 6 test the Transformer with a mixture of simultaneous challenges. In both of the challenging processes the Transformer explains about 60\% of the gap between the Kalman filter and the oracle filter for the seed, Table~\ref{tab:results5and6}. Across seeds the Transformer's results vary. The average result in Table~\ref{tab:results} is skewed downwards due to a very poor performance on a few of the seeds. On other seeds the Transformer managed to perform quite well with reasonable variation across training runs.

\begin{table}[!htbp]
    \caption{Comparison to all benchmark models.}
    \small
    \begin{tabular}{c?cc?cccc}
        R$^2$ & Transformer & Transformer max & Kalman & Kalman max & Oracle & Mean(y) \\
       \hline
       Process 5  & 0.88 & 0.89 & 0.80 & 0.84 & 0.91 & 0.90 \\
       Process 6  & 0.81 & 0.82 & 0.82 & 0.85 & 0.93 & 0.92
    \end{tabular}
    \label{tab:results5and6comp}
\end{table}

\begin{table}[!htbp]
    \centering
    \caption{Comparison to Kalman filter baseline.}
    \small
    \begin{tabular}{c?ccccc}
        Performance & Fit & Fit max & Gain & Corr & Loss  \\
       \hline
       Process 5  &  28.2 (8.1) & 33.1 (2.8) & 57.3 & 0.941 & 0.2016 \\
       Process 6  &  25.7 (21.9) & 30.3 (21.9) & 59.3 & 0.9 & 0.0959
    \end{tabular}
    \label{tab:results5and6}
\end{table}

\begin{figure}[!htbp]
  \centering
  \subfloat[a][Data]{\hspace{-0.5cm}\includegraphics{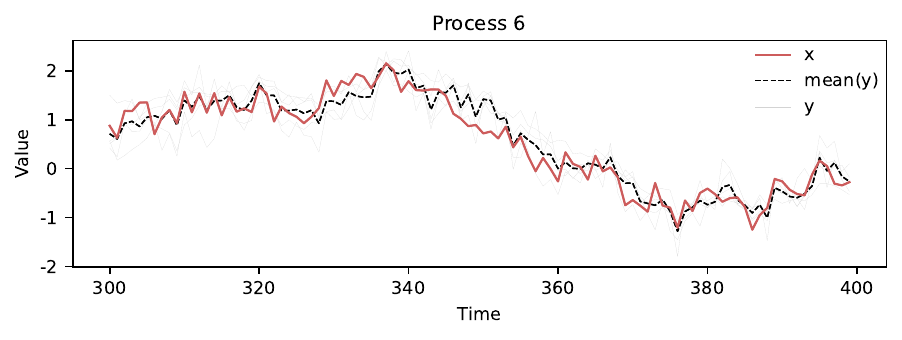}} \\
  \subfloat[b][Factor estimates]{\hspace{-0.5cm}\includegraphics{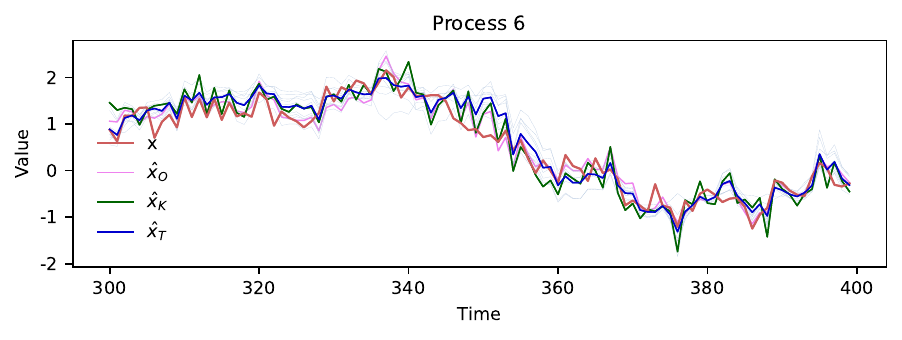}} \\
  \subfloat[c][Real time projections]{\hspace{-0.6cm}\includegraphics{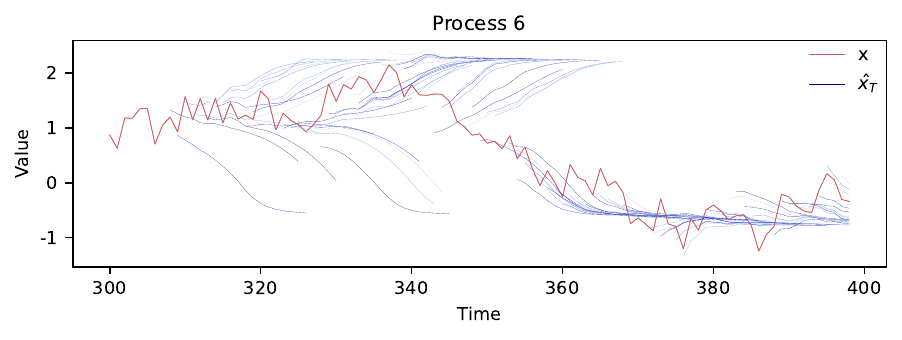}}
  \caption{(b) Transformer's factor estimate tracks the true factor reasonably. (c) The sudden change in the direction of the factor projections indicate a regime switch.}
  \label{fig1:proc6_1}
\end{figure}

The real time projections in Figure~\ref{fig1:proc6_1} reveals two directions to which the factor is expected to move, corresponding to the means of the two regimes. The Transformer correctly identifies a regime switch around period 350 and starts to project the factor downwards. The Transformer also seems to predict the regime change prematurely for a few times. Interestingly, the regime shift predictions do not result from a simple cutoff in terms of estimated factor value, as some of the downward trajectories begin from a higher starting point than some of the upward projections. The projection is therefore based on some local dynamic pattern which is mostly accurate.
\begin{figure}[H]
  \centering
  \subfloat[Residual stream]{\hspace{-0.5cm}\includegraphics[scale=1]{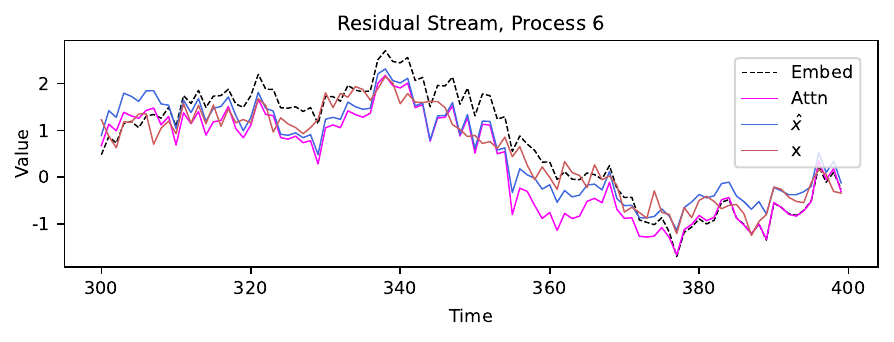}}  \vspace{-0.4cm}\\
  \subfloat[Learning through epochs]{\hspace{-0.5cm}\includegraphics[scale=1]{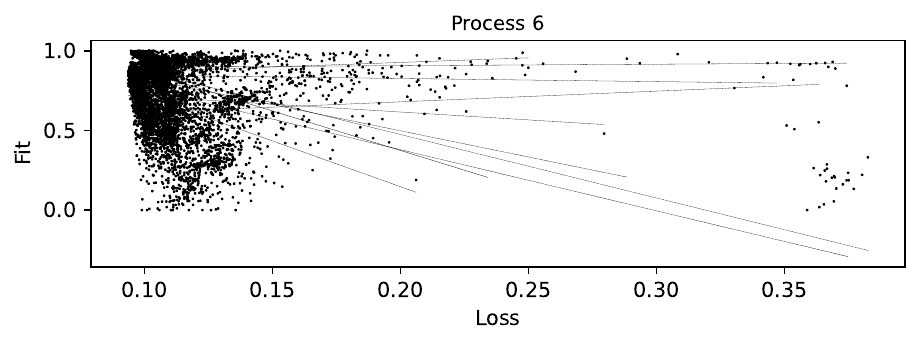}}
  \caption{(a) Transformer's internal factor representations at different stages. (b) Most training runs show an increasing accuracy while validation loss shrinks through training.}
  \label{fig1:proc6_2}
\end{figure}

\begin{figure}[H]
  \centering
  \subfloat[a][Growth regime]{\hspace{-0.5cm}\includegraphics[scale=1]{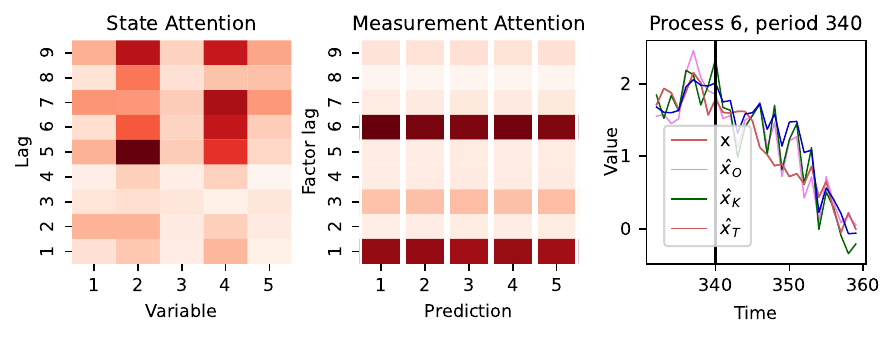}} \\
  \subfloat[b][Downturn regime]{\hspace{-0.5cm}\includegraphics[scale=1]{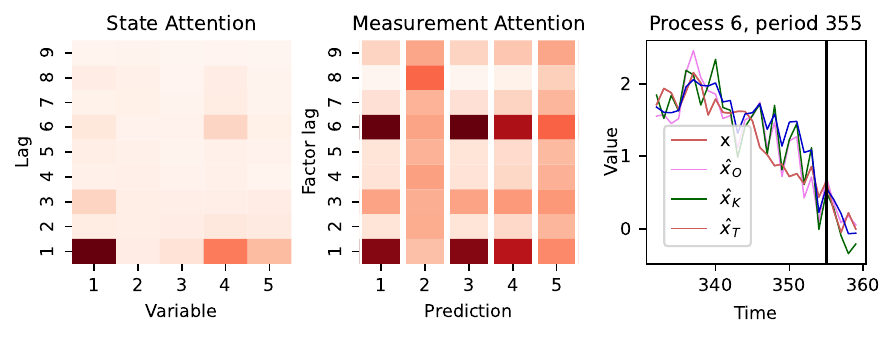}}
  \caption{Attention snapshots before and after a regime switch.}
  \label{fig1:proc6_3}
\end{figure}

In Figure~\ref{fig1:proc6_3} the State Attention patterns look very different in the two regimes. At the peak of the growth regime in (a) the Transformer pays most attention to multiple past values of variables 2 and 4. In the direct aftermath of a switch to downturn, panel (b), only the most recent values of variables 1, 4 and 5 receive attention. Also the Measurement Attention pattern changes. During the growth regime the Transformer has seemingly found a way to use the information in factor lags 1 and 6 for the predictions. In the new regime, at least during transition, some of the predictions have started to utilize information from the whole factor series.

\begin{figure}[H]
  \centering
  \subfloat[State Attention for variables]{\includegraphics{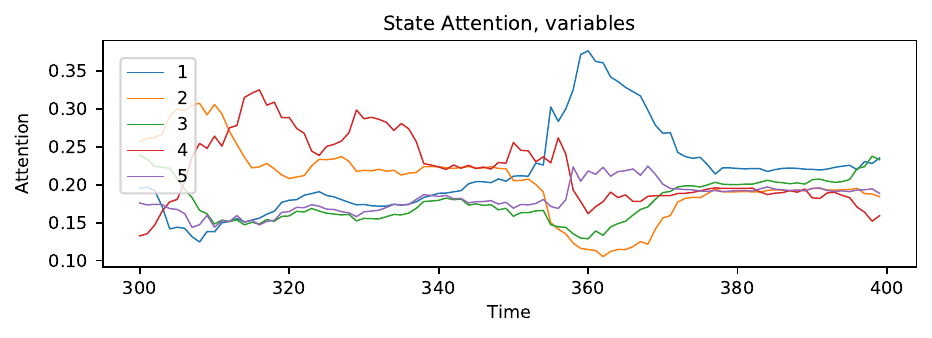}}  \\
  \subfloat[State Attention for lags]{\includegraphics{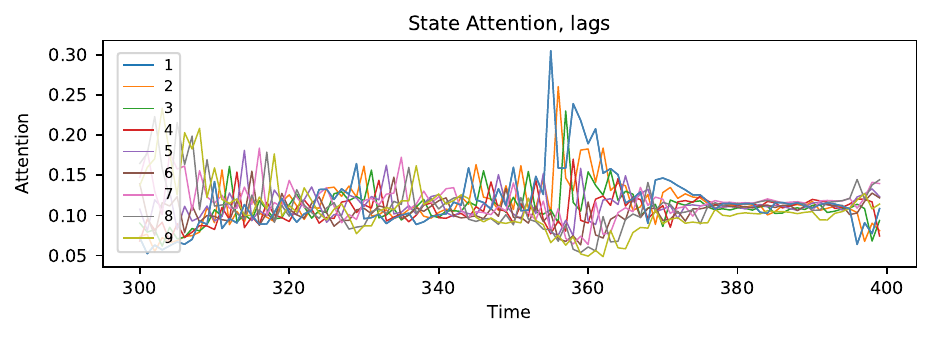}} \\
  \subfloat[Measurement Attention for lags]{\includegraphics{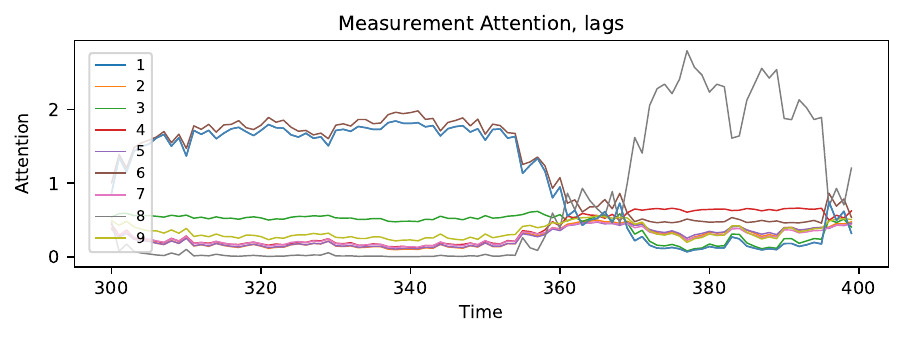}}
  \caption{An abrupt change in the Attention patterns reveals a regime switch.} \label{fig1:proc6_4}
\end{figure}

Figure~\ref{fig1:proc6_4} show a clear change in the attention patterns around period 350, hinting at a regime switch. Panel (a) shows a change in the composition of variables used for constructing the factor estimate. Variable 1 was unimportant in the first regime, and now becomes the most important conveyor of information about the factor's movements, especially during the transitory period. Panels (b) and (c) show a temporary change in the time-allocation of attention at the moment of the switch. The State Attention strongly increases the focus to the most recent variable values at the time of the regime shift. This moment remains important through the lags in subsequent periods. After the transitory period the underlying process reaches the new mean and the Attention patterns stabilize.

\subsection{Impact of prior information} \label{subsec:priorimpact}

The results in Subsection~\ref{subsec:results} were obtained using prior information in the training loss function with a 60\% weight. Functioning of prior information is explained in Subsection~\ref{subsec:loss}. In order to analyze the impact of prior information on the results, the Monte Carlo is repeated twice more: With a training loss function consisting only of prediction errors and no prior information, and then with a training loss function consisting only of prior information and no prediction errors. The 60\% weight prior information is found to substantially improve the accuracy and consistency of the results.

Without prior information, $\lambda=0$, the loss function in equation~\eqref{eq:loss} reduces to
\begin{align}
    L(\hat y, y) = \frac{1}{k} \sum_{i=1}^{k} | \hat y_{t+1, i} - y_{t+1, i} |
\end{align}

The accuracy of factor estimates and the consistency of results are reduced markedly in the absence of the prior information. The reason for this is probably the small number of training examples. The prior information gives useful structure to the Transformer's parameter space. Using a weak prior is in some ways analogous to the situation, where the Transformer is pre-trained with large amounts of data from a process that the prior-model resembles. Table~\ref{tab:results_prior} summarizes the results.

When only prior information is used to guide the estimation, $\lambda=1$, the loss function in equation~\eqref{eq:loss} becomes
\begin{align}
    L(\hat x_{T}, \hat x_{K}) =  \frac{1}{P} \sum_{t=1}^{P} |\hat x_{T,t+1} - \hat x_{K,t+1} |
\end{align}

The Transformer learns to estimate the dynamic factors for each dataset, as if it was a linear--Gaussian model.
\begin{table}[!htbp]
    \captionsetup{width=0.7\linewidth}
    \caption{Training without prior information leads to less accurate and less consistent results. By using only the Kalman prior in estimation, the Transformer learns to emulate a linear--Gaussian model. The best result can be found in between.}
    \small
    \centering
    \begin{tabular}{c?ccc}
       Fit\% & $\lambda=0$ & $\lambda=0.6$ & $\lambda=1$ \\
       \hline
       Process 2  & 6.1 (8.3) & 25.0 (4.3) & 1.4 (0.5) \\
       Process 4  & 33.3 (14.8) & 49.2 (23.3) & 1.0 (1.8)   \\
       Process 6  & 8.9 (49.2) & 25.7 (21.9) & -3.2 (3.7) 
    \end{tabular}
    \label{tab:results_prior}
\end{table}

The Transformer learns to emulate the arbitrary restrictions of a conventional model, merely by observing examples of the factor estimates from that model. Conventional factor models can be seen as special cases of the Transformer. The Monte Carlo experiment suggests that the results of a conventional model could in some cases be improved if it was estimated using the Transformer. The Transformer can also be used for testing how much the data seems to rebel against the restrictions of a conventional model, such as the linearity postulate by the Kalman filter.

\begin{figure}[!htbp]
\begin{tabular}{ccc}
    \textit{No prior information} & \textit{60\% prior information} & \textit{Only prior information} \\
    \hspace{-1.2cm}
     \includegraphics[scale=1]{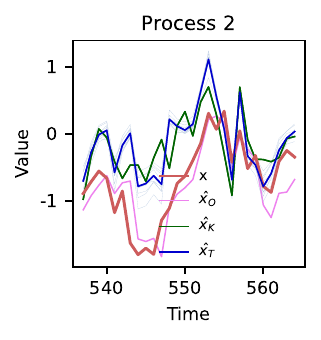}\hspace{-0.6cm} &  \includegraphics[scale=1]{Figures/process2_box.pdf}\hspace{-0.6cm} & 
    \includegraphics[scale=1]{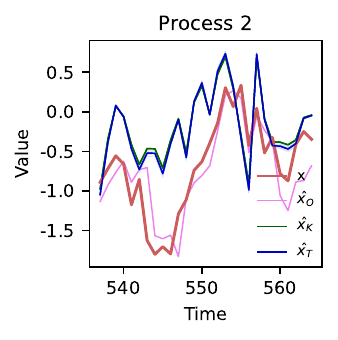} \vspace{-0.3cm} \\
     Fit 6.1\% (8.3) & Fit 25.0\% (4.3) & Fit 1.4\% (0.5) \\
    
    \hspace{-1.8cm}
    \includegraphics[scale=1]{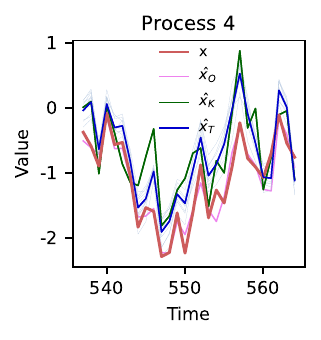}\hspace{-0.6cm} &  \includegraphics[scale=1]{Figures/process4_box.pdf}\hspace{-0.6cm} & 
    \includegraphics[scale=1]{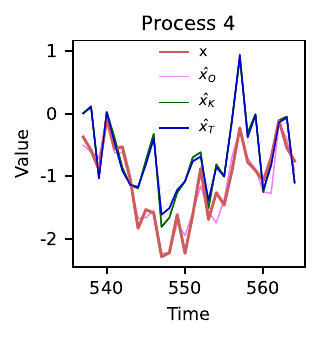}\vspace{-0.3cm} \\
     Fit 33.3\% (14.8) & Fit 49.2\% (23.3) & Fit 1.0\% (1.8) \\

    \hspace{-1.8cm}
    \includegraphics[scale=1]{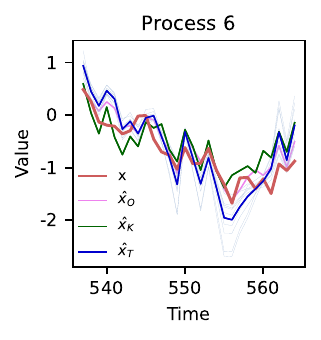}\hspace{-0.6cm} &  \includegraphics[scale=1]{Figures/process6_box.pdf}\hspace{-0.6cm} & 
    \includegraphics[scale=1]{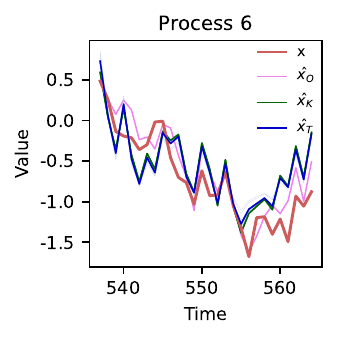} \vspace{-0.3cm}\\
     Fit 8.9\% (49.2) & Fit 25.7\% (21.9) & Fit -3.2\% (3.7) \\
\end{tabular}
\caption{Comparing the factor estimates for processes 2, 4 and 6 on rows, under different amounts of prior information on columns.}
\end{figure}

\textcite{GouletCoulombe2025} shows that there is a conceptual link between linear regression and the Attention mechanism, where the linear coefficients can be interpreted as a static Attention map. Building on this link, I visualize the Attention patterns for the three cases above to highlight the difference that prior information brings about. When the Transformer is trained using only prior information and different conventional factor models, it is possible to evaluate the Attention patterns implied by those models, and to compare them to the Transformer's less restricted patterns.

\begin{figure}[H]
\begin{tabular}{ccc}
    \textit{No prior information} & \textit{60\% prior information} & \textit{Only prior information} \\
    \hspace{-1.6cm}
     \includegraphics[scale=0.9]{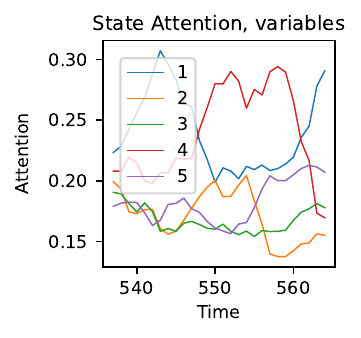}\hspace{-0.9cm} &  \includegraphics[scale=0.9]{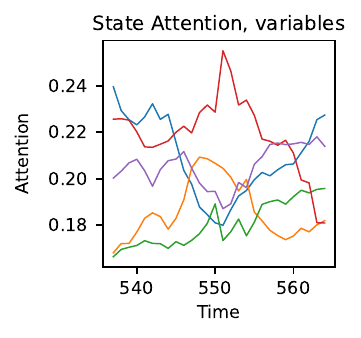}\hspace{-0.9cm} & 
    \includegraphics[scale=0.9]{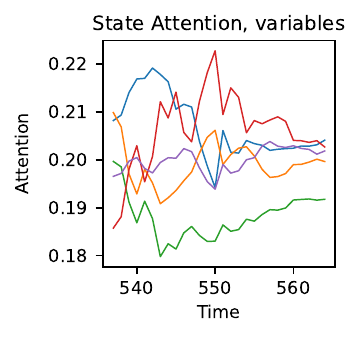} \vspace{-0.3cm} \\    
    \hspace{-1.8cm}
    \includegraphics[scale=0.9]{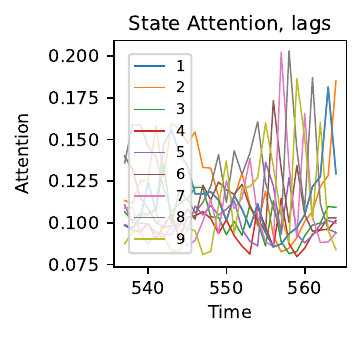}\hspace{-0.9cm} &  \includegraphics[scale=0.9]{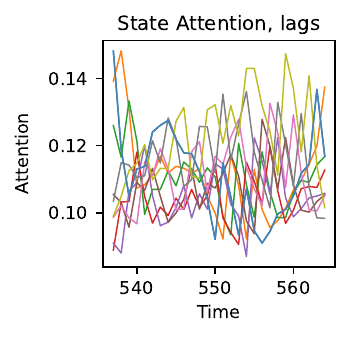}\hspace{-0.9cm} & 
    \includegraphics[scale=0.9]{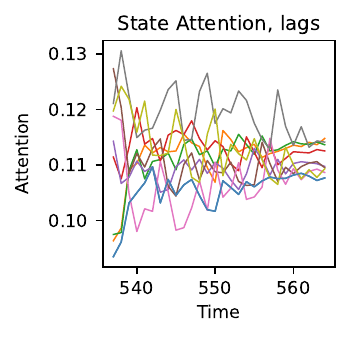}\vspace{-0.8cm}\\    
    \hspace{-1.8cm}
    \includegraphics[scale=0.9]{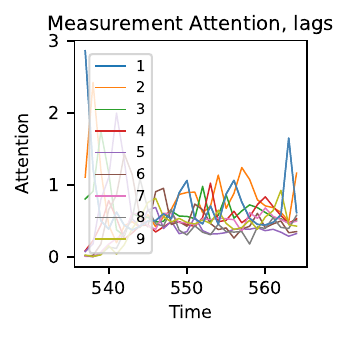}\hspace{-0.9cm} &  \includegraphics[scale=0.9]{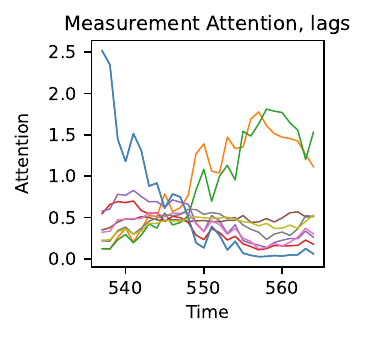}\hspace{-0.6cm} & 
    \includegraphics[scale=0.9]{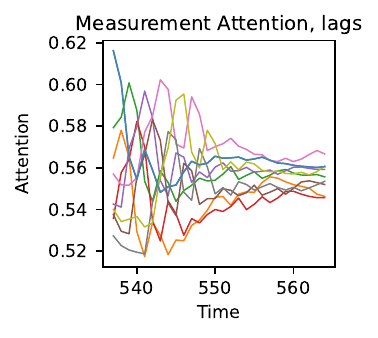}\vspace{-0.3cm}
\end{tabular}
\caption{Comparing the Attention patterns for process 6 under different levels of prior information. The rightmost column of figures show how the Attention patterns of a linear--Gaussian factor model would look like. Rows visualize the same Attention modules under different levels of prior information.}
\label{fig1:priorComparison2}
\end{figure}

Figure~\ref{fig1:priorComparison2} compares the Attention patterns for Process 6 under different strengths of prior information regularization. The rightmost column show how the Kalman filter's Attention patterns look like. Even though Kalman filter has time invariant linear coefficients, the Attention patterns can still change over time due to measurement errors and heteroskedastic shocks. Row one shows the relative importance of different variables on the factor estimate. The Attention pattern remains somewhat similar regardless of the prior information: The importance of variables 1 and 4 seem to correlate negatively with each other and the variable 3 is generally less important for determining the factor estimate. The impact of prior information can be better seen in the time allocation of Attention for the variable and for the factor lags, on rows two and three.


\section{Coincident Index for the U.S. economy} \label{sec:empirical}

I use the Transformer to estimate a measure of real activity for the U.S. economy, from variables related to industrial production, sales, labor income, and hours worked. These variables were proposed for constructing a coincident index in a seminal study by \textcite{stock1989new}. The resulting factor is regarded to coincide with the business cycle. The Transformer's coincident index is compared to a baseline linear--Gaussian coincident index, which follows the methodology in \textcite{stock1989new}.

To compare the results from both models, I focus on three recent recession episodes: The Covid crisis, The 2008 financial crisis, and the recession of 1990. The Transformer's factor estimate behaves quite similarly to the Kalman filter's factor estimate during the Covid recession. During the 2008 crisis and the recession of 1990, the Transformer dampens the volatility somewhat compared to the baseline. Both models tend to agree with the recession dates by the Business Cycle Dating Committee of the National Bureau of Economic Research (NBER).

\subsection{Data}

The time series dataset consists of $i=1,2,\dots, 4$ variables, with $t=1,2,\dots,670$ monthly observations from 1967 to 2025. The four variables, describing the macroeconomy and labor markets in the United States, are the same as those which \textcite{stock1989new} used to construct a coincident index for the U.S. economy. All of the variables are seasonally adjusted, expressed as logarithmic differences, and standardized variable-wise to the same scale, $y_{ti} = \frac{\text{y}_{ti}-\bar {\textbf{y}_i}}{\text{sd}(\text{y}_i)}$.

\textit{Production} is a monthly index of industrial production, covering a large part of the variation in national output. It includes manufacturing, mining, and electric and gas utilities. Data is from the Board of Governors of the Federal Reserve System (US). More information about the construction of the industrial production index can be found in the Appendix.

\textit{Sales} includes the real manufacturing and trade sales, expressed in millions of Chained 2017 Dollars. Data is from Federal Reserve Bank of St. Louis.

\textit{Income} is the real personal income excluding current transfer receipts, expressed in billions of chained 2017 dollars, annual rate. Data is from the U.S. Bureau of Economic Analysis.

\textit{Hours} worked consists of average weekly hours of production and nonsupervisory employees in Manufacturing. Only those hours are counted which were paid for, divided by the number of workers receiving payments. Non-supervisory employees include for example lawyers and working-supervisors such as group leaders. The data comes from U.S. Bureau of Labor Statistics. More information about the determination of hours worked can be found in the Appendix.
\begin{table}[H]
    \centering
    \caption{Variables used in \textcite{stock1989new} to construct a coincident index for the U.S. economy. Values are standardized logarithmic differences.}
    \begin{tabular}{l|c}
        Variable & Explanation  \\
        \hline 
        Production & industrial production, total, index 2017=100  \\
        Sales & Real manufacturing and trade sales, total \\
        Income & Real personal income, total less transfer payments  \\
        Hours & Hours worked in production 
    \end{tabular}
    \label{tab:variables}
\end{table}

The Business Cycle Dating Committee under the National Bureau of Economic Research (NBER) publishes a retrospective analysis of expansions and recessions in the U.S. economy. The recession dates are used to analyze, whether the factor estimates from the Transformer and Kalman models declined during a crisis event as expected.
\begin{figure}[H]
    \centering
    \includegraphics[width=\linewidth]{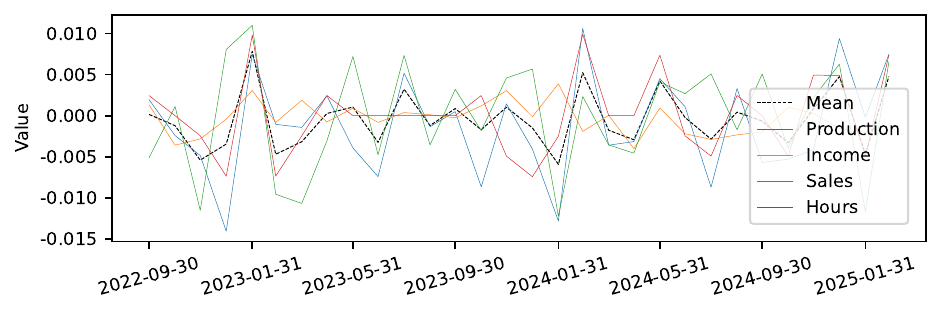}
    \caption{A snapshot into the U.S. macro data. The whole series for 1967--2025 is visualized in the Appendix.}
\end{figure}

The data is divided into non-overlapping training (80\%) and validation (20\%) sets by hand. The gradients and parameter updates are computed from the training data, whereas the validation data is only used to monitor how well the results generalize out-of-sample. Both training and validation sets include segments of time periods from different eras, and they both include influential crisis events. Within the training and validation sets, a rolling window is applied to create input matrices with all possible information sets. Each matrix consists of $P+1$ consecutive time periods, first $P$ of which are used to predict the most recent time period. Because only the last period's predictions for the observable variables are used in the loss function, it is important to create overlapping context windows with the rolling window.

\begin{figure}[H]
    \centering
    \includegraphics[width=\linewidth]{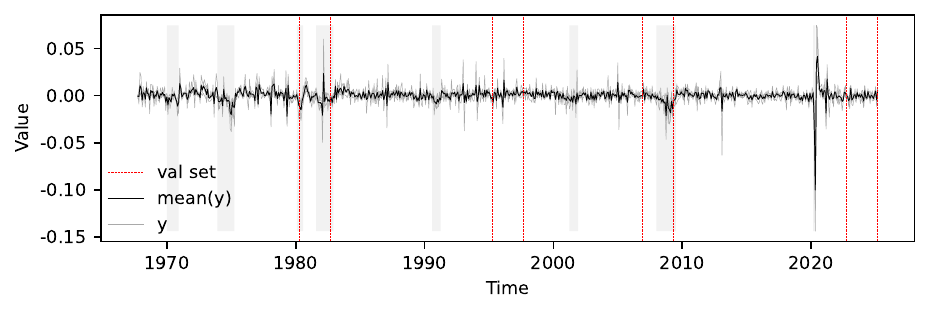}
    \caption{Training and validation groups are chosen so that both include impactful recent recessions. The 2008-crisis is in the validation set but the Covid crisis is not. NBER-recession periods denoted with grey bars.}
\end{figure}

\subsection{Kalman filter} \label{subsec:kalman_sw}

The benchmark dynamic factor in this study is estimated with the following state space model, using the Kalman filter. This analysis is in line with the classical analysis by \textcite{stock1989new}, which uses the same data and a similar procedure to derive a coincident index.

State-transition equation for the dynamic factor $x_t\in\mathbb R$ is given by
\begin{align}
    x_t &= \mu_x + \alpha x_{t-1} + \varepsilon_t && \varepsilon_t \sim \mathcal N(0, \sigma_x^2)
\end{align}

and the measurement equation for the four observable variables is
\begin{align}
    \text{Production}_t &= \mu_1 + \beta_1 x_t + u_{1,t} && u_{1,t} \sim \mathcal N(0, \sigma_y^2)\\
    \text{Sales}_t &= \mu_2 + \beta_2 x_t + u_{2,t} && u_{2,t} \sim \mathcal N(0, \sigma_y^2)\\
    \text{Income}_t &= \mu_3 + \beta_3 x_t + u_{3,t} && u_{3,t} \sim \mathcal N(0, \sigma_y^2)\\
    \text{Hours}_t &= \mu_4 + \beta_4 x_t + u_{4,t} && u_{4,t} \sim \mathcal N(0, \sigma_y^2)
\end{align}

Parameters $\mu_x, \mu_i, \alpha, \beta_i  \in \mathbb R$ for $i\in \{1,\dots,4\}$, $\sigma_x, \sigma_y \in \mathbb R^+$ are estimated with maximum likelihood. The Kalman filter is then used with the estimated model to obtain the factor estimate $\hat x_K$. For simplicity, the result from this process will be referred to as the factor estimate by the Kalman filter. The specific model used in \textcite{stock1989new} had more lags. I used the same simple one-lag linear factor model to enable better comparison to the Monte Carlo.

\subsection{Results} \label{sec:analysis}

This subsection focuses on three recent recessions, to showcase the similarities and differences between the Transformer and Kalman filter factor models. Each period's factor estimate from the Transformer is based on the look-back window of P=9, past three quarters of a year of observable variables. The Transformer's factor estimate is the average over 20 alternatives, each obtained with a similar training process but different initial parameter values. The factor estimates are scaled to match the baseline, to enable comparison. The hyperparameters are otherwise the same as in Table~\ref{tab:hyper}, except for a lower weight on prior information, $\lambda=0.2$, and lower regularization with $dr=0.1, L2=0.01$, due to the even smaller size of the empirical dataset.

The real time factor projections are generated with the trained Transformer, which had the lowest validation loss out of 20 runs. The predicted next period's values are used as the most recent lags on the next round, to recursively predict the next 6 months. This results in a forward-looking series of factor estimates, starting from each period. Because the real time projections are computed from one trained Transformer, the results may deviate from the average of all training runs.

The behavior of the Transformer is analyzed by visualizing the Attention matrices from the State and Measurement Encoders, describing the first period of each recession. Additionally, the distribution of Attention for variables and lags are followed through the recession.

\subsubsection{Covid crisis}

\begin{figure}[H]
  \centering
  \subfloat[a][Data]{\hspace{-0.5cm}\includegraphics[scale=0.93]{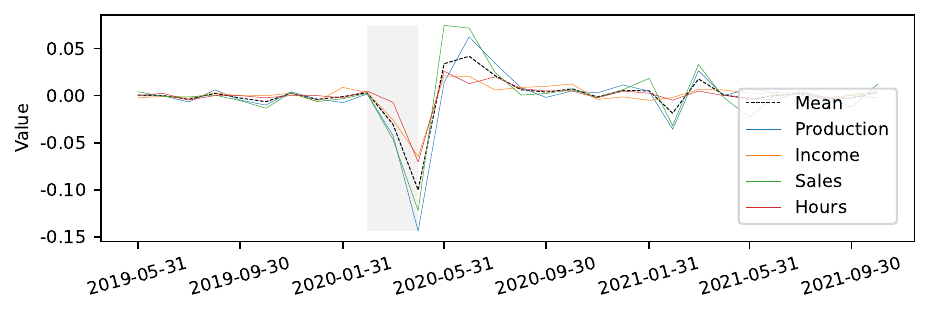}} \\ 
  \subfloat[b][Factor estimates]{\hspace{-0.5cm}\includegraphics[scale=0.93]{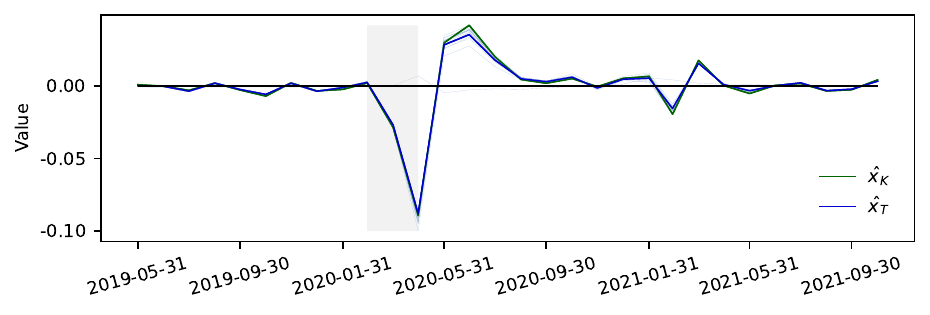}} \\ 
  \subfloat[c][Real time factor projections]{\hspace{-0.5cm}\includegraphics[scale=0.93]{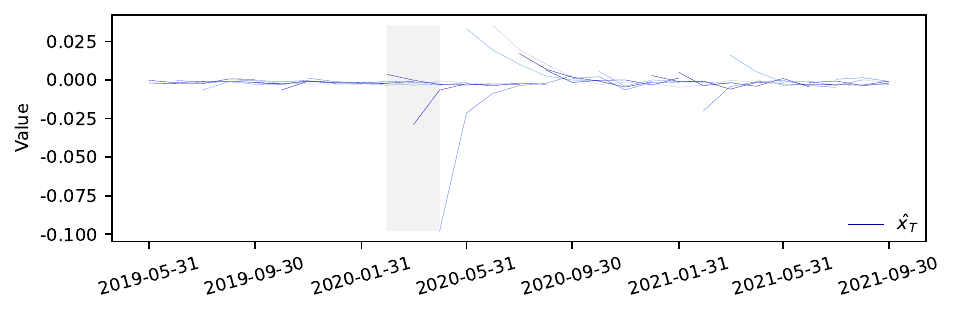}}
  \caption{Results for the Covid-crisis. The blue lines in (b) and (c) depict the Transformer factor estimates from different training runs and the thick blue line denotes their piecewise average. One of the training runs clearly converged to a suboptimal solution.}
  \label{fig:covid1}
\end{figure}

Figure~\ref{fig:covid1} (b) indicates that the factor estimates from both models behave very similarly during the Covid recession, with the Transformer estimating a slightly slower recovery after the initial response. The real time projections for the factor estimate in (c) show downward pressure on the first period of recession, even though the factor estimate for that period is still above zero.

\begin{figure}[H]
  \captionsetup[subfloat]{farskip=0pt,captionskip=0pt}
  \centering
  \subfloat[Snapshot]{\hspace{-0.5cm}\includegraphics[scale=0.93]{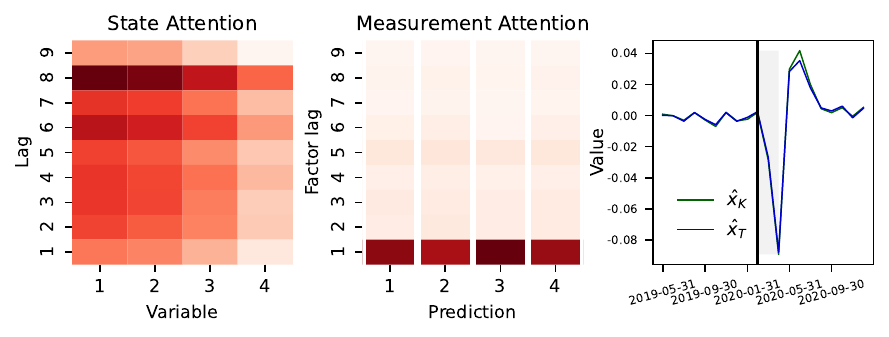}} \\
  \subfloat[State Attention]{\hspace{-0.5cm}\includegraphics[scale=0.93]{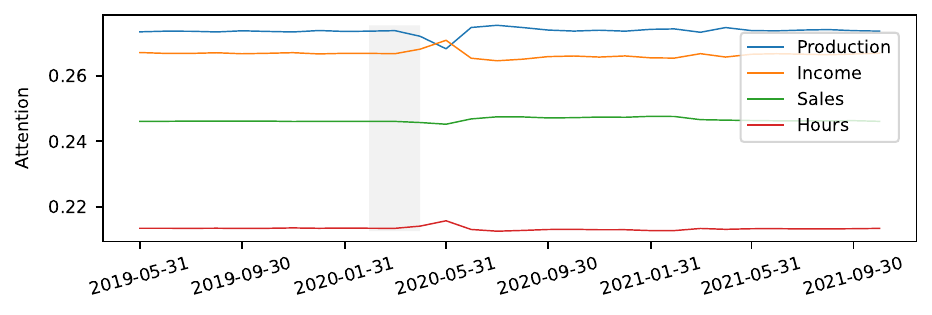}} \\
  \subfloat[State Attention]{\hspace{-0.5cm}\includegraphics[scale=0.93]{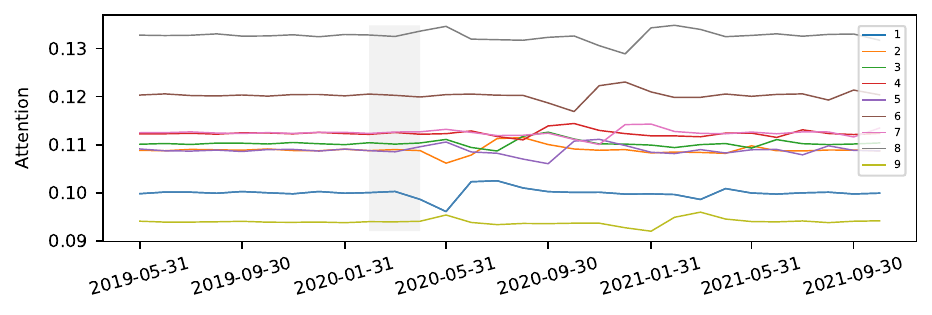}} \\
  \subfloat[Measurement Attention]{\hspace{-0.5cm}\includegraphics[scale=0.93]{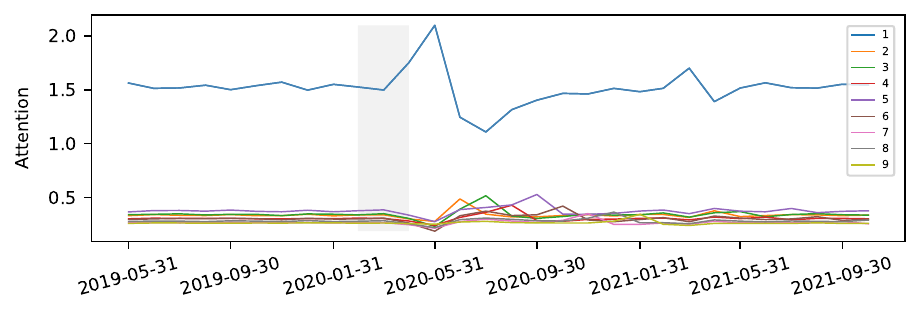}}
  \caption{The steady Attention patterns react to the Covid crisis.}
  \label{fig:covid2}
\end{figure}

Figure~\ref{fig:covid2} shows a steady distribution of Attention prior to Covid. From panel (a) we can see that what is regarded important are most lags of Production and Income, and some past lags of Sales and Hours. During the recession the pattern tweaks in panel (b), increasing the importance of income over production. Panel (c) shows temporal changes in Attention patterns during and after the recession.  The final panel (d) shows that the most recent factor estimate increases in importance during the turmoil, for predicting the next period values of the four observable variables.

\subsubsection{2008 financial crisis}

During the 2008 crisis, both coincident indices estimate a slower gradual decline, compared to the Covid recession.

\begin{figure}[H]
  \centering
  \subfloat[a][Data]{\hspace{-0.5cm}\includegraphics[scale=0.93]{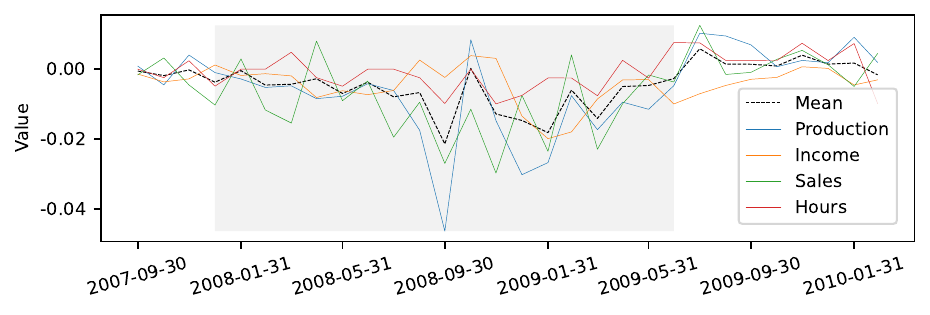}} \\
  \subfloat[b][Factor estimates]{\hspace{-0.5cm}\includegraphics[scale=0.93]{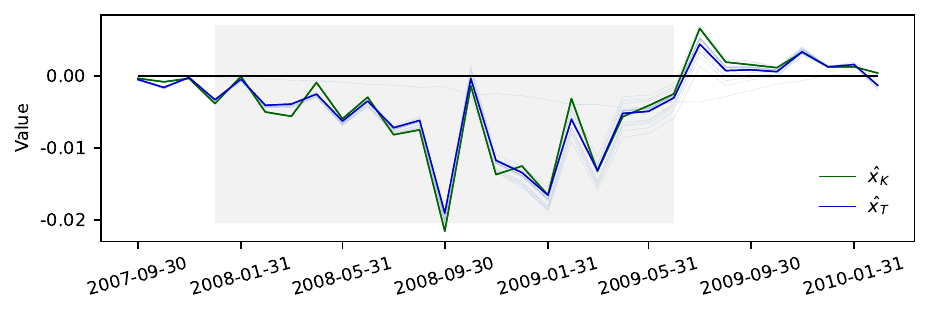}} \\
  \subfloat[c][Real time factor projections]{\hspace{-0.5cm}\includegraphics[scale=0.93]{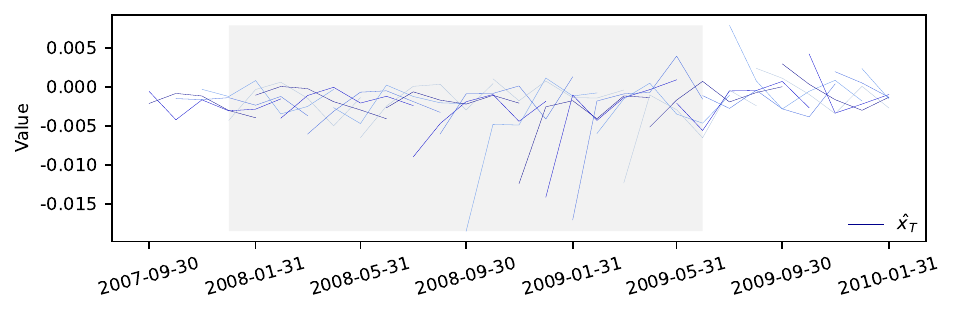}}
  \caption{Coincident indices during the 2008 crisis.}
  \label{fig:2008_1}
\end{figure}

\begin{figure}[H]
  \captionsetup[subfloat]{farskip=0pt,captionskip=0pt}
  \centering
  \subfloat[Snapshot]{\hspace{-0.5cm}\includegraphics[scale=0.93]{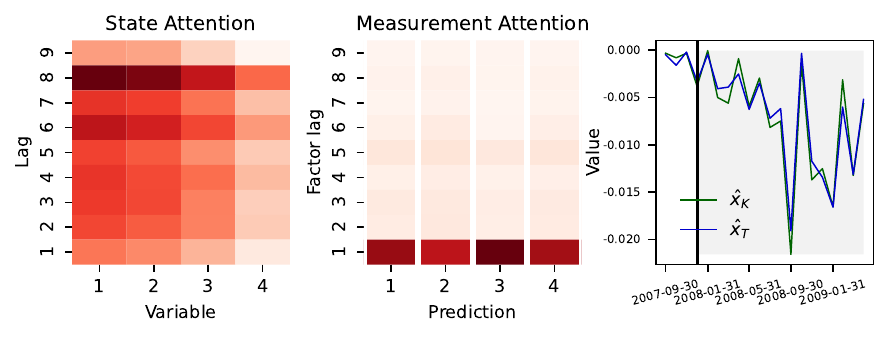}} \\
  \subfloat[State Attention]{\hspace{-0.5cm}\includegraphics[scale=0.93]{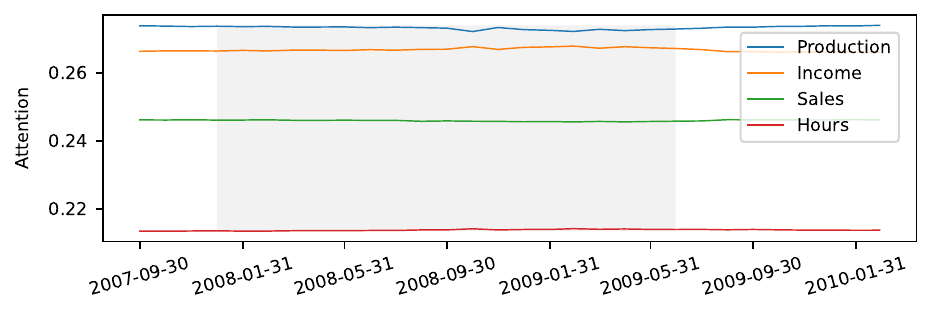}} \\
  \subfloat[State Attention]{\hspace{-0.5cm}\includegraphics[scale=0.93]{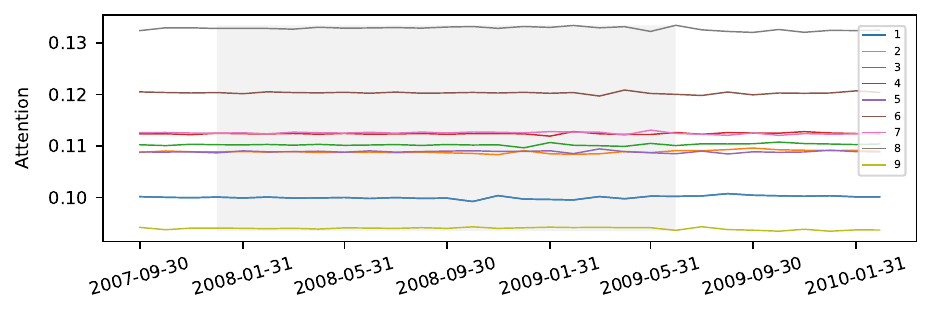}} \\
  \subfloat[Measurement Attention]{\hspace{-0.5cm}\includegraphics[scale=0.93]{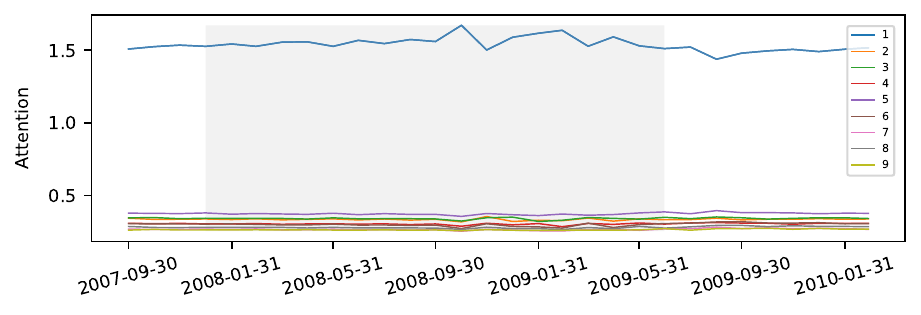}}
  \caption{The State Attention patterns are less eventful during the 2008 recession than during the Covid crisis, but the Measurement Attention is a bit more lively. The relative importance of variables stays similar during all of the crises.}
  \label{fig:2008_2}
\end{figure}

\subsubsection{1990 crisis}

The final event study is the recession of the early 1990s in the U.S. The Transformer's coincident index deviates more from the baseline during this episode than during the more recent crises, mostly with smaller local fluctuations.

\begin{figure}[H]
  \centering
  \subfloat[a][Data]{\hspace{-0.5cm}\includegraphics[scale=0.93]{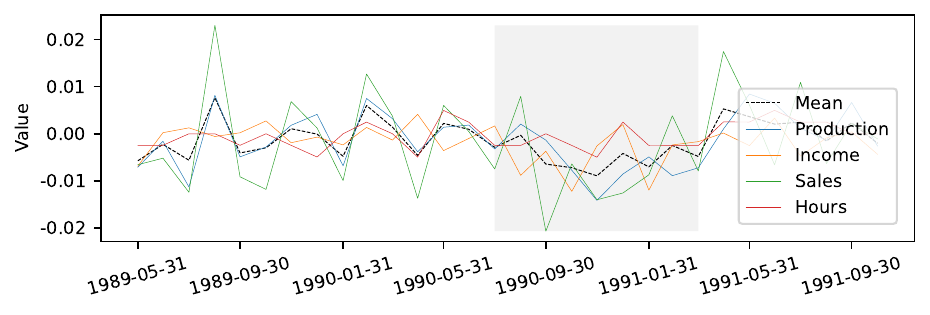}} \\
  \subfloat[b][Factor estimates]{\hspace{-0.5cm}\includegraphics[scale=0.93]{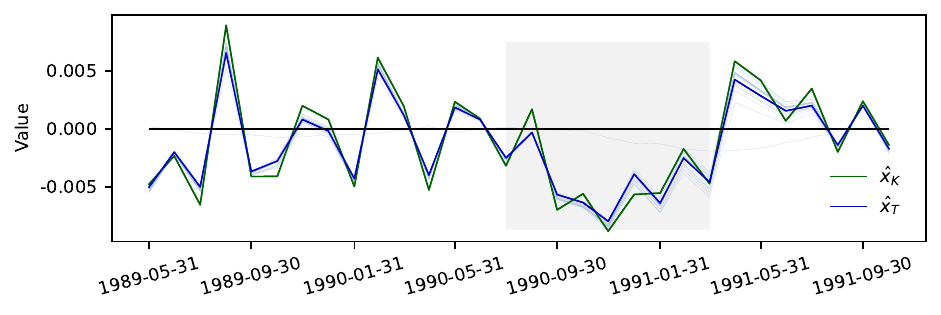}} \\
  \subfloat[c][Real time factor projections]{\hspace{-0.5cm}\includegraphics[scale=0.93]{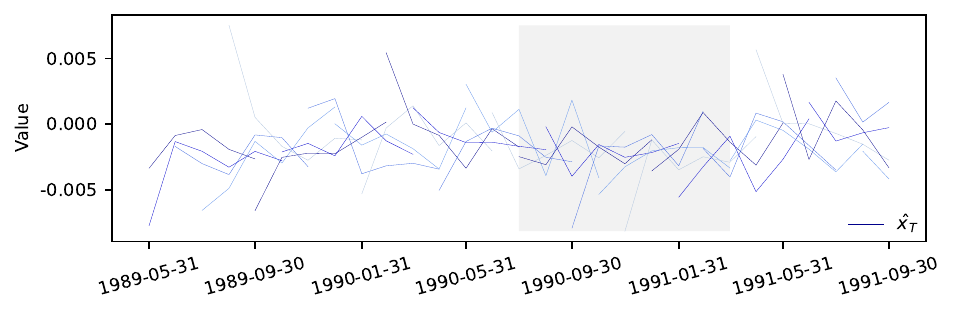}}
  \caption{Results for the 1990 crisis.}
  \label{fig:1990_1}
\end{figure}

\begin{figure}[H]
  \captionsetup[subfloat]{farskip=0pt,captionskip=0pt}
  \centering
  \subfloat[Snapshot]{\hspace{-0.5cm}\includegraphics[scale=0.93]{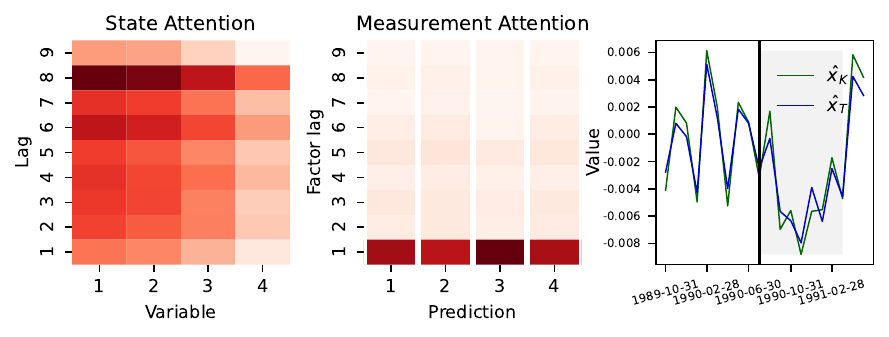}} \\
  \subfloat[State Attention]{\hspace{-0.5cm}\includegraphics[scale=0.93]{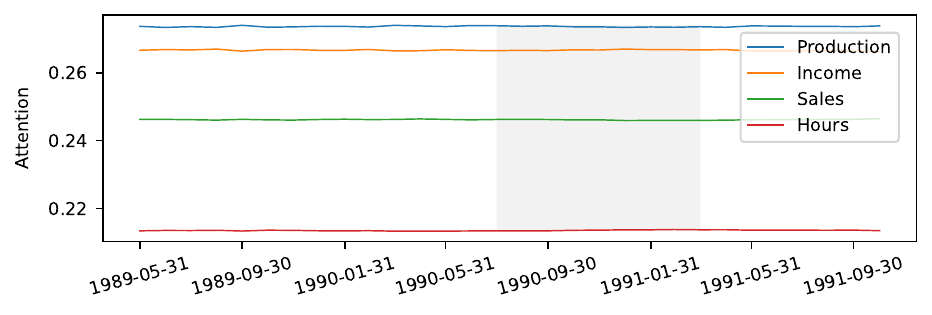}} \\
  \subfloat[State Attention]{\hspace{-0.5cm}\includegraphics[scale=0.93]{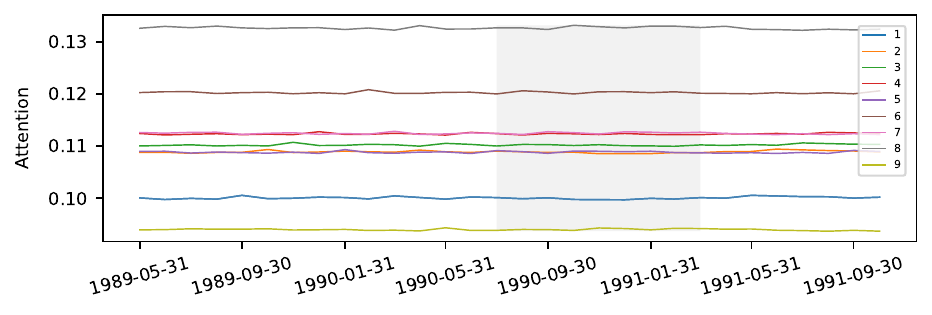}} \\
  \subfloat[Measurement Attention]{\hspace{-0.5cm}\includegraphics[scale=0.93]{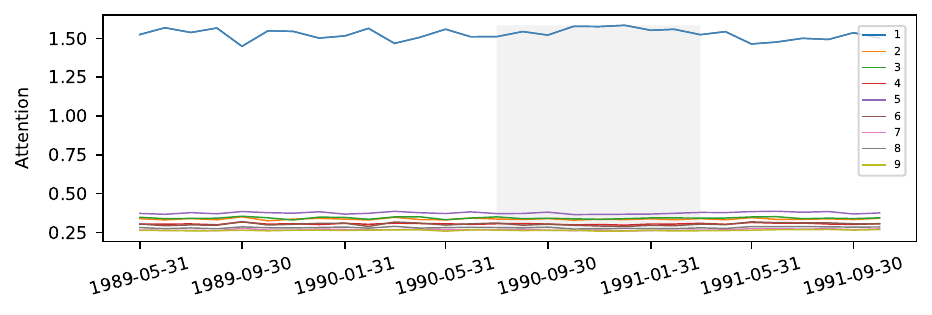}}
  \caption{Attention patterns during the 1990 crisis are similar to those during the 2008 crisis.}
  \label{fig:1990_2}
\end{figure}

\section{Discussion} \label{sec:discussion}

It is difficult to select a particular model specification for analyzing real-world data. Yet, this is the standard challenge that has to be dealt with in most macroeconomic applications. It is typical to select a linear model, or some well-known simple nonlinear model, as a reasonable first approximation. Then, based on an initial analysis and several metrics, the model can be tweaked so that it accommodates the data better. Machine learning makes it possible to approach the problem from a different angle, by reducing the level of involvement needed from the researcher in model selection. The Transformer presented here tries to make the best of both worlds by allowing different identifying assumptions to be imposed on the data softly, while remaining open to findings that disobey the assumptions.

The Transformer can entertain a large set of hypotheses, or models, that can explain the data. This strength can also be a weakness in some situations. Many complicated models could explain the same small set of observations without generalizing out-of-sample. Therefore a lot of training examples are needed to falsify wrong complex hypotheses. As the space of potential models is very large, the Transformer can get stuck in a bad local optimum. It is useful to constrict and guide the search for models on small macroeconomic datasets, by prioritizing on where to look first. The Kalman filter is able to learn good approximations more easily from smaller samples, because it is restricted to only look for linear relations. Linear--Gaussian model is often a good approximation also for nonlinear and non-Gaussian factors, enabling the use of the Kalman filter as prior information. The way to use prior information, proposed in this study, pulls the Transformer towards an area in the parameter space, from where it is more useful to search for nonlinear improvements for the factor estimate. In effect, this gives an informed hyper-prior for the model-selection task. Adding even more structure from economic theory and from conventional models can help Transformers reach new levels of usefulness in macroeconomic analysis in the future.

Despite the good performance in most of the Monte Carlo datasets, the results are not yet conclusive of the Transformer's power and reach. The processes were kept somewhat simple and well behaved, partly to enable getting consistent results with the baseline linear factor model with the Kalman filter. By introducing stronger nonlinearities and heavier tails, the estimation procedure with Kalman filter becomes unreliable. Therefore, the current datasets can still be approximated linearly to a high degree. This can be seen from the results obtained with Kalman filter, by using the true linear parameters without estimation, yielding often a high $R^2$ not too far from that of the oracle filter. Surprisingly, the mean of the observable variables also often tracks the true factor robustly. As the mean is used to initialize the Transformer's factor representation, it could have improved the Transformer's results by offering a good starting point. However, in Process 2 the "Transformer max" managed to surpass both of the baselines, the estimated Kalman filter and the mean of data, when the early stopping was conducted based on factor accuracy on the test set. This shows that the Transformer is not bounded by the initialization or prior information. But improving the performance might require developing new tools to early stop the training more optimally.

The Monte Carlo results could be further improved by choosing a more appropriate prior for each dataset. The prior information can be of great use in training, but perhaps also lead astray, by pulling the Transformer towards a sub-optimal area in its parameter space. The Transformer seems to perform especially well compared to the Kalman filter, when the measurement equation is not linear--Gaussian. On the other hand, Kalman filter can often reach a high accuracy by relying on the linear measurements, if only the state-transition deviates from linear--Gaussian. The linear--Gaussian prior information can be easily replaced with a more suitable factor model, if such can be inferred from a preliminary analysis of the data.

Another way to improve the results is by tuning the hyperparameters separately for each dataset. Tuning has potentially a drastic impact on the results. The tuning step was deliberately omitted in this study for the sake of simplicity. This Monte Carlo setup also made it reasonable to use the same Kalman filter as a baseline for all datasets, even though better filtering tools exist for the datasets 2--6. The linear baseline model can also be updated in many ways, for example by adding lags and altering the filtering procedure. Using a better benchmark as prior information could also improve the Transformer's results.

One path forward is to pre-train the Transformer on a large corpus of multivariate time series data, and then fine-tune it with the macroeconomic data. The datasets used for pre-training only need to be loosely similar to the applications in mind. The order by which different datasets are introduced to the Transformer in pre-training should follow an ascending level of complexity - for example, by first teaching the Transformer simple simulated VARMA(1,1) dynamics, moving then to weather data with mild stochasticity, and finally using macro-financial datasets with evolving data generating processes and wild stochasticity. The Transformer learns general and some particular features of time series data, and also what is not typical for dynamic sequences. Pre-training finds an informed hyper-prior to the model selection, restricting and directing the subsequent fine-tuning.

This study applies the Transformer to monitoring the real economy. The policy-analysis can be improved in several ways, to make it more useful in practice. In order to see how the factor estimate would have looked like exactly in the past at period $t$, each estimate should be obtained from a Transformer that was trained only with data up to t. The vintage data should not include revisions. In the current study the baseline Kalman filter, also used as prior regularization, was applied to the whole dataset. To be more thorough, the Kalman filter should be estimated only from training data to avoid leakage.

All of the validation data could be slowly transferred to the training dataset as the training progresses, to utilize all of the information in the small dataset. If this is done gradually, while reducing the learning rate at the same time, the risk of overfitting should be smaller than if all of the data were allocated in the training set from the beginning.

The Transformer's coincident index acts reasonably, but there is room for improvement. The coincident index behaves somewhat similarly to the process 3 of the Monte Carlo experiment, with nonlinear state-transition but linear measurements. The Attention patterns show some peculiarities, such as a constant higher attention to some of the past lags. It is possible that the Transformer has found an idiosyncratic way to distill useful information into particular token representations. Or that the particular scheme used for partitioning the data into training and validation sets had an impact on the factor estimate. Due to the double-Encoder architecture, the Transformer cannot learn from predicting the first $P$ periods of each training segment, as it needs the full context window of $P$ lags to operate. Using a rolling window to create overlapping information sets does help, but the first $P$ periods of each training segment still go unobserved by design. Using a masked Measurement Decoder with teacher forcing could be useful, if it is needed to thinly slice and shuffle the time series data to get representative training and validation sets.

\section{Conclusion} \label{sec:conclusion}

This study proposes a Transformer network for flexible estimation of nonlinear dynamic factors. The architecture, determining how inputs are transformed into outputs, is designed for analyzing and interpreting latent variables with macroeconomic time series data. The accuracy on small datasets is improved by using the identifying restrictions of conventional factor models, but more softly as prior-information regularization. Prior information is supplied to the loss function with a weight between zero and one, to guide the Transformer's training. I demonstrate the impact of prior-information regularization using a linear factor model with the Kalman filter, but any conventional factor model can be used instead. The choice of granular tokens makes the Transformer resemble a nonlinear vector autoregression. This helps interpret the results with the Attention mechanism, which indicates the most important input variables and their lags for the factor estimate at any given moment. Attention matrices are used to observe how the attention pattern for variables and their lags evolve over time, indicating for example regime switches.

In the proof-of-concept Monte Carlo experiment the Transformer was able to estimate factors more accurately than the linear factor model with the Kalman filter, in many cases where the process behind the datasets deviated from linear--Gaussian.

The Transformer is used to estimate a dynamic factor from the U.S. macroeconomic data. The factor is interpreted as a measure of real economic activity, coincident with the business cycle. The results are compared to those obtained with a linear factor model, and to the retrospective NBER--determined recession dates. The Transformer generally agrees with the baseline coincident index and the NBER recession bars during the Covid crisis, with some deviations during the 2008-09 recession and the early-1990s downturn.

\printbibliography[heading=subbibintoc, title={References}]

\section*{Appendix A} \label{appendix}
\vspace{-0.5cm}


\begin{figure}[H]
  \centering
  \subfloat{\includegraphics[scale=1]{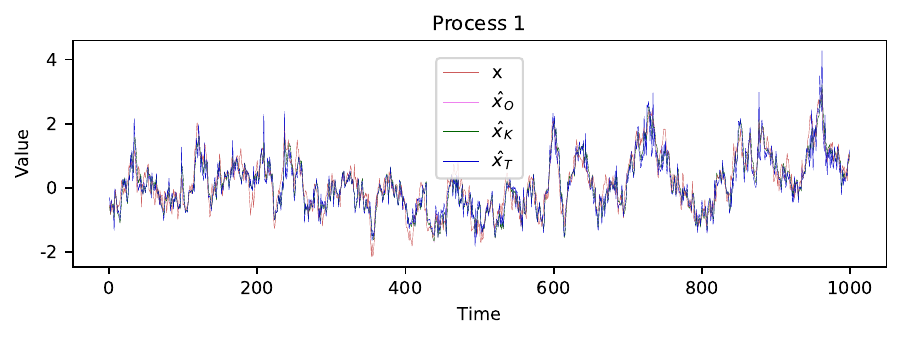}}\\[-0cm]
  \subfloat{\includegraphics[scale=1]{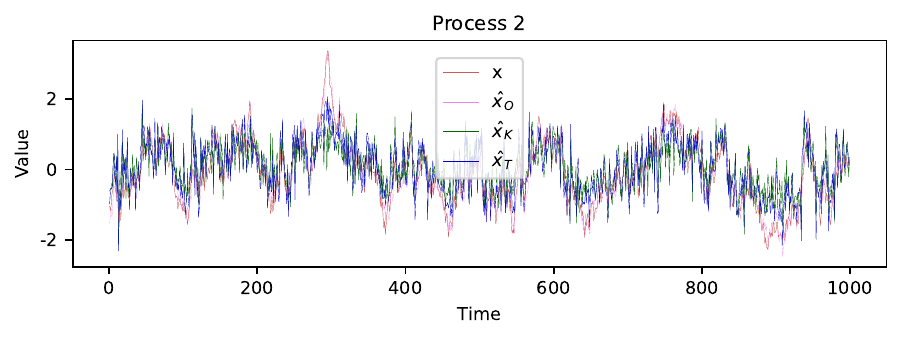}}\\[-0cm]
  \subfloat{\includegraphics[scale=1]{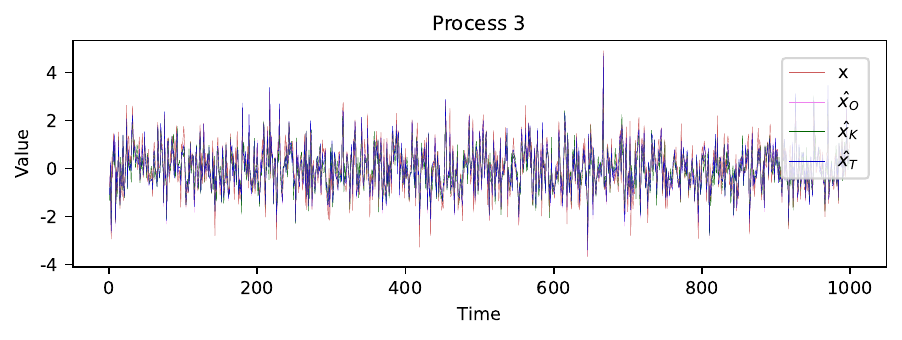}}\\[-0cm]
  \caption{Processes 1--3, comparison of factor estimates in the whole 1000 observation test set. All figures have vector graphics, enabling zooming in with high resolution.} \label{fig:long_1to4}
\end{figure}

\begin{figure}[H]
  \centering
    \subfloat{\includegraphics[scale=1]{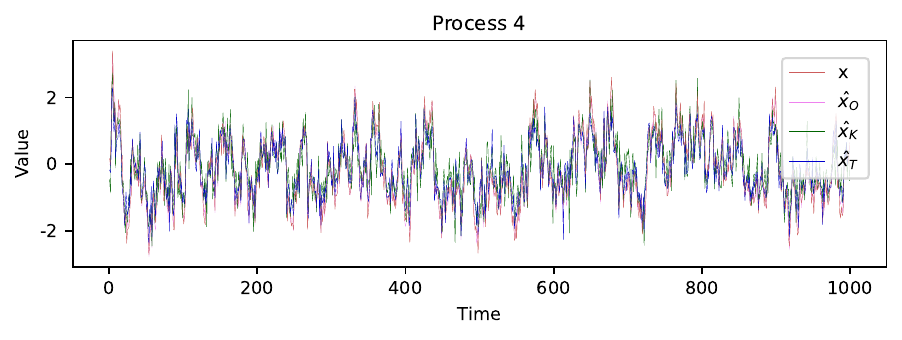}}\\[-0cm] 
  \subfloat{\includegraphics[scale=1]{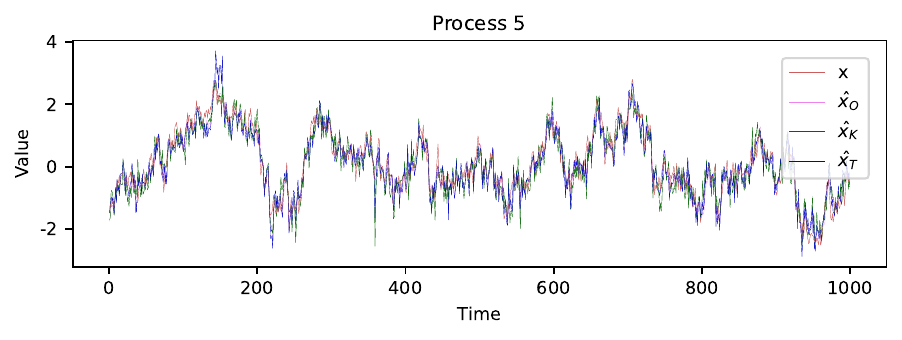}}\\[-0cm] 
  \subfloat{\includegraphics[scale=1]{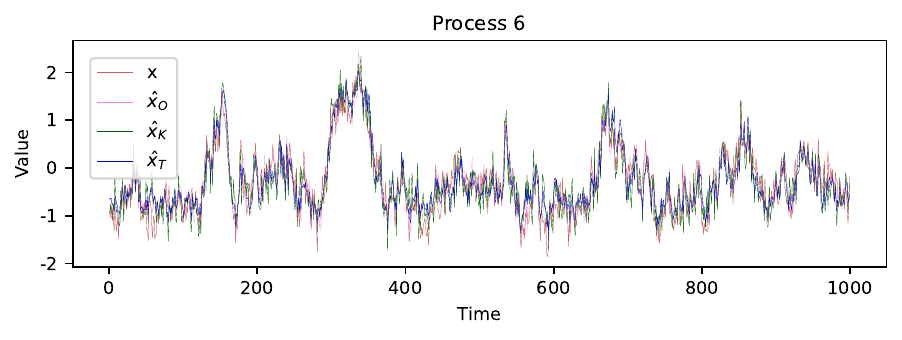}}\\[-0cm]
  \caption{Processes 4--6: Factor estimates.} \label{fig:long_5to6}
\end{figure}

\begin{table}[!htbp]
    \caption{Relative accuracy measure Fit for all seeds of the Monte Carlo in Section~\ref{sec:montecarlo}. Some seed were discarded because of a very poor performance by the estimated Kalman filter baseline.}
    \small
    \centering
    \begin{tabular}{c?cccccccccc}
       Fit\% & 1 & 2 & 3 & 4 & 5 & 6 & 7 & 8 & 9 & 10 \\
       \hline
       Process 1  & -29.0 & -59.6 & -45.1 & 5.0 & -56.9 & -19.6 & -51.3 & -38.2 & -52.1 & -35.9 \\
       Process 2  & 32.8 & 45.5 & 22.7 & 10.9 & 26.0 & 41.4 & 32.2 & 48.6  & - & - \\
       Process 3  & 45.2 & 42.8 & 45.8 & 45.7 & 46.6 & 45.1 & 26.7 & 44.7 & - & - \\
       Process 4  & 59.6 & 42.9 & 57.5 & 35.8 & 56.4 & 54.6 & 54.5 & 55.9 & -18.7 & 54.0   \\
       Process 5  & -84.7 & 28.4 & 44.3 & 27.4 & 13.5 & -108.5 & -11.6 & 20.6 & - & -  \\
       Process 6  & 38.5 & 43.8 & -15.8 & -6.0 & -35.9 & 30.0 & 21.7 & -20.3 & -14.6 & -
    \end{tabular}
    \label{tab1:results_all_seeds}
\end{table}


\begin{figure}[H]
  \centering
  \subfloat{\includegraphics[scale=1]{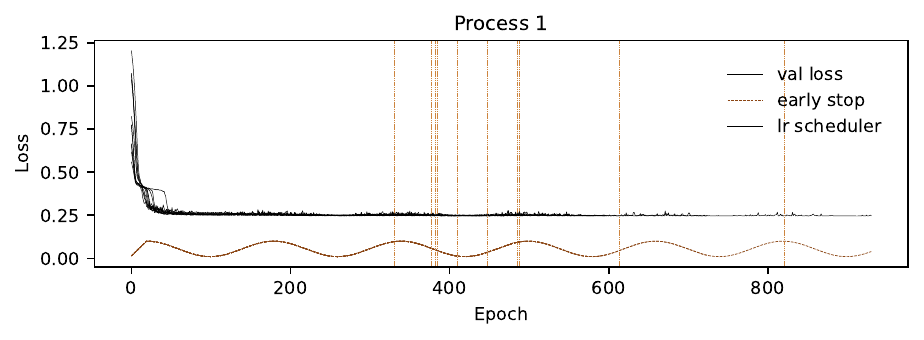}}\\[-0.2cm] 
  \subfloat{\includegraphics[scale=1]{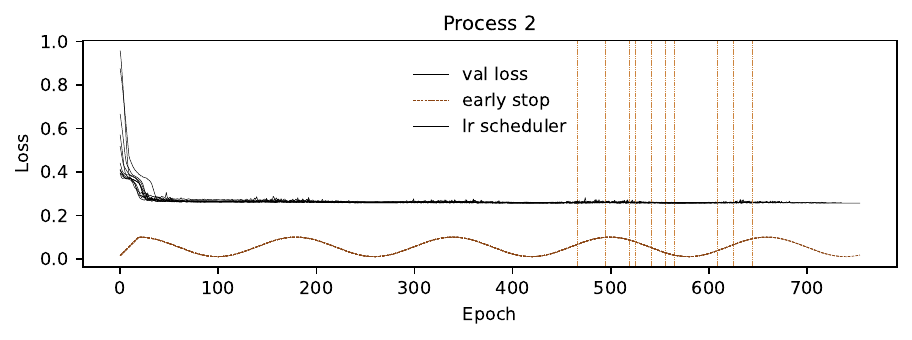}}\\[-0.2cm] 
  \subfloat{\includegraphics[scale=1]{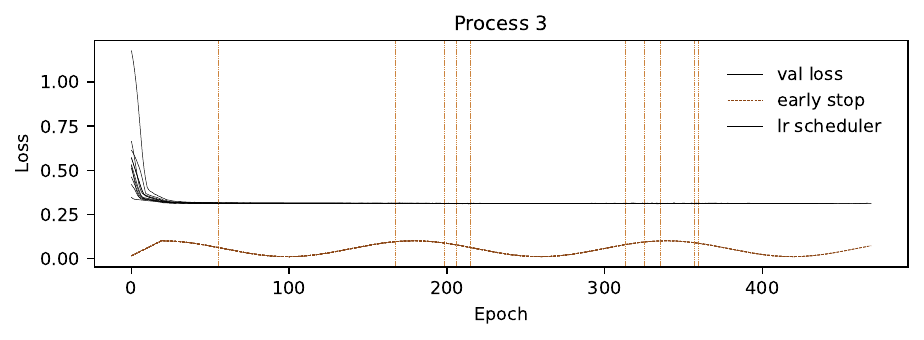}}\\[-0.2cm] 
  \caption{Processes 1-3: Losses, early stops and learning rate schedulers. The validation loss, used as early stopping criterion, includes only prediction errors without prior information.} \label{fig1:loss_1to4}
\end{figure}

\begin{figure}[H]
  \centering
    \subfloat{\includegraphics[scale=1]{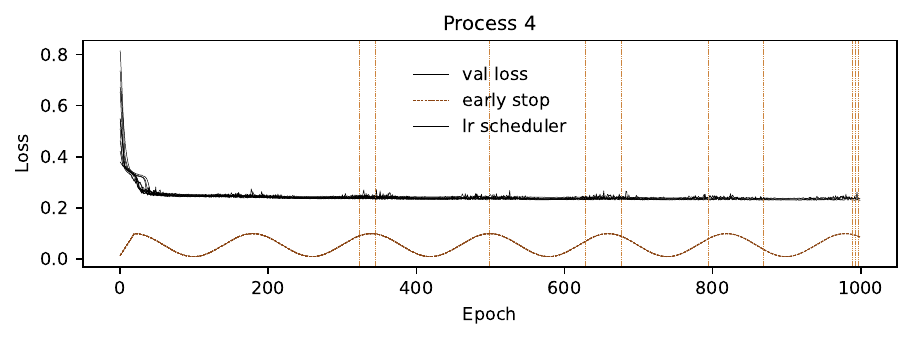}}\\[-0cm]
  \subfloat{\includegraphics[scale=1]{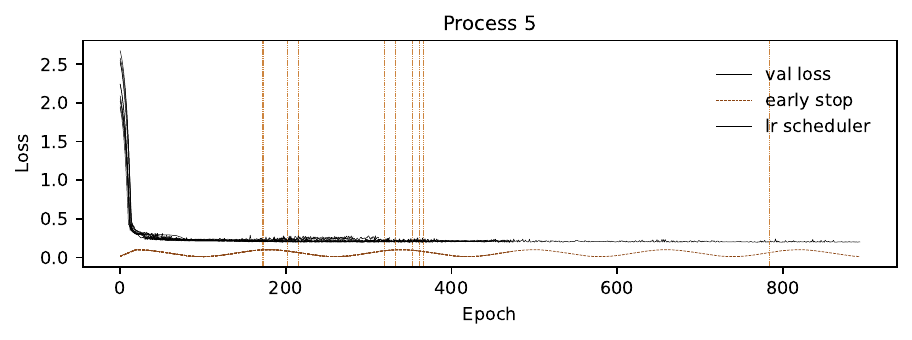}}\\[-0cm] 
  \subfloat{\includegraphics[scale=1]{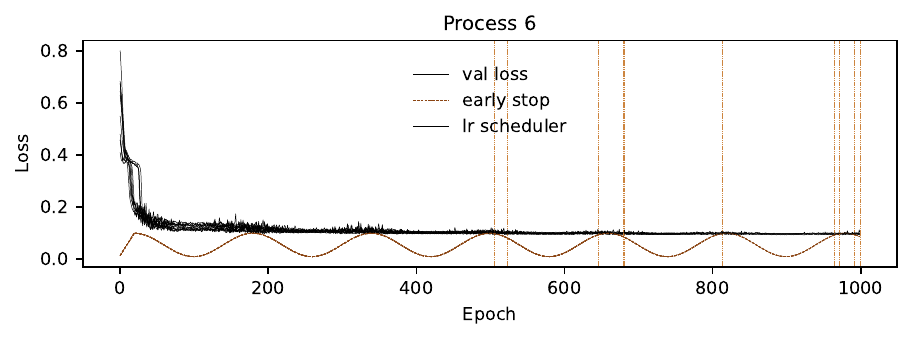}}\\[0cm]
  \caption{Processes 4--6: Losses, early stops and learning rate schedulers.} \label{fig:loss_5to6}
\end{figure}

\begin{figure}[H]
    \centering
    \includegraphics[width=1\linewidth]{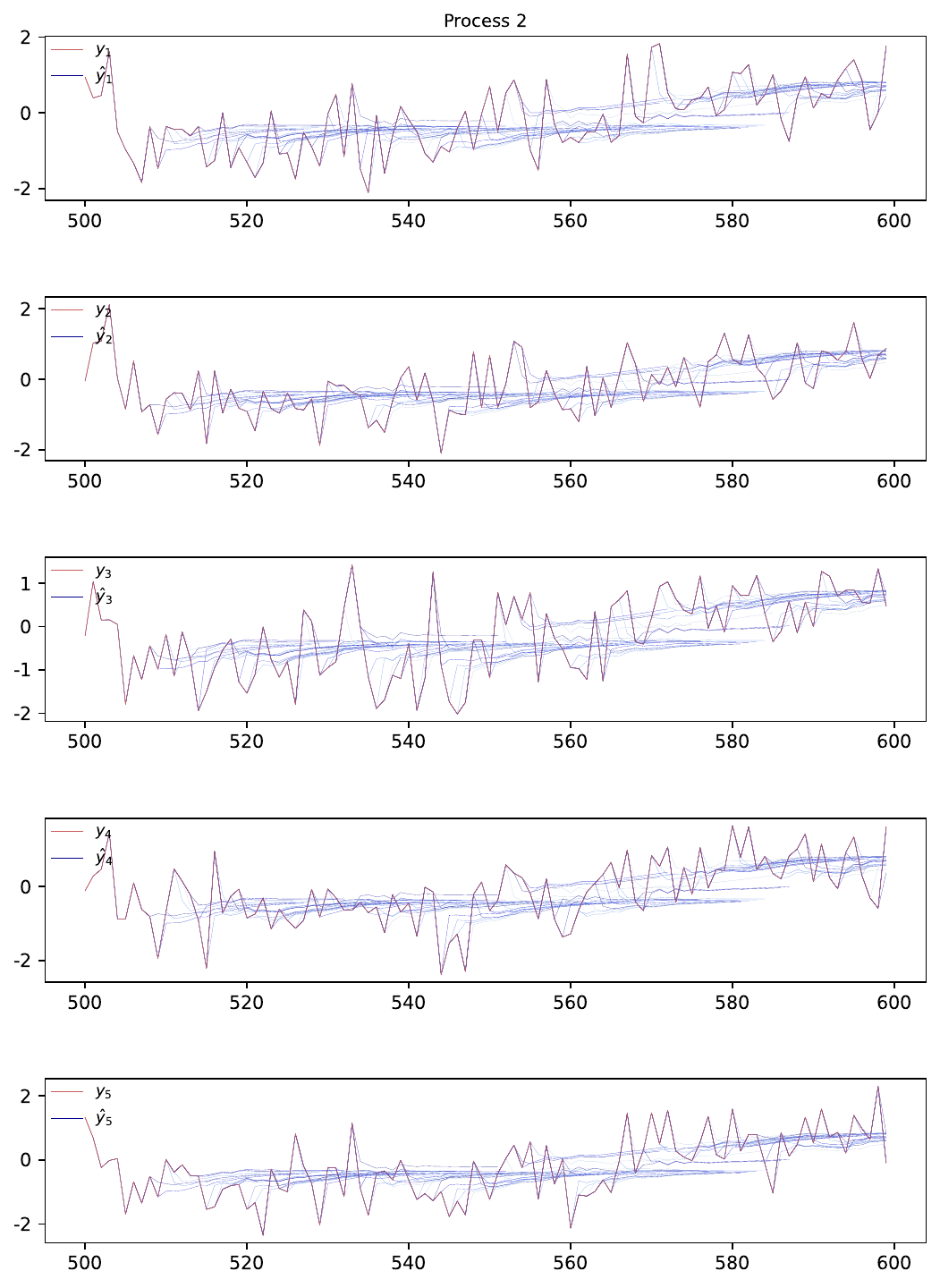}
    \caption{Process 2: Real time 20-period ahead prediction tentacles for the 5 observable variables from the Measurement Encoder. The simulated processes were relatively simple, and so are the dynamics that can be predicted in absence of knowledge about the future shocks.}
    \label{fig:pred2}
\end{figure}

\begin{figure}[H]
    \centering
    \includegraphics[width=1\linewidth]{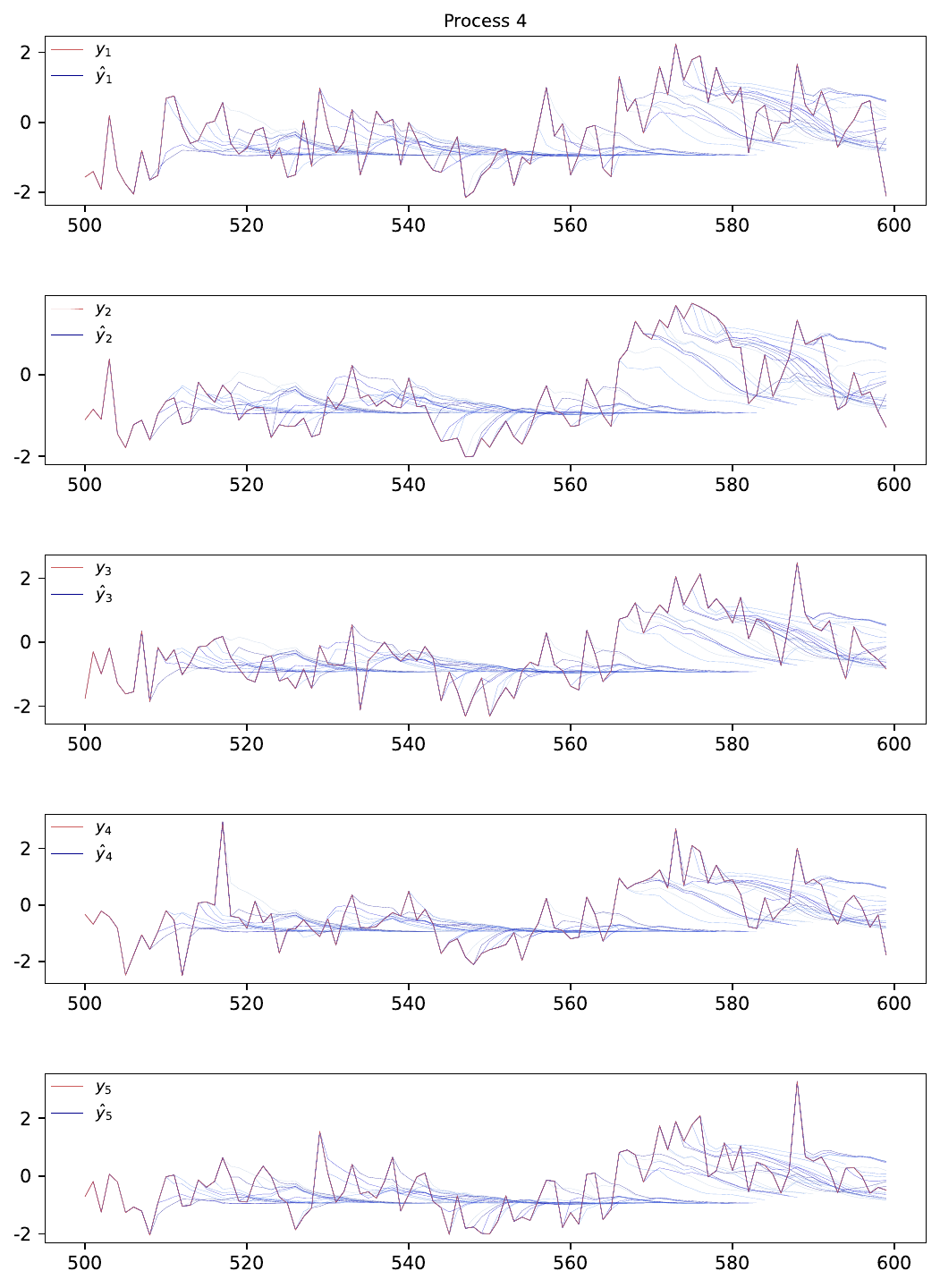}
    \caption{Process 4: 20-period ahead prediction tentacles from the Measurement Encoder.}
    \label{fig:pred4}
\end{figure}

\begin{figure}[H]
    \centering
    \includegraphics[width=1\linewidth]{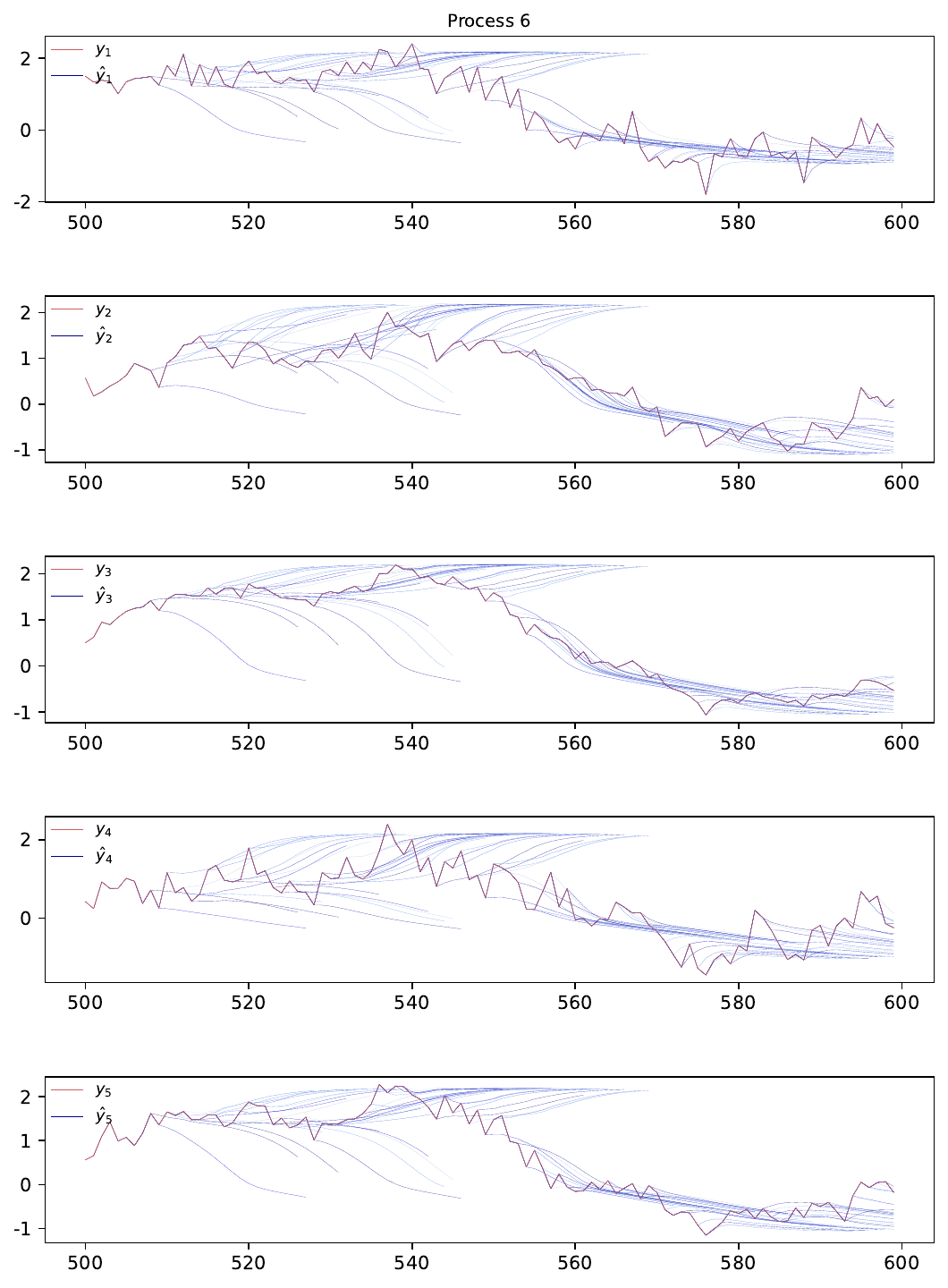}
    \caption{Process 6: 20-period ahead prediction tentacles for the observable variables.}
    \label{fig:pred6}
\end{figure}

\begin{figure}[H]
  \centering
  \subfloat[State Attention for variables over time]{\begin{tabular}{cc}
    \hspace{-0.65cm}
     \includegraphics[scale=1]{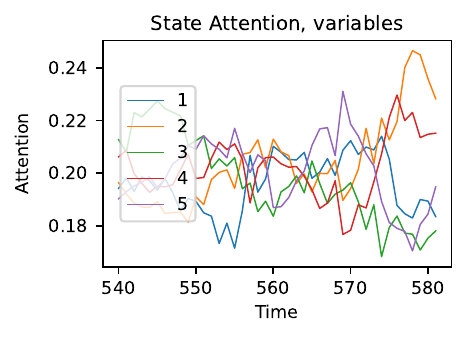}\hspace{-0.65cm} &  \includegraphics[scale=1]{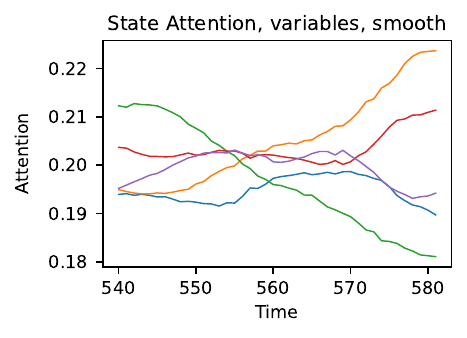}
     \end{tabular}}  \vspace{-0.4cm}\\
  \subfloat[State Attention for lags over time]{\begin{tabular}{cc}
    \hspace{-0.65cm}
     \includegraphics[scale=1]{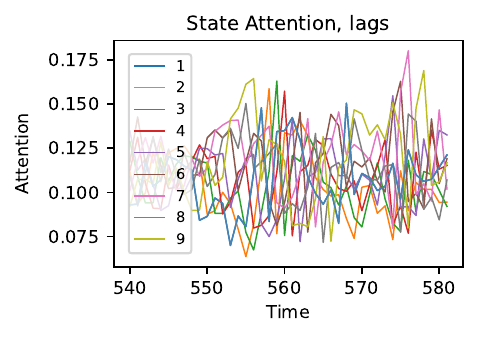}\hspace{-0.65cm} &  \includegraphics[scale=1]{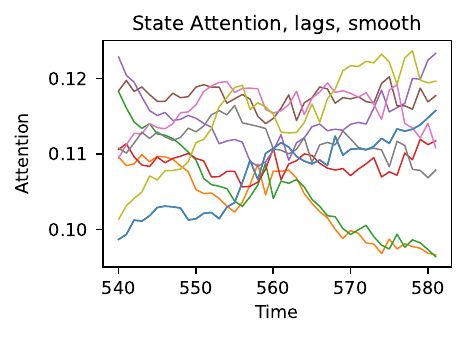}
     \end{tabular}} \vspace{-0.4cm}\\
  \subfloat[Measurement Attention for lags over time]{\begin{tabular}{cc}
    \hspace{-0.65cm}
     \includegraphics[scale=1]{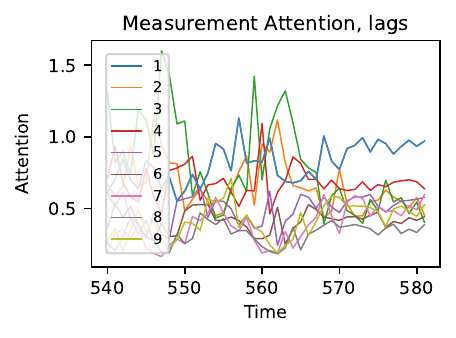}\hspace{-0.65cm} &  \includegraphics[scale=1]{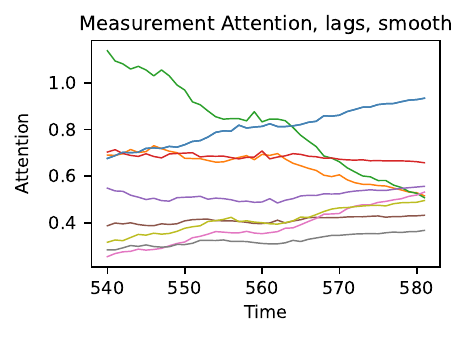}
     \end{tabular}}
  \caption{An example of how Attention patterns can be enhanced using an exponential smoother, to improve interpretability at the cost of losing some information.}
  \label{fig:process2AttnSmooth}
\end{figure}

\section*{Appendix B} \label{appendixA}

\subsection*{FRED Notes on hours worked and industrial production:}

\textit{Hours worked.} Average weekly hours refer to the average hours per worker for which pay was received, and they can differ from standard or scheduled hours. Factors such as unpaid absenteeism, labor turnover, part-time work, and stoppages cause average weekly hours to be lower than scheduled hours of work for an establishment. Group averages further reflect changes in the workweek of component industries. Average weekly hours are the total weekly hours divided by the employees paid for those hours.

Production and related employees include working supervisors and all nonsupervisory employees (including group leaders and trainees) engaged in fabricating, processing, assembling, inspecting, receiving, storing, handling, packing, warehousing, shipping, trucking, hauling, maintenance, repair, janitorial, guard services, product development, auxiliary production for the plant's own use (for example, power plant), recordkeeping, and other services closely associated with the above production operations.

Nonsupervisory employees include those individuals in private, service-providing industries who are not above the working-supervisor level. This group includes individuals such as office and clerical workers, repairers, salespersons, operators, drivers, physicians, lawyers, accountants, nurses, social workers, research aides, teachers, drafters, photographers, beauticians, musicians, restaurant workers, custodial workers, attendants, line installers and repairers, laborers, janitors, guards, and other employees at similar occupational levels whose services are closely associated with those of the employees listed.

The series comes from the 'Current Employment Statistics (Establishment Survey).' The source code is: CES3000000007. \url{https://fred.stlouisfed.org/series/AWHMAN}.\\

\textit{Industrial production.} The Federal Reserve's monthly index of industrial production and the related capacity indexes and capacity utilization rates cover manufacturing, mining, and electric and gas utilities. The industrial sector, together with construction, accounts for the bulk of the variation in national output over the course of the business cycle. The industrial detail provided by these measures helps illuminate structural developments in the economy. The industrial production (IP) index measures the real output of all relevant establishments located in the United States, regardless of their ownership, but not those located in U.S. territories. \url{https://fred.stlouisfed.org/series/INDPRO}.

\subsection*{NBER notes on business cycle dating:}

The NBER’s Business Cycle Dating Committee maintains a chronology of U.S. business cycles. The chronology identifies the months of peaks and troughs of economic activity. Recessions are the periods between a peak and a trough. By convention, the NBER classifies the peak month as the last month of the expansion and the trough month as the last month of the recession. The NBER's definition emphasizes that a recession involves a significant decline in economic activity that is spread across the economy and lasts more than a few months.

Because a recession must influence the economy broadly and not be confined to one sector, the committee emphasizes economy-wide measures of economic activity. The determination of the months of peaks and troughs is based on a range of monthly measures of aggregate real economic activity published by the federal statistical agencies. There is no fixed rule about what measures contribute information to the process or how they are weighted in the decisions.

The committee's approach to determining the dates of turning points is retrospective. In making its peak and trough announcements, it waits until sufficient data are available to avoid the need for major revisions to the business cycle chronology.

Read more on \url{https://www.nber.org/research/business-cycle-dating}.

\begin{figure}[H]
    \centering
    \includegraphics[width=\linewidth]{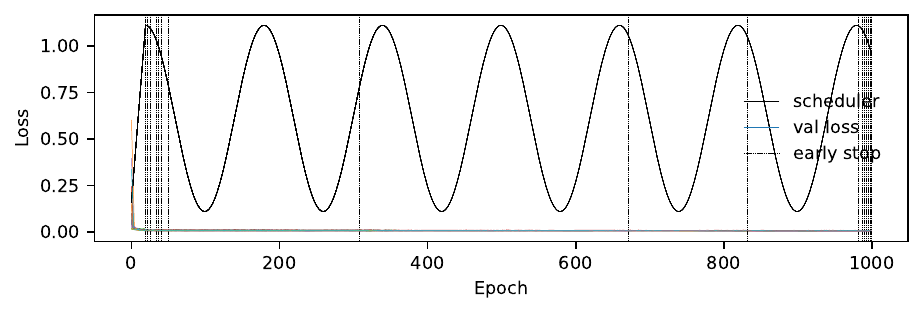}
    \caption{Most of the training runs continued to the second and third scheduler cycle.}
    \label{fig:training}
\end{figure}

\subsection*{Results for the whole dataset}

\begin{figure}[H]
    \centering
    \includegraphics[width=\linewidth]{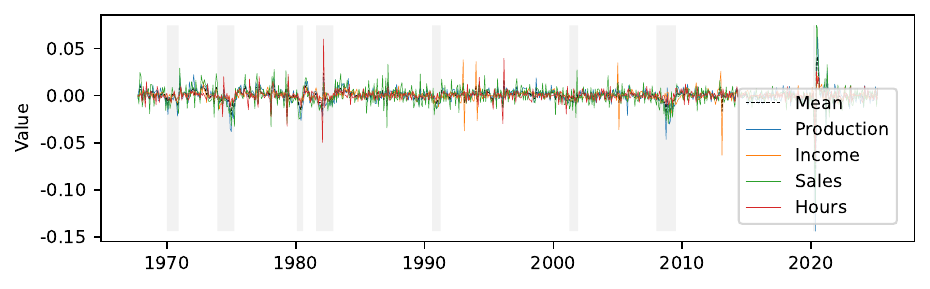}
    \caption{Time series for the four observable variables for 1967--2025.}
    \label{fig:data_all}
\end{figure}
\begin{figure}[H]
    \centering
    \includegraphics[width=\linewidth]{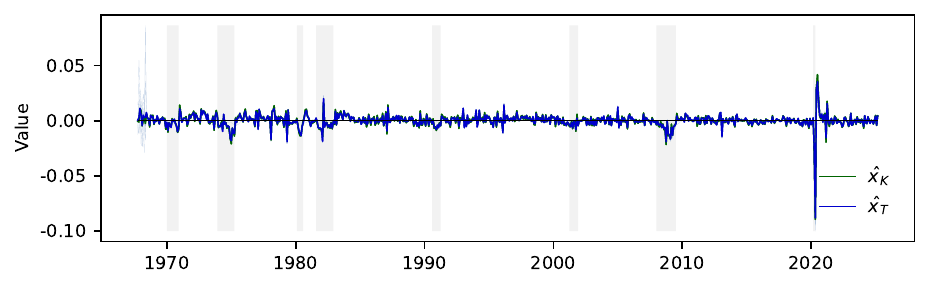}
    \caption{Comparison of factor estimates for the whole data. The figure is zoomable with vector graphics. NBER-recession bars are indicated with grey.}
\end{figure}
\begin{figure}[H]
    \centering
    \includegraphics[width=\linewidth]{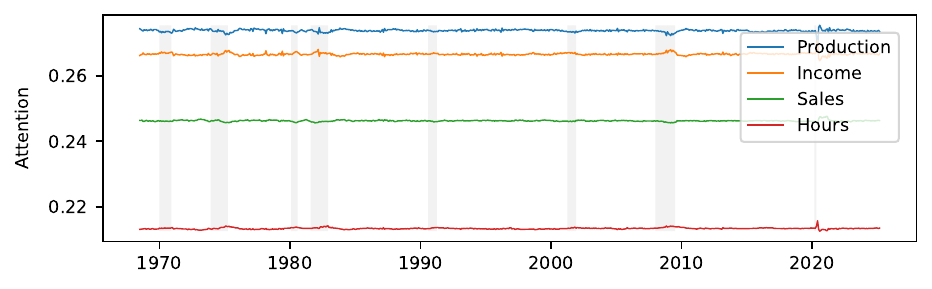}
    \caption{State Attention patterns for variables over time.}
\end{figure}
\begin{figure}[H]
    \centering
    \includegraphics[width=\linewidth]{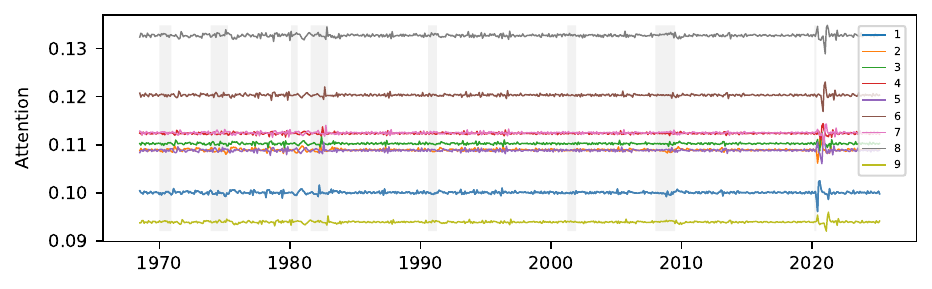}
    \caption{State Attention patterns for lags.}
\end{figure}
\begin{figure}[H]
    \centering
    \includegraphics[width=\linewidth]{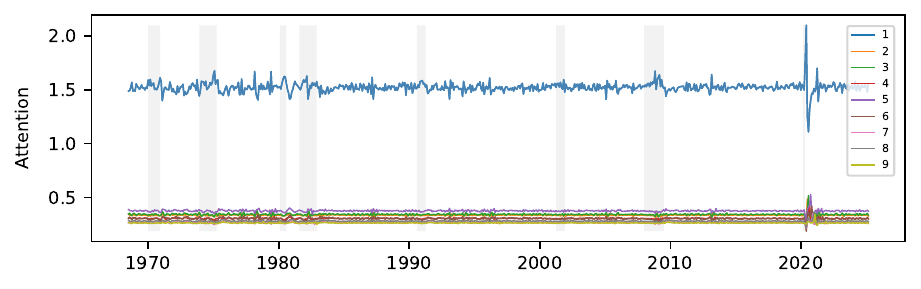}
    \caption{Measurement Attention patterns for factor lags.}
\end{figure}
\begin{figure}[H]
    \centering
    \includegraphics[width=\linewidth]{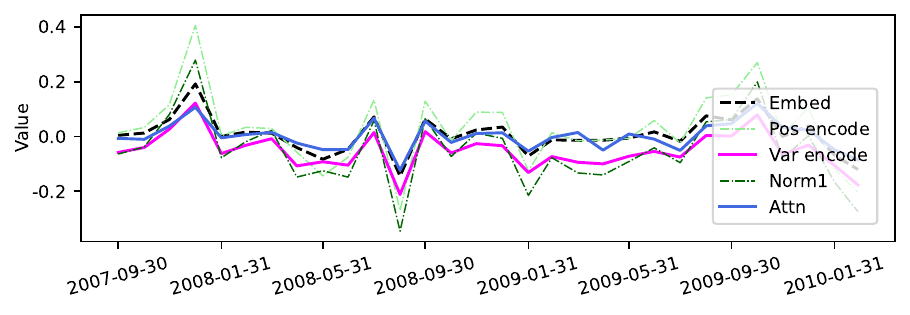}
    \caption{Visualizing the residual stream helps evaluate the impact of each operation on the outcome.}
\end{figure}

\subsection*{Real-time predictions for the observable variables}
\begin{figure}[H]
    \centering
    \includegraphics[width=\linewidth]{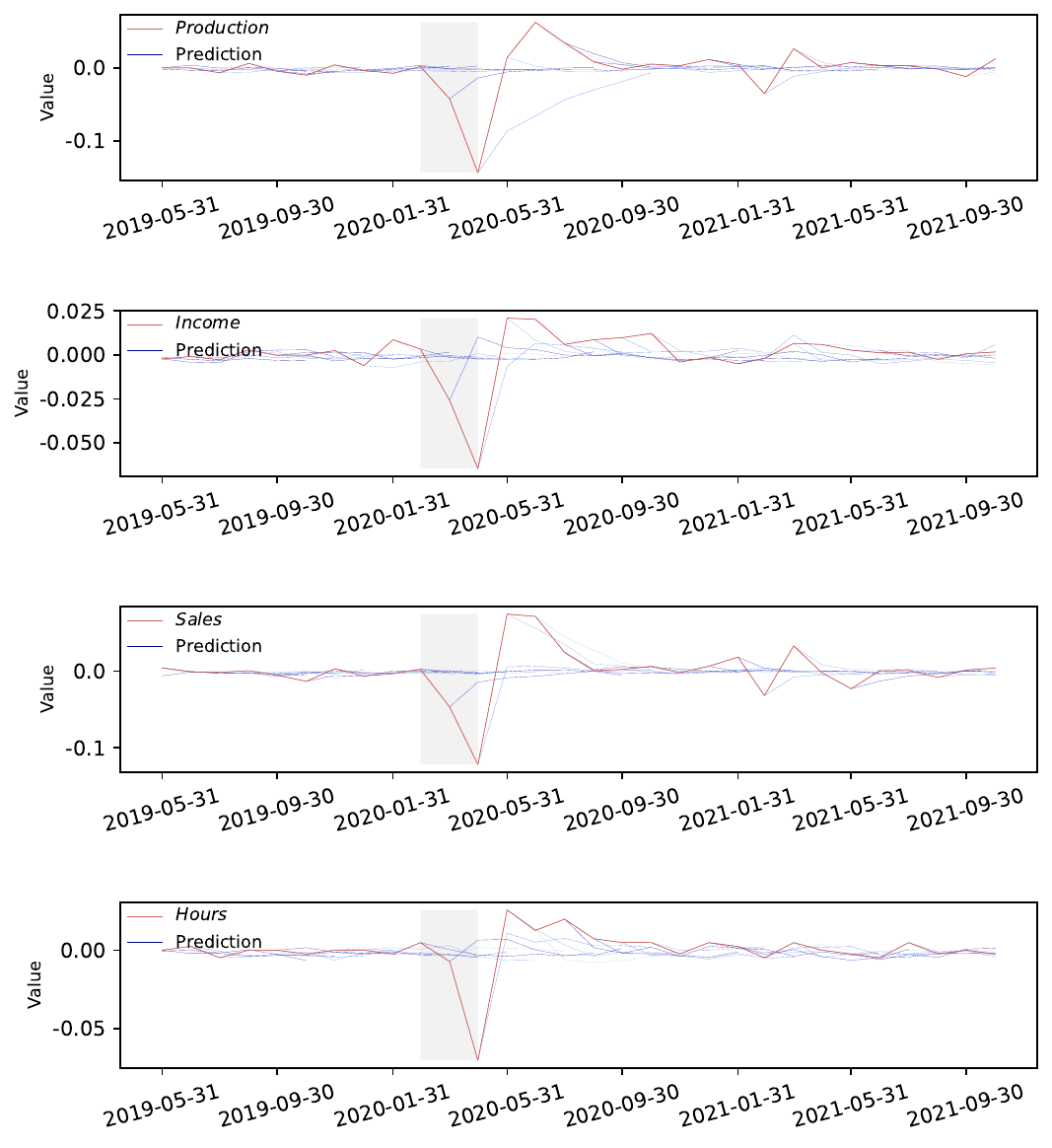}
    \caption{Real-time predictions for the observable variables during the Covid crisis.}
    \label{fig:tentacle_y_covid}
\end{figure}
\begin{figure}[H]
    \centering
    \includegraphics[width=\linewidth]{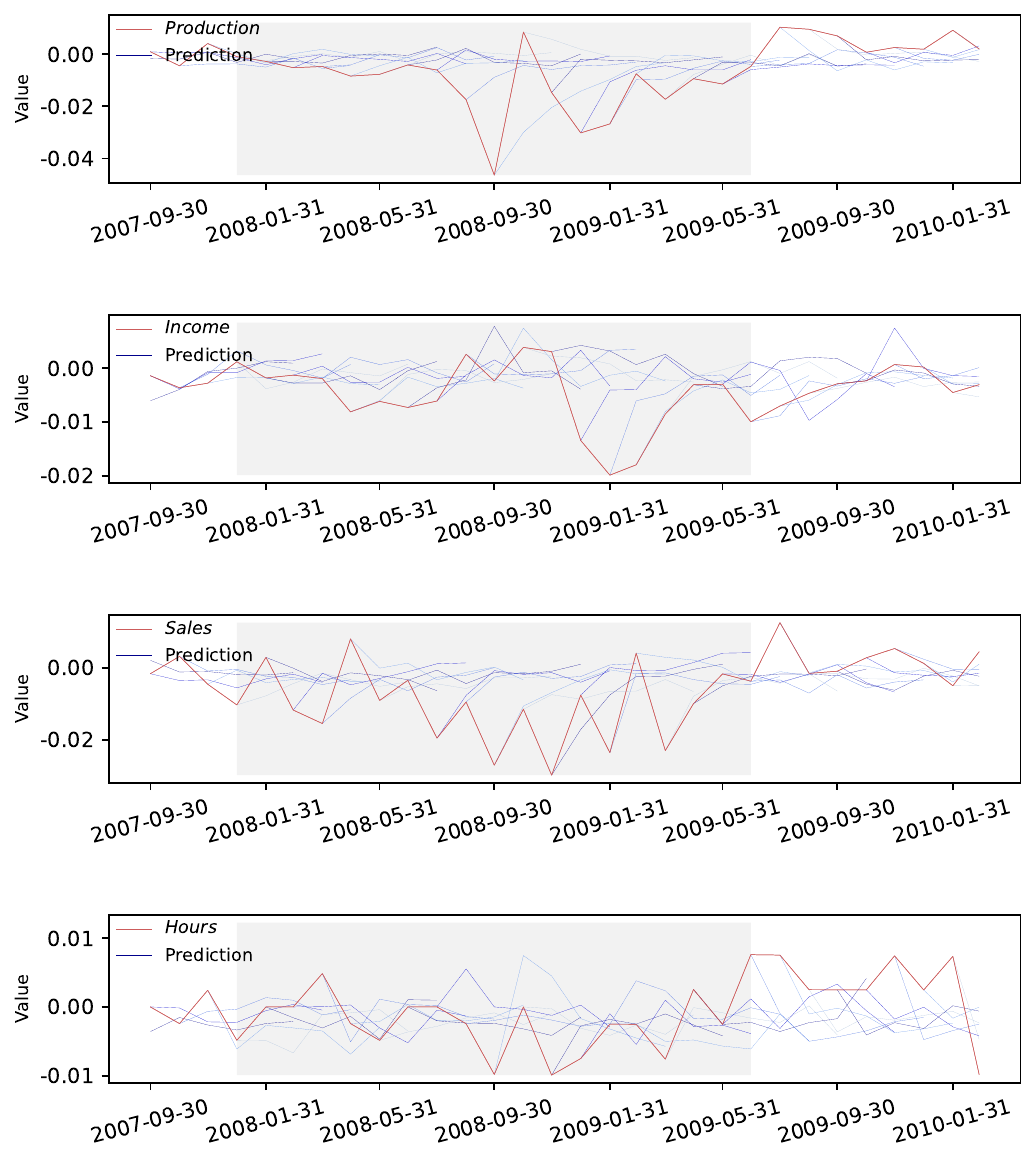}
    \caption{Real-time predictions for the observable variables during the 2008 crisis.}
    \label{fig:tentacle_y_2008}
\end{figure}
\begin{figure}[H]
    \centering
    \includegraphics[width=\linewidth]{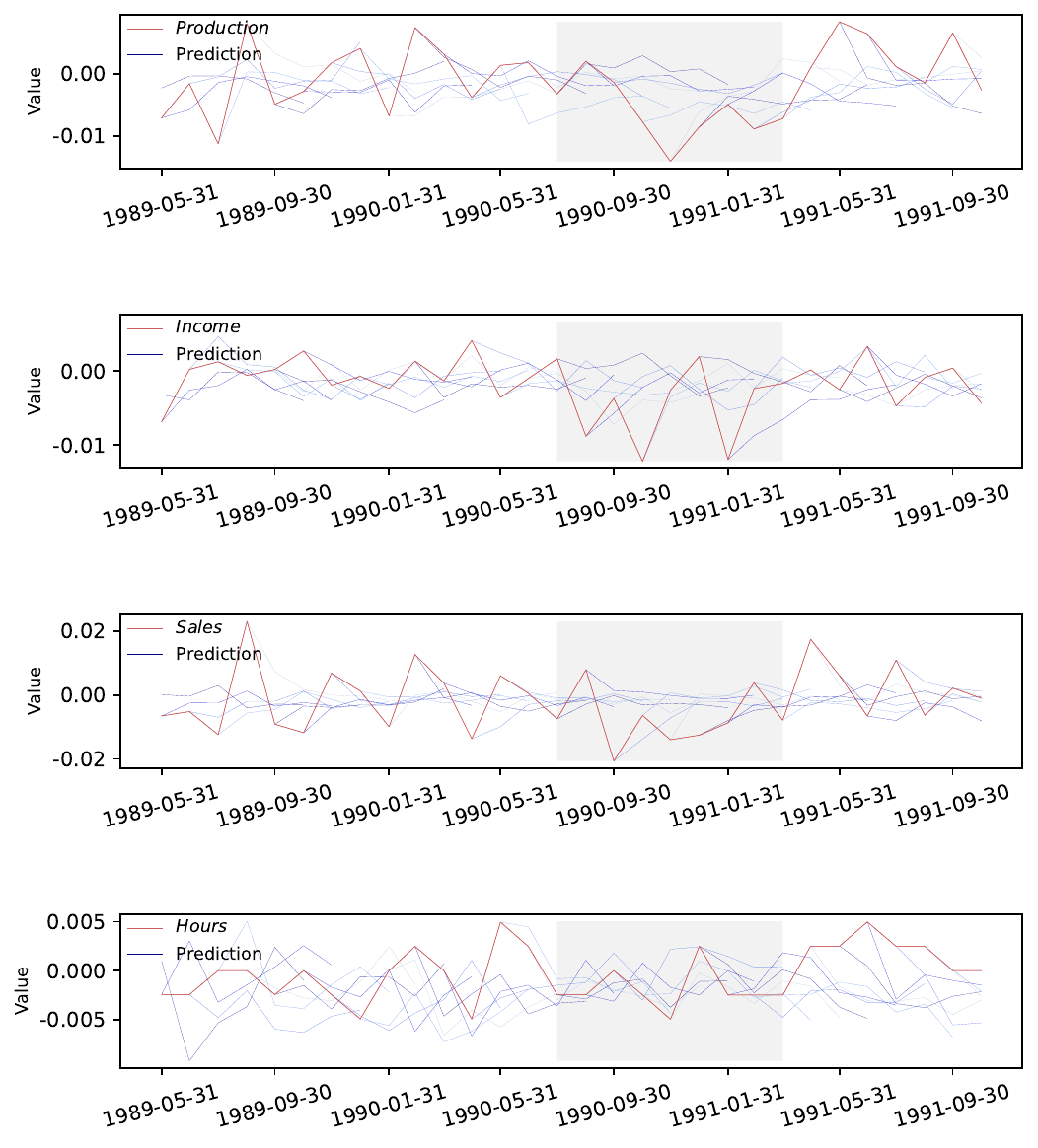}
    \caption{Real-time predictions for the observable variables during the 1990 crisis.}
    \label{fig:tentacle_y_1990}
\end{figure}

\end{document}